\documentclass[fleqn,usenatbib]{mnras}

\usepackage{newtxtext,newtxmath}

\usepackage[T1]{fontenc}
\usepackage{ae,aecompl}

\usepackage{graphicx}	
\usepackage{amsmath}	
\usepackage{amssymb}	
\hypersetup{
    colorlinks,
    citecolor=blue, 
    filecolor=black,
    linkcolor=blue, 
    urlcolor=red
}
\usepackage{geometry}
\usepackage{pdflscape}
\usepackage{caption}
\usepackage{subcaption}
\usepackage{longtable}
\usepackage{array}
\newcolumntype{H}{>{\setbox0=\hbox\bgroup}c<{\egroup}@{}} 

\usepackage{siunitx}
\DeclareSIUnit[number-unit-product = {}] \jansky{Jy}

\title[Search for high-z radio galaxies]{A search for faint high-redshift radio galaxy candidates at 150 MHz}
\author[A. Saxena et al.]{A. Saxena$^1$\thanks{E-mail:
saxena@strw.leidenuniv.nl}, P. Jagannathan$^2$, H. J. A. R{\"o}ttgering$^1$, P. N. Best$^3$, H. T. Intema$^1$,
\newauthor
M. Zhang$^1$, K. J. Duncan$^1$, C. L. Carilli$^{2, 4}$, G. K. Miley$^1$\\ \\
$^1$Leiden Observatory, Leiden University, P.O. Box 9513, 2300 RA Leiden, The Netherlands\\
$^2$National Radio Astronomy Observatory, 1003 Lopezville Road, Socorro, NM 87801-0387, USA\\
$^3$Institute for Astronomy, University of Edinburgh, Royal Observatory, Blackford Hill, Edinburgh EH9 3HJ, UK\\
$^4$Cavendish Astrophysics Group, University of Cambridge, Cambridge CB3 0HE, UK}

\date{Accepted 2018 January 15. Received 2018 January 15; in original form 2017 November 16}

\pubyear{2017}

\DeclareCaptionFormat{cont}{#1 (cont.)#2#3\par}

\begin{document}
\label{firstpage}
\pagerange{\pageref{firstpage}--\pageref{lastpage}}
\maketitle

\begin{abstract}
Ultra-steep spectrum (USS) radio sources are good tracers of powerful radio galaxies at $z > 2$. Identification of even a single bright radio galaxy at $z > 6$ can be used to detect redshifted 21cm absorption due to neutral hydrogen in the intervening IGM. Here we describe a new sample of high-redshift radio galaxy (HzRG) candidates constructed from the TGSS ADR1 survey at 150 MHz. We employ USS selection ($\alpha \le -1.3$) in $\sim10000$ square degrees, in combination with strict size selection and non-detections in all-sky optical and infrared surveys. We apply flux density cuts that probe a unique parameter space in flux density ($50 < S_{\textrm{150}} < 200$ mJy) to build a sample of 32 HzRG candidates. Follow-up Karl G. Jansky Very Large Array (VLA) observations at 1.4 GHz with an average beam size of $1.3$ arcseconds ($''$) revealed $\sim 48\%$ of sources to have a single radio component. P-band (370 MHz) imaging of 17 of these sources revealed a flattening radio SED for ten sources at low frequencies, which is expected from compact HzRGs. Two of our sources lie in fields where deeper multi-wavelength photometry and ancillary radio data are available and for one of these we find a best-fit photo-z of $4.8 \pm 2.0$. The other source has $z_{\textrm{phot}}=1.4 \pm 0.1$ and a small angular size ($3.7''$), which could be associated with an obscured star forming galaxy or with a `dead' elliptical. One USS radio source not part of the HzRG sample but observed with the VLA nonetheless is revealed to be a candidate giant radio galaxy with a host galaxy photo-z of $1.8\pm0.5$, indicating a size of 875 kpc.
\end{abstract}


\section{Introduction} 
\label{sec:introduction}
The discovery of high-redshift radio galaxies (HzRGs) has enabled extensive studies of large scale structure and galaxy evolution. HzRGs are among the most massive galaxies in the universe and are thought to be the progenitors of the massive ellipticals we observe today. HzRGs contain large amounts of dust and gas \citep{bes98b, arc01, car02b, reu04, deb10}, are dominated by an old stellar population \citep{bes98, roc04, sey07} and seen to have high star formation rates \citep{wil03, mcl04, mil06, vil07, sey08}. Likely due to their large stellar masses, HzRGs are often found to be located in the centre of clusters and proto-clusters \citep{pen00, ven02, rot03, mil04, hat11, ors16}. \citet{mil08} summarise the properties of radio galaxies and their environments in their extensive review.

Radio galaxies at $z>6$ can be used to study the epoch of reionisation (EoR) i.e. the epoch at which the universe makes a phase transition from being neutral to ionised. A bright radio source in the EoR provides a perfect background against which absorption by the neutral intergalactic medium can be observed. The 21cm hyper-fine transition line is redshifted into the low-frequency regime at $z\ge6$ ($\nu < 200$ MHz) and the HI absorption features in the continuum of a bright radio source can be observed using current and next-generation telescopes such as the Low Frequency Array\footnote{www.lofar.org, \citep{lofar}} (LOFAR), The Murchison Widefield Array\footnote{www.mwatelescope.org \citep{tin13}} (MWA) and the Square Kilometer Array\footnote{www.skatelescope.org \citep{ska}} \citep{car02, fur02, xu09, mac12, ewa14, cia15}. Detection of even a single bright radio source at $z\ge6$ enables studying the processes responsible for reionsing the universe in unparalleled detail.  

Over the years, the highest redshift of a known radio galaxy has increased gradually. Starting with a detection at redshift of $z=0.45$ \citep{min60}, it took another two decades for the discovery of radio galaxies out to $z \sim 1$ \citep{spi77, spi80}. It was in the 90s that significant progress was made towards pushing the highest known redshift of radio galaxies, resulting in the discovery of the currently highest redshift radio galaxy known, at $z=5.19$ \citep{vbr99}. With state-of-the-art radio telescopes such as the Giant Metrewave Radio Telescope\footnote{http://www.gmrt.ncra.tifr.res.in} (GMRT) and LOFAR, it is now possible to go deeper than ever before at low-frequencies, opening up a new parameter space for HzRG searches.

Most searches for HzRGs have relied on exploiting the ultra-steep spectrum (USS; spectral index, $\alpha < -1.3$, where for frequency $\nu$ and flux density $S_\nu$, $\alpha$ is given by $S_{\nu} \propto \nu^\alpha$) search technique. \citet{tie79} and \citet{blu79} found that optical identification fraction went down significantly for radio sources with steeper spectral indices. This was attributed to USS sources having higher redshifts and it is now generally observed that the spectral index of radio galaxies steepens at higher redshifts \citep[][and references therein]{ker12}. The true physical details of why this happens, however, are still under debate but the widely held view is that the $z-\alpha$ correlation is a result of a concave radio spectrum, i.e. flattening of the spectral index at lower radio frequencies, coupled with a radio K-correction. We refer the reader to \citet{mil08} for a review on the possible causes of spectral index steepening at higher redshifts. Regardless, this technique has proven to be very successful in identifying HzRGs \citep{rot94, cha96, blu99, deb00, afo11}. USS selection, however, does not produce complete samples of HzRGs as the spectral index cut naturally excludes part of the complete sample \citep[see][]{jar09}. Since the goal of this study is to extend the search for radio galaxies to higher redshifts and not necessarily produce complete samples, we still rely on USS selection for the purposes of this paper.

In addition to having an ultra-steep spectrum, HzRGs are also expected to have compact radio morphologies and the statistical decrease of angular sizes of radio sources with redshift has been known for a long time \citep{mil68, nil93, nee95, dal02}. More recently, \citet{mor17} show that the decrease in size with redshift is also observed in deep radio observations at low frequencies. These compact radio sources of a possible high-$z$ nature are often referred to as compact steep spectrum (CSS) sources and are generally a few arcseconds across. It is believed that CSS sources are in early stages of their lifetime, i.e. young \citep{fan95}. \citet{blu99b} showed that there is an `inevitable youthfulness' associated with radio sources at high redshifts in a way that all observable HzRGs must be seen when their lobes are less than $10^7$ years old. Further, \citet{sax17} modelled the compactness of radio galaxies at $z\sim6$ due to the denser cosmic microwave background at higher redshifts leading to increased inverse Compton losses, thereby shortening their lifetimes. Therefore, selecting compact radio sources as potential HzRG candidates is well justified.

Lastly, to optimise the selection of candidate HzRGs from large all-sky surveys, deep optical and infrared datasets over large areas on the sky are ideal. For example, \citet{ker12} showed that a $K$-band cut of 18.5 recovers almost all the currently known high-z radio sources and near infrared selection is very successful at isolating high-z radio galaxies than any selection based on radio properties alone. Therefore, selecting bright radio sources with steep spectral indices and small angular sizes that are also blank in the relatively shallow all-sky optical and/or infrared surveys has the potential effectively detect promising HzRGs candidates \citep{gar08, mid11, ker12}.

Perhaps one of the most well-studied sample of radio galaxies was put together by \citet{sey07} and \citet{deb10}, where they studied a sample of $\sim70$ known high redshift radio galaxies to obtain detailed SEDs over a large wavelength range. Their sample, however, probes only the most powerful radio galaxies ($L_{\textrm{500MHz}} > 10^{27}$ W Hz$^{-1}$). Very little is known about the properties of less powerful radio galaxies as the samples of ultra-steep spectrum radio sources have mostly come from shallow all-sky surveys. In this work we present a sample of newly identified ultra-steep spectrum radio sources at 150 MHz using the new, deeper low-frequency survey carried out using the GMRT called TGSS ADR1 \citep{int17}. Our sample aims to extend the studies of radio galaxies to lower radio powers. Probing this unique parameter space is essential to study the overall evolution of radio galaxies across cosmic time, as well as to reveal any underlying differences between the most and less powerful radio galaxies. Probing fainter radio flux densities is also a way to look for powerful radio galaxies at some of the highest redshifts. 

In Section \ref{sec:datasets} we describe the data sets used to identify targets that probe a unique parameter space at low radio frequencies and describe our target selection methods. In Section \ref{sec:vla} we present our follow-up VLA A-configuration observations at 1.4 GHz and 370 MHz of targets in our final sample. We describe the data reduction and imaging procedures. These observations yield accurate flux densities, morphologies and spectral indices. In Section \ref{sec:mw} we present the multi-wavelength properties of sources in our sample and comment on a few individual sources that happen to lie in famous extragalactic fields and have a wealth of multi-wavelength data available. We discuss classes of radio sources that display an ultra-steep spectral index and that could potentially be present in our sample in Section \ref{sec:discussion}. Finally, we summarise our findings in Section \ref{sec:summary}.

Throughout this paper we assume a flat $\Lambda$CDM cosmology with $H_0 = 67.8$ km/s/Mpc and $\Omega_m = 0.307$. These parameters are taken from the first Planck cosmological data release \citep{pla14}.

\section{Radio data sets and sample selection}
\label{sec:datasets}
\subsection{TIFR GMRT Sky Survey (TGSS) Alternative Data Release at 150 MHz}
The high-redshift radio galaxy (HzRG) candidates presented in this work are selected from the independent reanalysis of archival TIFR GMRT Sky Survey (TGSS) data called the First Alternative Data Release (ADR1; \citealt{int17}). This is a survey of the 150 MHz radio sky using the Giant Metrewave Radio Telescope (GMRT) in India. The observations were centred at a frequency of 147.5 MHz with a bandwidth of 16.7 MHz.

This data release covers 36,900 sq. degrees of the sky between declinations of -53$^\circ$ and +90$^\circ$. Each pointing has an integration time of 15 minutes and covers roughly 7 sq. degrees of the sky. The majority of pointings have a background noise level below 5 mJy/beam and a resolution of 25 arcseconds ($''$). The overall astrometric accuracy is better than 2$''$. 

\subsection{VLA FIRST and NVSS at 1.4 GHz}
The Faint Images of the Radio Sky at Twenty centimetres (FIRST) is a 1.4 GHz survey of the sky covering 10,000 square degrees \citep{Becker1995, White1997}. This survey was carried out using the NRAO Karl G. Jansky Very Large Array (VLA) in B-configuration acquiring 3 minute snapshots using $2\times7$ 3-MHz channels centred around 1365 and 1435 MHz. The FIRST survey has a typical noise level of 0.15 mJy/beam and a resolution of 5$''$. The astrometric reference frame of the radio maps is accurate to $0.05''$ with individual sources having 90\% confidence error circles of 1$''$ down to the survey threshold. There are $\sim90$ sources per sq. degree at the 1 mJy/beam detection threshold.

The NRAO VLA Sky Survey (NVSS) is a continuum survey at 1.4 GHz, which covers the entire sky north of $-40^\circ$ declination \citep{con98}. The survey has a resolution of 45$''$ and a completeness limit of about 2.5 mJy. The larger restoring beam makes this survey more sensitive to extended emission compared to FIRST, where the latter performs better for point sources. Therefore, FIRST and NVSS are highly complimentary to each other. 

\subsection{Target selection}
To select HzRG candidates for this study, we use the region on the sky in TGSS ADR1 that is coincident with the footprint of the VLA FIRST survey at 1.4 GHz covering roughly 10000 sq. degrees. Further, we only include sources above a declination of $+10^\circ$ to aid eventual follow-up observations with LOFAR. These conditions give us an RA and Dec range of approximately $7\textrm{h}<\textrm{RA}<18\textrm{h}$ and $10^\circ<\textrm{Dec}<70^\circ$.
\begin{figure}
	\centering
    \vspace{10pt}
	\includegraphics[scale=0.45]{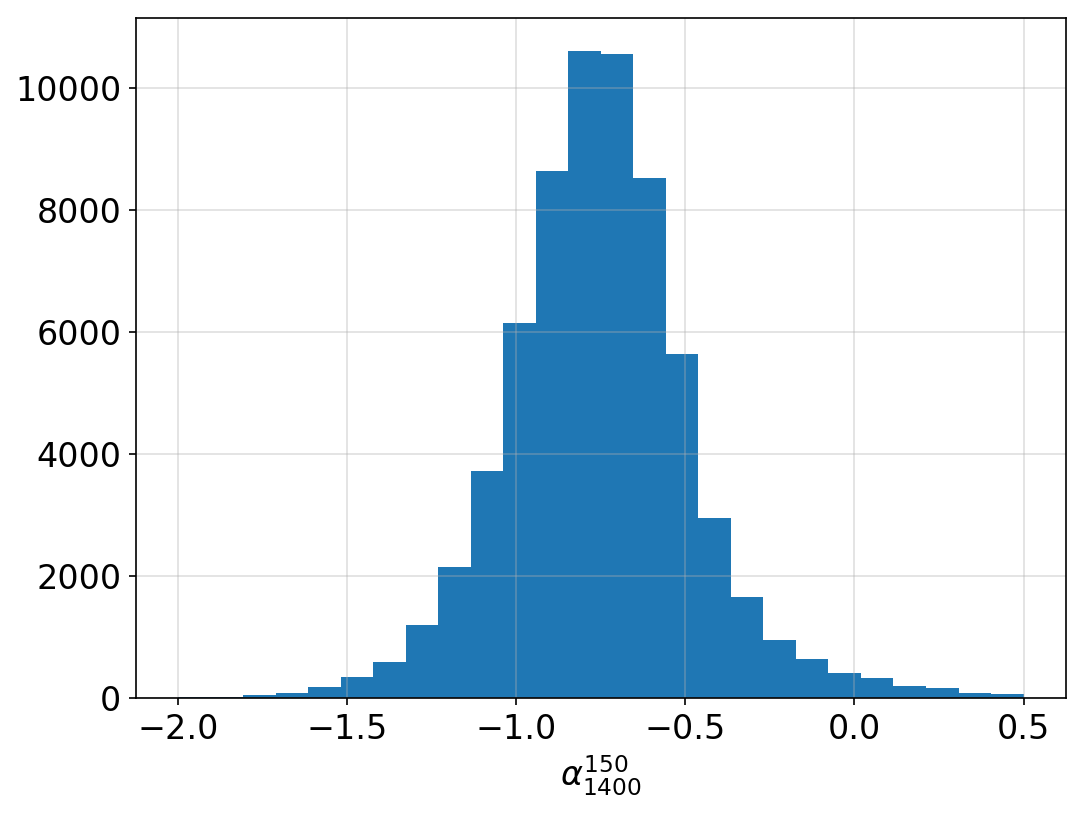}
	\caption{Spectral index distribution of all unresolved TGSS ADR1 sources matched with FIRST. The distribution peaks at $-0.75$, which is consistent with many previous studies of spectral index distribution. For further analysis in this study, we only consider sources with $\alpha^{150}_{1400}<-1.3$.}
	\label{fig:alpha}
\end{figure}

\subsubsection{Angular size restriction}
We introduce a size restriction when selecting potential HzRGs from TGSS ADR1 by selecting only compact, single component radio sources. The angular resolution of ADR1 is 25$''$ but to account for errors, we use a more conservative size restriction by requiring the catalogued major axis to be $<28''$. Next, all selected sources are matched with the FIRST catalogue using the position of the peak pixel in ADR1 and a conservative matching radius of 15$''$, which ensures that the peak pixel of any matched counterpart lies within the TGSS beam size of $\sim25''$. A conservative radius for single component radio sources should overcome the positional uncertainties introduced due to different resolutions of the two surveys. After matching, another size restriction is introduced by only retaining sources that are compact in FIRST, which could include either point sources or compact, multiple component sources. This is done by selecting only sources with a deconvolved major axis of less than 10$''$ in the FIRST catalogue. This leaves us with sources with angular sizes in the range $5''-10''$ in FIRST, in line with the expected HzRG sizes \citep{mil08, sax17}. To ensure robust detections at 1.4 GHz, we only select sources with at least $8\sigma$ detection in FIRST, i.e. sources brighter than 1.2 mJy at 1.4 GHz. This results in a sample of 65,996 sources. Next, we determine spectral indices of all the ADR1-FIRST matched sources between 150 MHz and 1.4 GHz. The spectral index distribution of the matched unresolved sources is shown in Figure \ref{fig:alpha}.

\subsubsection{Ultra-steep spectral index selection}
\begin{figure}
	\centering
    \vspace{10pt}
	\includegraphics[scale=0.43]{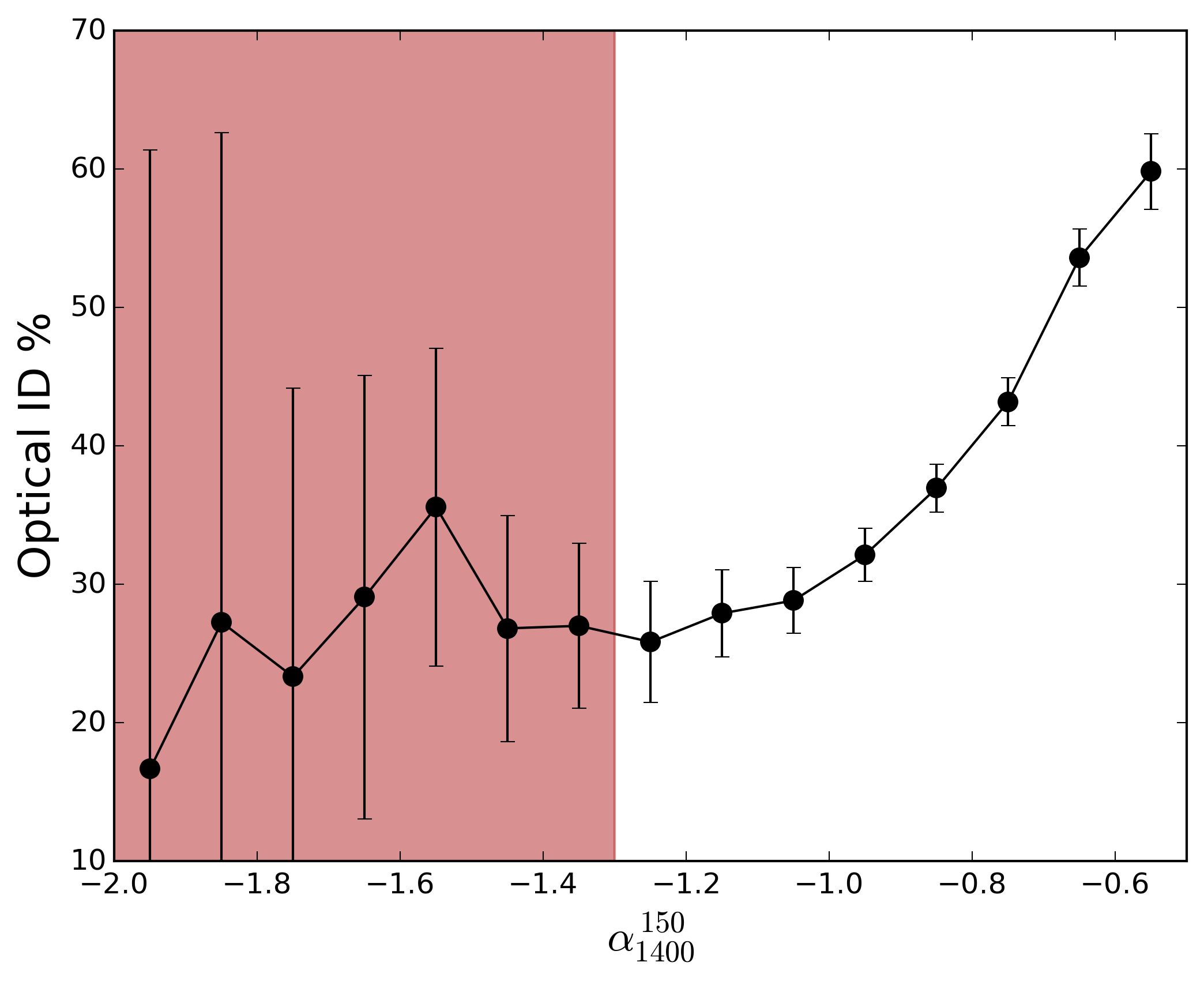}
	\caption{Optical ID fraction (using the band-merged SDSS DR12 catalogue) as a function of radio spectral index with Poissonian uncertainties in each spectral index bin. Clearly the ID fraction for sources with steeper spectral indices is lower. Since SDSS and PS1 probe mainly the low redshift universe, optical identification will be more common for lower redshift galaxies and the decreasing ID fraction for sources with steeper spectral indices can be interpreted as a redshift-spectral index correlation.}
	\label{fig:idfrac}
\end{figure}

Owing to the success of selecting sources with steep spectral indices to identify HzRGS, we employ ultra-steep spectrum (USS) selection in ADR1 by selecting sources with spectral indices between 150 MHz and 1.4 GHz steeper than $-1.3$ ($\alpha^{150}_{1400} < -1.3$). This results in 1564 sources, which is roughly 2\% of the total number of matched, unresolved sources initially part of the sample. To demonstrate that this technique increases the probability of finding high-redshift radio sources, we show the optical ID (SDSS) fraction as a function of spectral index for all TGSS-FIRST matched sources in Figure \ref{fig:idfrac}. Any source detected in FIRST with an optical counterpart within 5$''$ from its peak pixel was assumed to have a potential optical ID. We elaborate on the optical and infrared counterpart identification process in greater detail in the following sections. Clearly, the ID fraction decreases with increasingly steeper spectral indices, suggesting that the hosts of USS sources are fainter than the detection limit in an optical survey complete to a certain flux limit and therefore, are likely to be more distant. Towards the steepest spectral indices ($\alpha < -1.6$) we suffer from low number statistics. The overall trend can be interpreted due to a $z-\alpha$ relation \citep{mil08}.

A point to be noted is that the TGSS ADR1 is prone to small flux density variations on scales of several degrees, owing to the observing strategy implemented \citep{int17}. \citet{deg17} found this variation to bias the spectral indices in certain regions of the sky by values $<0.06$, which was also reported by \citet{hur17}. For the purposes of this study, however, this result does not significantly influence the selection as the spectral indices of sources of interest would be expected to remain ultra-steep even after taking the variation into account. 

\subsubsection{Flux limits -- probing a unique parameter space}
Next, we apply an upper flux limit at \SI{1.4}{\giga\hertz}, such that none of our sources could have been detected in previous large area HzRG searches using the ultra-steep spectrum technique, such as the \citet{deb00} sample that consists of 669 USS sources with $S_{\textrm{1400}} > 10$ mJy. The deepest lowest frequency data in their analysis comes from the Westerbork Northern Sky Survey (WENSS) at \SI{92}{\centi\meter} (\SI{325}{\mega\hertz}; \citealt{ren97}). We select only sources with $S_{1.4} < 10$ mJy in FIRST that ensures that none of these sources would have been detected in WENSS and therefore, were missed by the \citet{deb00} sample. These flux limits result in the flux densities at 150 MHz lying between 50 and 200 mJy, which ensures robust detections in TGSS ADR1. Applying these flux limits resulted in 817 sources remaining in our sample.

The high resolution provided by the FIRST survey is prone to missing extended emission from radio sources, as this emission is `resolved out' from the final image. Therefore, to ensure that the sources that we have selected are truly compact at 1.4 GHz, we look for their counterparts in the lower resolution NVSS survey at the same wavelength. We use a more conservative matching radius of 20$''$ to account for positional offsets of the peak pixel owing to the difference in resolution between the two surveys. The minimum flux density requirement in FIRST ensures that all our sources must also be detected in NVSS (although many of them are very faint). Enforcing a match in NVSS gives the added advantage of eliminating variable sources from our sample, since FIRST and NVSS were carried out at different epochs and having consistent flux densities in both these surveys should dramatically reduce chances of variability. After matching, we compare the integrated flux densities in FIRST and NVSS and keep only sources with comparable flux densities, i.e., $S_{\textrm{FIRST}}/S_{\textrm{NVSS}} = 0.76 - 1.3$, roughly within $1\sigma$ agreement. This further reduced the sample size to 588 sources. 

\subsubsection{Optical and infrared non detections}
As mentioned before, searching for radio sources with no optical and/or faint infrared counterparts greatly assists in isolating high-redshift radio sources. Therefore, we focus on only those sources in the remaining sample that do not have any obvious counterparts in available all-sky optical and infrared surveys. To do this, we look for optical counterparts for radio sources in the Sloan Digital Sky Survey Data Release 12 (SDSS; \citealt{ala15}) band-merged catalogue, which covers the entire FIRST footprint in the \textit{u, g, r, i} and \textit{z} bands and in the PanSTARRS1 survey catalogue (PS1; \citealt{ps1}), providing photometry in \textit{g, r, i, z} and \textit{y} bands over 3$\pi$ steradian of the sky. 

We use the FIRST positions for optical counterpart identification with a strict matching radius of 5$''$, which is the resolution of the FIRST survey. It is important to note that we are not interested in the true optical counterpart for the purposes of this study. Finding a counterpart within a conservative matching radius of $5''$ is a way to reject radio sources that are not blank in the all-sky optical surveys. As the redshift distribution of galaxies in SDSS peaks at around $z=0.3$ and there are very few galaxies at $z>1$ even at the faintest end \citep{she12}, we use optical detections to rule out low redshift radio galaxies and for further analysis, we only select sources with no potential optical counterparts in both SDSS and PS1.

It is possible that intrinsically redder objects at $z>0.3$ may be missed by SDSS and therefore, we match the remaining sources with the ALLWISE survey, which builds upon the work of the \textit{Wide-field Infrared Survey Explorer (WISE}; \citealt{wri10}\textit{)} mission covering the entire sky at wavelengths of 3.4, 4.6, 12 and 22 \SI{}{\micro\meter}. The median photometric redshift for faint red objects with $R<20.0$ peaks at around $z=0.3$ \citep{bil14} and may be slightly higher for even fainter red objects, so we retain only those sources in the sample with no counterparts in ALLWISE either, using a matching radius of $6''$, owing to the lower resolution of ALLWISE. This ensures that none of our radio selected sources have optical or infrared counterparts in the available all-sky surveys and are very likely not low redshift radio galaxies.

\begin{table}
	\centering
	\caption{Number of sources and fraction of the total matched sources between TGSS ADR1 and FIRST at every stage of selection. We start our selection by only considering unresolved sources in TGSS ADR1.}
	\begin{tabular}{l c c}
	\hline
		Selection step						&	No. sources	&	Fraction \\
	\hline
	\hline
		Unresolved TGSS-FIRST matches		&	65,996 		&	100\%	\\
		$\alpha^{150}_{1400} < -1.3$			&	1564			&	2\%		\\
		Flux limits $+$ consistent NVSS		&	588			&	0.9\%	\\
		No SDSS and PS1 $+$ no WISE		&	32			&	0.05\%	\\
	\hline		
	\end{tabular}
	\label{tab:selection}
\end{table}

Final visual inspection of all sources that satisfied the above mentioned criteria was carried out to ensure that they are a) unresolved or barely resolved in FIRST, b) truly compact and c) have no optical or infrared counterparts. At times the FIRST image contained an additional fainter radio component very close to the peak pixel that was reported in the catalogue. In such cases, we measure the total flux density of the radio emission and update the FIRST flux for spectral index calculation. In some cases the FIRST flux was taken out of the $S_{\text{FIRST}}$/$S_{\text{NVSS}} = 0.76-1.3$ selection range, but these sources were nevertheless retained. Such isolated cases, however, did not result in the recalculated spectral indices becoming less steep than -1.3. There was one source in particular (USS422) that had an ultra-steep spectral index from FIRST ($\alpha = -1.8$) but was relatively flat when compared with NVSS ($\alpha = -1.0$). This source was unresolved in both FIRST and NVSS. This meant that this particular source did not strictly adhere to the HzRG selection criteria, but the intriguing nature of radio emission suggesting large scale diffuse emission led us to include this source in the follow-up VLA observations presented in the next sections. 
\begin{figure}
	\centering
    \vspace{-10pt}
	\includegraphics[scale=0.45]{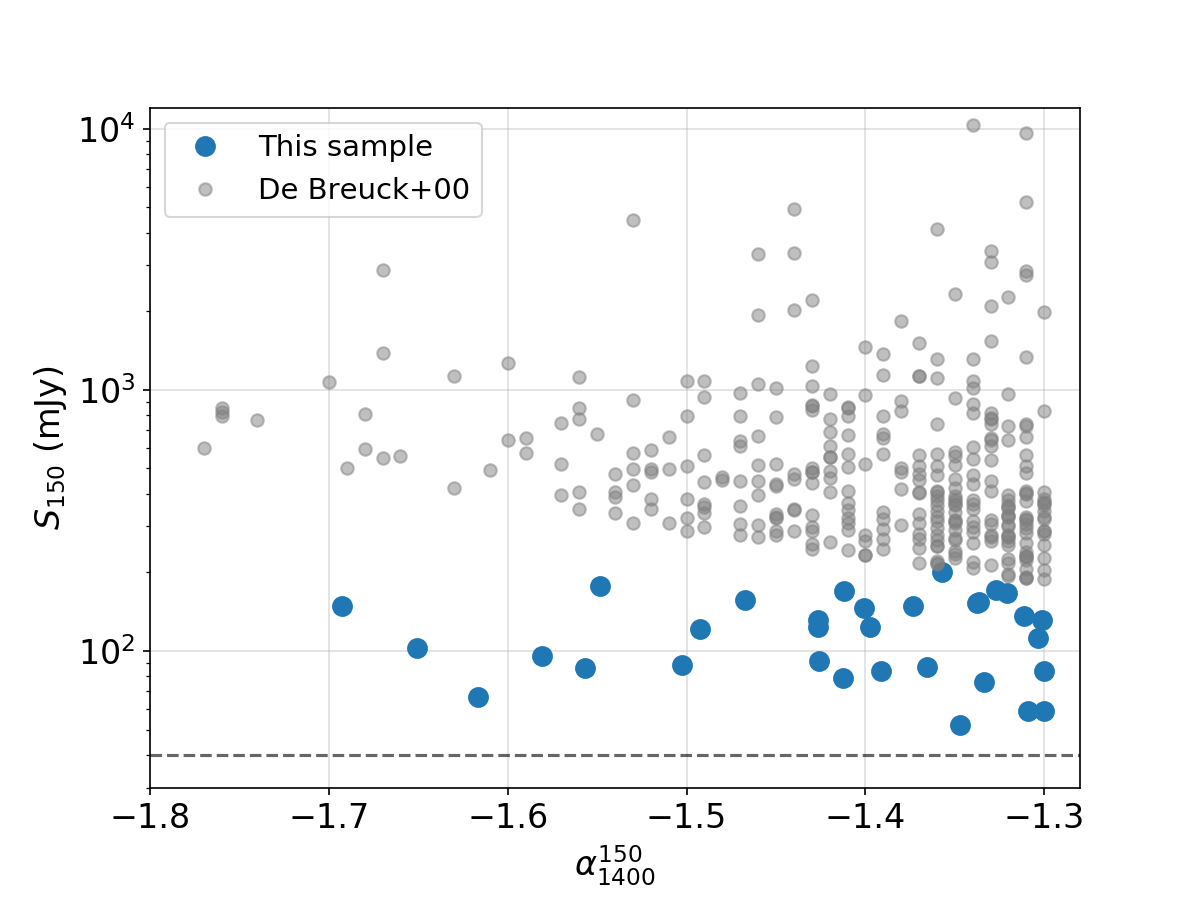}
	\vspace{-10pt}
	\caption{The flux-spectral index parameter space probed by our sample (blue points). Also shown in grey are USS candidates from \citet{deb00} with their flux densities scaled to 150 MHz. The dashed line is the $8\sigma$ detection limit in TGSS ADR1 that we enforced to ensure robust detections. Clearly, our sample goes fainter in flux density and probes a completely new parameter space that has the potential to discover highest redshift radio galaxies.}
	\vspace{-5pt}
	\label{fig:fluxspecindex}
\end{figure}

The strict selection criteria resulted in 32 promising HzRG candidates.  In Figure \ref{fig:fluxspecindex} our sample probes a completely new parameter space aimed at detecting radio galaxies in or very close to the epoch of reionisation. The grey points are sources from \citet{deb00}, where we have used the reported spectral index for each source between 325 and 1400 \SI{}{\mega\hertz} to scale their flux densities to 150 MHz. We show the fraction of the initial sample remaining after each selection step in Table \ref{tab:selection}.

The convention for working names used for sources in this study is of the form `USSXY' where XY is the identification number that was assigned to radio sources in an early version of the TGSS ADR1 source catalogue. We have chosen to continue using these working names for all practical purposes in this paper, but also give the official ADR1 catalogue names in the source table presented. The identification numbers are not necessarily correlated with any other property of the sources.

\section{VLA L and P-band observations}
\label{sec:vla}
\subsection{Objectives}
The primary objective of follow-up VLA observations (Project ID: VLA/16B-309) in A-configuration was to obtain accurate radio positions and morphologies (L-band) and look for signs of spectral curvature (P-band) for sources in our final sample. With an average resolution of $1.3''$ provided by the L-band (1.4 GHz) it is possible to get sub-arcsecond positional accuracy for the identification of the galaxy hosting the radio source. This further enables blind spectroscopy at the radio position without requiring the need to identify a counterpart in deep optical/IR images. Due to constraints on target visibility, we observed 31 out of 32 HzRG candidates and the potential diffuse source (USS422) at 1.4 GHz. The one source that was left out lies in the equatorial strip that is part of the FIRST Southern Sky Coverage and could not be accommodated in the observing blocks containing sources from the Northern Sky Coverage. Owing to time and visibility constraints, we could only observed 17 sources in P-band (370 MHz).

\subsection{Observations and data reduction}
The observations were carried out in A-configuration. The sample in L-band was observed in two blocks of 4 hours each, separated by 2 hours in LST to optimise UV coverage for scans on each source. The typical on-source time was 5 minutes in each observing block, giving a total of 10 minutes of exposure per source. The first block was observed on 28 November 2016 and the second on 28 December 2016. The observations were taken in standard L-band configuration using 27 antennas with 1024 MHz bandwidth (16 sub-bands $\times$ 64 MHz). The flux densities were calibrated using the primary VLA calibrator 3C286 observed in the middle of the observing block. Groups of 3 or 4 sources in RA were created and a suitable phase calibrator was identified for each RA group.

The P-band observations were carried out on 25 January 2017. These were taken in standard configuration with a bandwidth of 200 MHz. The exposures were much shorter, with a single 70 second on-source exposure per source. This short exposure should be enough to detect all targets with a SNR > 10, assuming the spectral index remains more or less constant at intermediate frequencies between 150 and 1400 MHz. 

Initial calibration and data reduction was carried out using the Common Astronomy Software Applications (CASA) package. The VLA data reduction pipeline provided by NRAO\footnote{available at \\ https://science.nrao.edu/facilities/vla/data-processing/pipeline/} was used to flag and calibrate the measurement sets. The CASA command \textit{rflag} was implemented on the calibrated datasets to flag most of the remaining radio frequency interference (RFI). Final images were produced using the \textit{clean} command in CASA, using either uniform or `robust' weighting scheme. For L-band images we used a cell size varying between $0.25-0.4''$ (to ensure that at least 3-5 pixels fall within the beam size). For P-band the chosen cell size was $1.2''$. The final images have an average resolution of $1.3''$ in L-band and $5.6''$ in P-band. The images have an average rms level of \SI{60}{\micro\jansky}/beam in L-band and \SI{1.8}{\milli\jansky}/beam in P-band, which is close to the expected theoretical noise levels. The ionosphere for P-band observations was fairly stable and ionospheric TEC corrections were applied using the \textit{gencal} task in CASA.

\subsection{Flux densities and morphologies}
Flux densities and angular sizes of all sources imaged were then measured using PyBDSF\footnote{http://www.astron.nl/citt/pybdsf/index.html} (the Python Blob Detector and Source Finder), which is a tool designed to identify sources in a radio image and extract meaningful properties. We use a threshold of 5$\sigma$ to identify `islands' of pixels surrounding the peak pixel. These identified pixels are then fitted with Gaussian components to determine flux densities, positions and de-convolved angular sizes. More details about the source finding algorithm can be found in the PyBDSF documentation.

\begin{figure}
	\centering
    \vspace{-10pt}
    \includegraphics[scale=0.45]{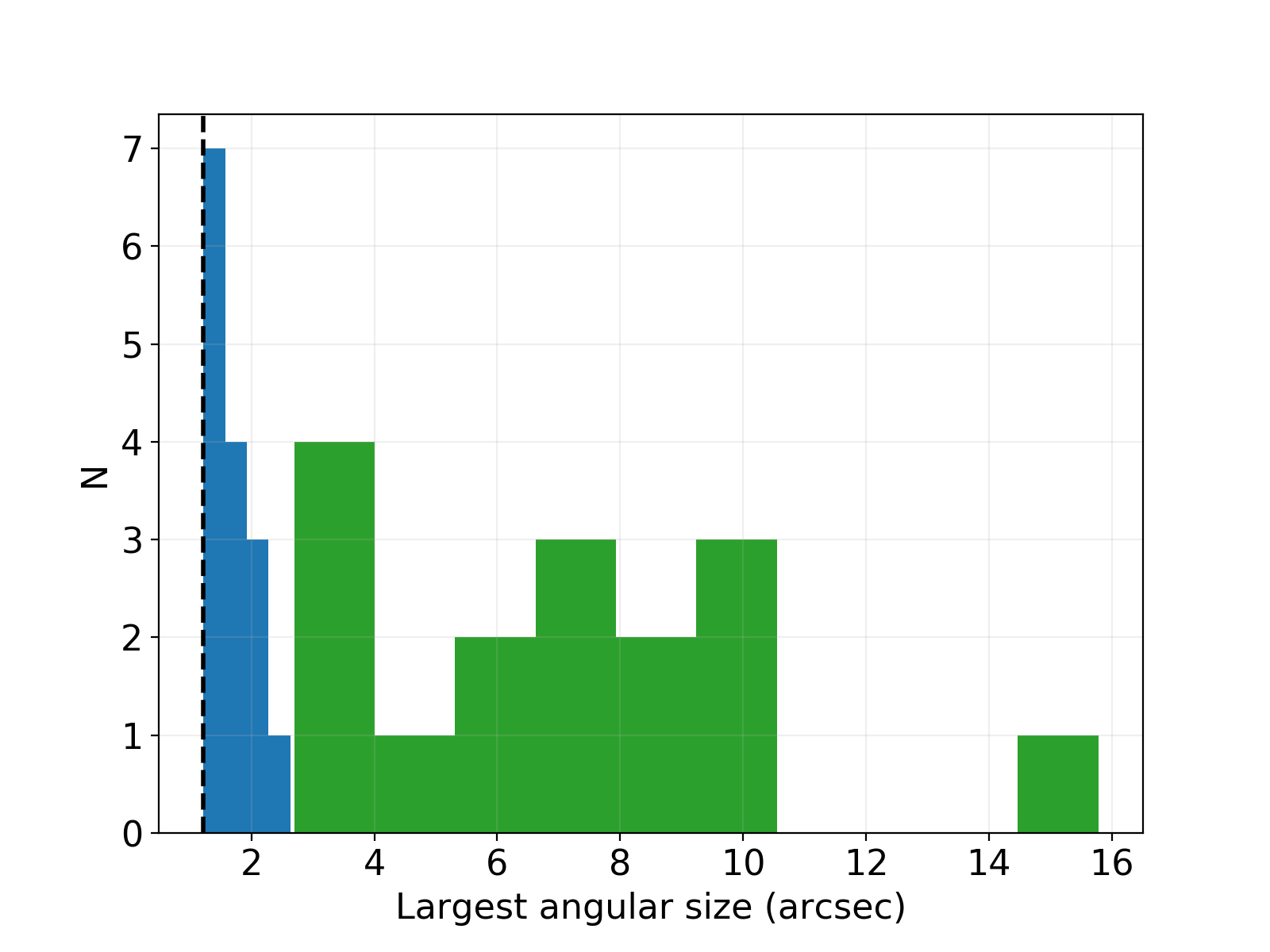}
	\caption{Largest angular size (LAS) distribution determined from our VLA observations of targets in the final sample with an average beam size of 1.3$''$. We find that 15 out of the 31 candidate HzRGs observed are fitted with a single component (blue). There is a wide range of LAS values for sources fit with multiple components (green).}
	\label{fig:sizedist}
\end{figure}

In Table \ref{tab:vlaflux} we report the source properties derived from the L-band images using PyBDSF with the above mentioned parameters. For sources that are fitted with multiple gaussian components by PyBDSF, we report the flux density of each component separately. Angular sizes of sources with multiple components are determined by measuring the separation between the peak positions of the components. The distribution of angular sizes is shown in Figure \ref{fig:sizedist}. We note that the angular sizes of a large majority of our sources are compact (LAS $\le 10''$, as expected from our selection criteria)  and in line with predictions for HzRGs by \citet{sax17} assuming $z>2$, which is promising. We find that 15 of the observed 32 sources (57\% of the sample) are fitted with only one Gaussian component by PyBDSF. A further 14 sources (41\%) were identified to have two components, one source with three components and another source with four components. For the source USS422, the wide-field image revealed the presence of a candidate giant radio galaxy (GRG). Only one hotspot of this candidate GRG was included in our sample. The angular separation between the two lobes of this galaxy is $\sim 1.7$ arcminute ($'$). We discuss the properties of this source in Section 4. 

\begin{figure}
	\centering
	\vspace{10pt}
    \includegraphics[scale=0.44]{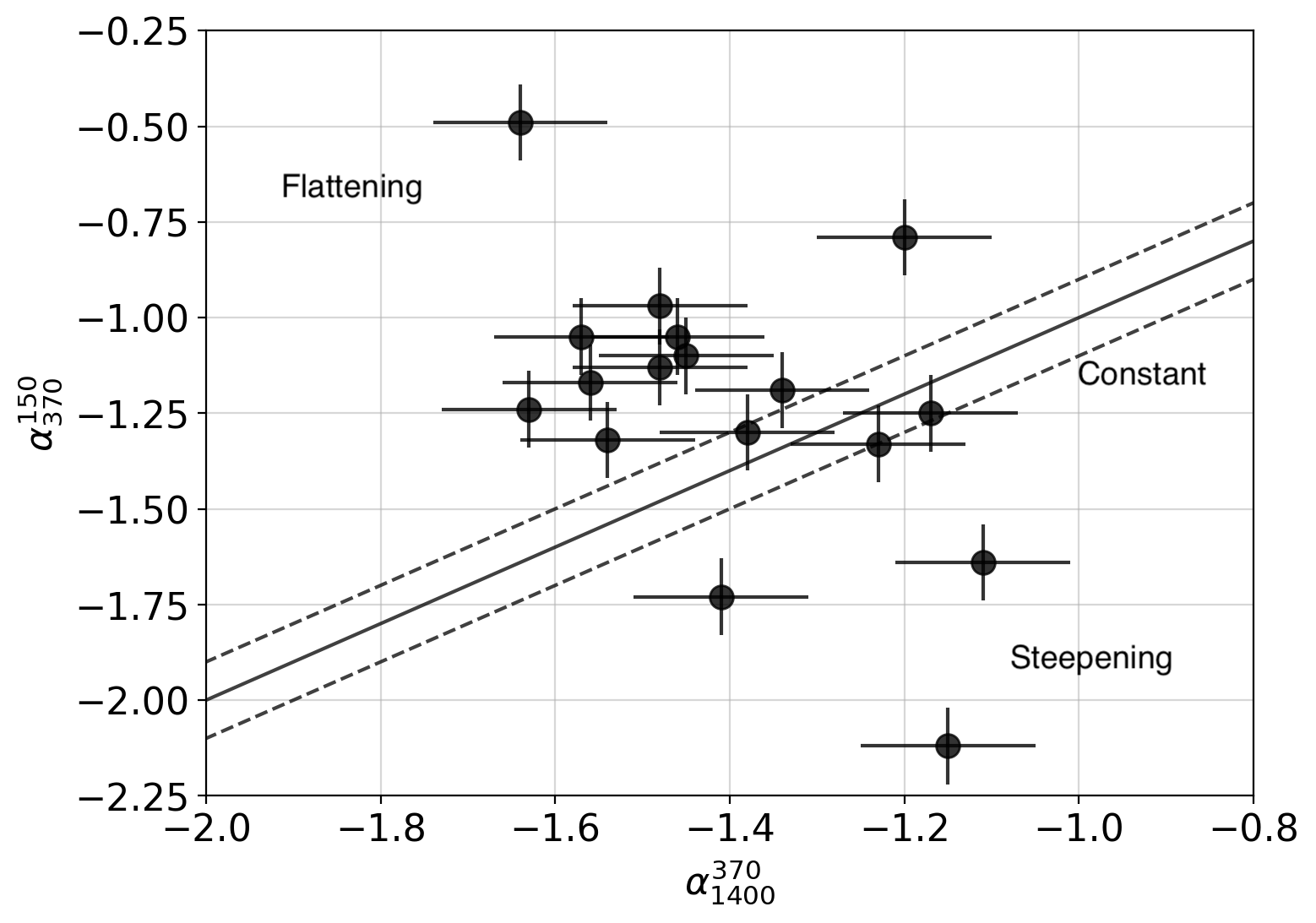}
	\caption{A radio colour-colour plot, showing spectral indices from 370 to 1400 \SI{}{\mega\hertz} on x-axis and 150 to 370 \SI{}{\mega\hertz} on y-axis. The solid line marks constant spectral index at both radio colours, with the dashed lines representing errors of 0.1. We mark the regions where we observe `Flattening', `Steepening' and `Constant' spectral indices for sources observed in both L- and P-bands.}
	\label{fig:radiocolours}
\end{figure}

We report flux densities measured at 370 MHz for sources that were also observed in the P-band in Table \ref{tab:pband}. We then calculate the spectral indices and investigate the shape of the radio spectra constructed using data at frequencies of 150, 370 and 1400 MHz. The resulting radio `colour-colour' plot is shown in Figure \ref{fig:radiocolours}. We find evidence of spectral flattening at lower frequencies in the spectra of 10 out of 17 sources, which is expected from smaller sized HzRGs as they are younger and expand into a denser environment \citep{cal17}. Three sources show a steepening of their spectrum at low frequencies. The radio SED of four sources has a constant spectral index from 150 MHz all the way to 1.4 GHz, within the uncertainties.

\begin{table*}
\centering
\footnotesize
\caption{Flux densities and spectral indices for the subset of the sample also observed at 370 MHz (VLA P-band). The errors on the spectral index determination have been rounded off to the closest one decimal place value. In the last column we comment on the type of radio SED displayed by each source, where `Steepening' indicates a steeper spectral index at low frequencies, `Flattening' indicates flatter spectral index at low frequencies and `Constant' indicats a constantly steep spectral index, within the uncertainties. The sources are listed in order of Right Ascension in this table and the full source table with more details in shown in Appendix 1 (Table \ref{tab:vlaflux}).}
\begin{tabular}{c c c c c c c}
\hline
  \multicolumn{1}{c}{Name} &
  \multicolumn{1}{c}{S$_{\textrm{150}}$ (mJy)} &
  \multicolumn{1}{c}{S$_{\textrm{370}}^{\textrm{tot}}$ (mJy)} &
  \multicolumn{1}{c}{S$_{\textrm{1.4}}^{\textrm{tot}}$ (mJy)} &
  \multicolumn{1}{c}{$\alpha_{\textrm{370}}^{\textrm{150}}$} &
  \multicolumn{1}{c}{$\alpha_{\textrm{1.4}}^{\textrm{370}}$} &
  \multicolumn{1}{c}{Spectral type} \\
\hline \hline

USS206 & 84 $\pm$ 16 & 65.6 $\pm$ 4.0  & 7.3 $\pm$ 0.1 & $-0.49$ $\pm$ 0.1 & $-1.64$ $\pm$ 0.1 & Flattening \\
USS182 & 146 $\pm$ 30 & 38.4 $\pm$ 4.6 & 8.5 $\pm$ 0.1 & $-1.33$ $\pm$ 0.1 & $-1.23$ $\pm$ 0.1 & Constant\\
USS31 & 149 $\pm$ 30 & 35.4 $\pm$ 2.3 & 4.8 $\pm$ 0.1 & $-1.73$ $\pm$ 0.1 & $-1.41$ $\pm$ 0.1 & Steepening \\
USS188 & 166 $\pm$ 33 & 65.0 $\pm$ 2.5 & 9.2 $\pm$ 0.1 & $-1.05$ $\pm$ 0.1 & $-1.57$ $\pm$ 0.1 & Flattening \\
USS309 & 83 $\pm$ 17 & 41.3 $\pm$ 3.7  & 8.4 $\pm$ 0.1 & $-0.79$ $\pm$ 0.1 & $-1.20$ $\pm$ 0.1 & Flattening \\
USS27 & 111 $\pm$ 22 & 37.8 $\pm$ 3.1 & 7.6 $\pm$ 0.1 & $-1.25$ $\pm$ 0.1 & $-1.17$ $\pm$ 0.1 & Constant \\
USS36 & 123 $\pm$ 25 & 37.9 $\pm$ 3.3 & 6.0 $\pm$ 0.1 & $-1.30$ $\pm$ 0.1 & $-1.38$ $\pm$ 0.1 & Constant \\
USS273 & 154 $\pm$ 31 & 61.1 $\pm$ 3.1 & 7.4 $\pm$ 0.1 & $-1.05$ $\pm$ 0.1 & $-1.57$ $\pm$ 0.1 & Flattening  \\
USS483 & 89 $\pm$ 18 & 29.2 $\pm$ 2.8 & 3.3 $\pm$ 0.1 & $-1.24$ $\pm$ 0.1 & $-1.63$ $\pm$ 0.1 & Flattening \\ 
USS46 & 87 $\pm$ 17 & 35.6 $\pm$ 2.5 & 4.9 $\pm$ 0.1 & $-0.97$ $\pm$ 0.1 & $-1.48$ $\pm$ 0.1 & Flattening \\
USS410 & 96 $\pm$ 19 & 31.6 $\pm$ 2.6 & 3.7 $\pm$ 0.1 & $-1.32$ $\pm$ 0.1 & $-1.54$ $\pm$ 0.1 & Flattening \\
USS18 & 131 $\pm$ 26 & 46.1 $\pm$ 2.4 & 5.7 $\pm$ 0.1 & $-1.17$ $\pm$ 0.1 & $-1.56$ $\pm$ 0.1 & Flattening \\  
USS172 & 124 $\pm$ 25 & 42.0 $\pm$ 4.0 & 7.1 $\pm$ 0.1 & $-1.19$ $\pm$ 0.1 & $-1.34$ $\pm$ 0.1 & Constant \\
USS159 & 149 $\pm$ 30 & 56.4 $\pm$ 2.8 & 8.0 $\pm$ 0.1 & $-1.10$ $\pm$ 0.1 & $-1.45$ $\pm$ 0.1 & Flattening \\ 
USS253 & 156 $\pm$ 31 & 35.8 $\pm$ 3.0 & 8.1 $\pm$ 0.1 & $-1.64$ $\pm$ 0.1 & $-1.11$ $\pm$ 0.1 & Steepening \\ 
USS312 & 79 $\pm$ 16 & 11.6 $\pm$ 2.5 & 2.5 $\pm$ 0.1 & $-2.12$ $\pm$ 0.1 & $-1.15$ $\pm$ 0.1 & Steepening \\
USS268 & 92 $\pm$ 18 & 33.3 $\pm$ 2.2 & 4.6 $\pm$ 0.1 & $-1.13$ $\pm$ 0.1 & $-1.48$ $\pm$ 0.1 & Flattening \\ 

\hline\end{tabular}
\label{tab:pband}
\end{table*}

We show contour maps of all sources imaged with the VLA in Appendix B. The contours begin at 0.25 mJy and are a geometric progression of $\sqrt{2}$, i.e. for every two contours the flux increases by a factor of 2. The VLA beam used to image each source is shown in the bottom-left part of the contour map.

\section{Multi-wavelength properties and interesting sources}
\label{sec:mw}

\subsection{Optical and IR photometry}
To gain a better understanding of the average properties of galaxies in our sample, we stack the optical and infrared photometry from publicly available surveys. Stacking analysis works best for galaxies with similar properties and since our galaxies are expected to be distributed over a range of redshifts, stacking photometry and photometric redshift determination may not yield a very accurate description of our sample. However, in order to gain a broad idea of the potential redshift space that our sample is probing, we obtain a photometric redshift using stacking anyway.

To do this, we first create postage stamp cutouts in available optical and infrared bands for all 32 sources in our sample. These cutouts are centred on the expected position of the optical counterpart, which was the flux-weighted peak pixel position of the radio emission for single component sources and the position of the pixel that was most likely to represent the expected position of the host galaxy chosen through visual analysis for radio sources with two (or more) components. We then perform sigma-clipped stacking of all the cutouts in each optical and infrared band in the following way. We calculate sigma-clipped statistics, where pixels above 5-sigma were discarded after the initial calculation of the mean and median values for each image. We then mask the pixels that are brighter than $10\sigma$ (where $\sigma$ is the standard deviation obtained from sigma-clipped statistics) and then stack the images. We ensure that the pixels containing the expected position of the host galaxy are never masked in this process.

Aperture photometry is then performed using \textsc{photutils}\footnote{https://photutils.readthedocs.io/en/stable/index.html} on the central pixels of the stacked images using an aperture with diameter 2$''$. This aperture size should take care of any uncertainty introduced when selecting the expected position of the optical host galaxy. Further, we place 100 apertures of 2$''$ on the stacked image to obtain an estimate of the background flux or noise. When the central aperture flux is lower or equal to the noise estimated from random apertures, we report the 3$\sigma$ limit on the non-detection. We elect to use PanSTARRS1 (PS1; \citealt{cha16}) for redder optical bands (\textit{i, z, y}) as they are deeper than SDSS. We then perform aperture corrected photometry in the \textit{WISE} bands \textit{W1} and \textit{W2} using data from the ALLWISE release. The point-spread function of \textit{WISE} is roughly 6$''$ in \textit{W1} and \textit{W2} bands and we use an aperture of size 7.2$''$ for both these bands. The bands \textit{W3} and \textit{W4} are shallow and we do not see any evidence of detections after stacking. Therefore, we do not include them in this analysis. The magnitudes, both limits and detections, determined from the stacking analysis in the SDSS, PS1 and ALLWISE bands are shown in Table \ref{tab:hizesp_stacked}.
\begin{table}
\centering
\caption{Stacked photometry (in AB mags) of the final USS sample from SDSS, PS1 and ALLWISE surveys. The lower limits indicate 3$\sigma$ limits on the background flux for bands where the aperture flux is lower than or equal to the background flux.}
\begin{tabular}{c c}
	\hline
    Filter & Magnitude (AB) \\
    \hline 
    \hline
    SDSS $u$ & $>23.15$ \\
    SDSS $g$ & $>23.88$ \\
    SDSS $r$ & $>23.15$ \\
    PS1 $i$ & $22.8 \pm 0.2$ \\
    PS1 $z$ & $>22.6$ \\
    PS1 $y$ & $>21.5$ \\
    WISE $W1$ & $19.56 \pm 0.2$ \\
    WISE $W2$ & $19.05 \pm 0.2$\\
    \hline
\end{tabular}
\label{tab:hizesp_stacked}
\end{table}
    
We then convert the AB magnitudes to fluxes and derive a photometric redshift by fitting the \citet{bro14} SED template using \textsc{eazy} \citep{bra08}. The redshift grid used ranges from 1.5 to 6.5 with a step size of 0.05. After applying a Bayesian apparent magnitude prior, the best-fit SED gives a photometric redshift of 2.95, with one-sigma confidence levels ranging from $1.0-5.5$. We note that $z>6$ fits are within the 3-sigma confidence levels. We also find that fitting other SED templates using \textsc{eazy} gives us similar photo-z estimates.

\subsection{Comments on interesting sources in the sample}
In this section we explore the possible nature and properties of a few sources that happen to lie in well-studied extragalactic fields, resulting in the availability of ancillary data at different wavelengths.

\subsubsection{USS410 in Lockman Hole field}
\begin{figure}
	\centering
	\vspace{-10pt}
	\includegraphics[scale=0.45]{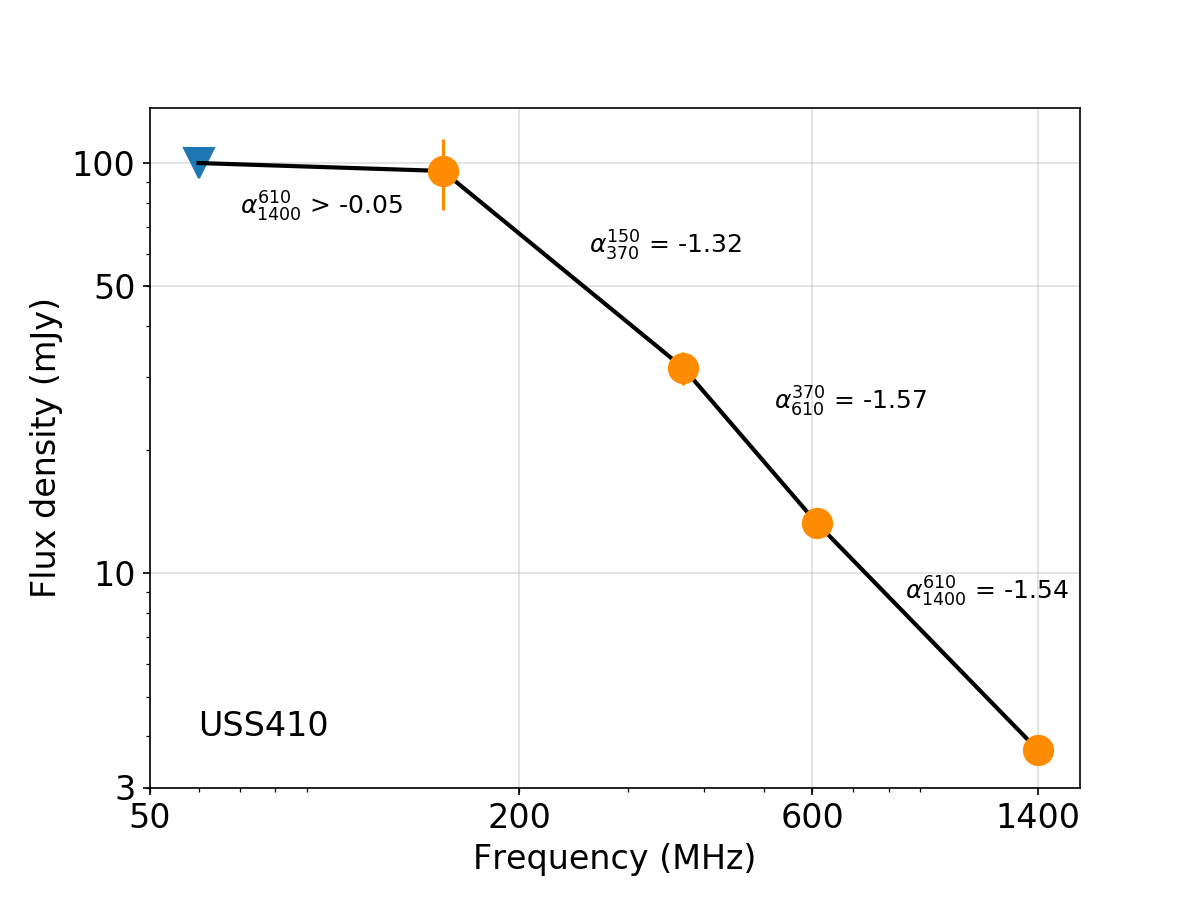}
    	\caption{The radio SED for USS410 from 60 MHz to 1400 MHz. Also shown are the spectral indices measured between data available for this source at various radio frequencies. The error bars on the higher frequency data points are smaller than the symbol size, and therefore may not be clearly visible. There is a clear spectral turnover between 60 \citep{mah16} and 150 MHz, whereas the spectral index remains consistently ultra-steep from 150 to 610 MHz \citep{gar10}, and from 610 to 1400 MHz. The potential high-z nature of the host galaxy (Figure \ref{fig:uss410_photoz}) makes USS410 a candidate peaked spectrum source at high redshift.}
    	\label{fig:uss410_radiosed}
\end{figure}
\begin{figure}
	\centering
    \vspace{-10pt}
    \includegraphics[scale=0.45]{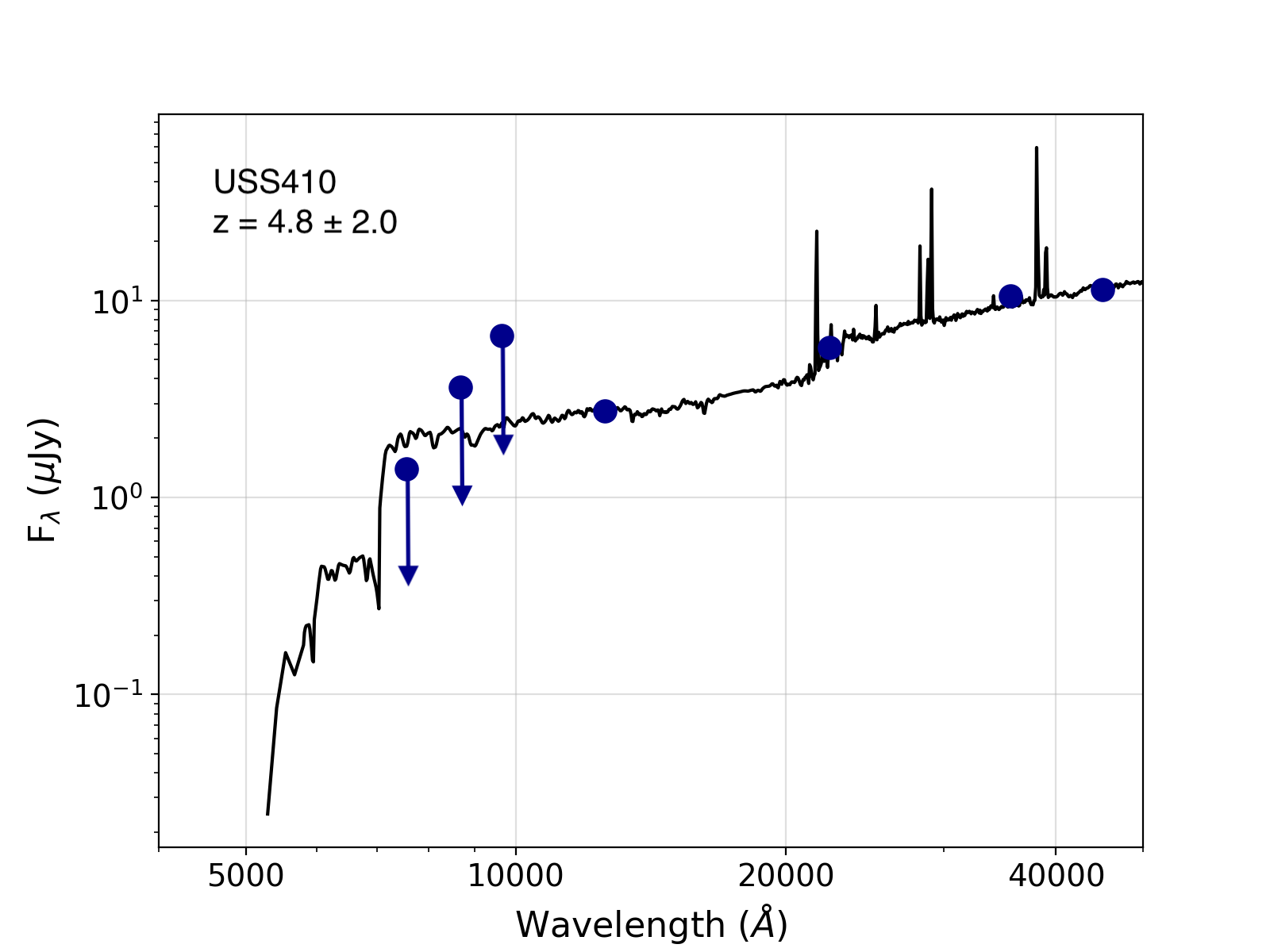}
    \caption{Best-fit SED (observed frame) with a photometric redshift of $z=4.8 \pm 2.0$ obtained for USS410 in the Lockman Hole field. Lower redshift solutions cannot formally be ruled out yet. SED templates by \citet{bro14} were used for the fitting. The photometry is obtained from upper limits from PS1 ($i, z, y$) and detections from UKIDSS DXS ($J, K$) and \textit{Spitzer} IRAC 3.6 and 4.5 \SI{}{\micro\meter}. The error bars on the detections are smaller than the symbol size and therefore, not visible. Fits resulting from excluding upper limits yield the same photometric redshift.}
    \label{fig:uss410_photoz}
\end{figure}
One source in our sample, USS410, lies in the Lockman Hole field. With the availability of deep LOFAR data at 150 MHz in this region \citep{mah16}, we compare the TGSS ADR1 and LOFAR 150 MHz flux densities for this source and find them to be within 5\% of each other. Additionally, LOFAR LBA (60 MHz) observations in this region did not detect this source down to a $5\sigma$ flux density limit of 100 mJy. This places its low-frequency spectral index at $\alpha_{150}^{60} > -0.05$, indicating a sharp spectral turnover at low frequencies \citep{mah16} and suggesting signs of synchrotron self-absorption or free-free absorption \citep[][and references therein]{cal15}. Deep radio observations are also available at 610 MHz \citep{gar10} with a resolution of $6''\times5''$ and USS410 has a detection with a flux density of $13.26\pm0.23$ mJy. With the addition of L- and P- band data that we obtained, we are able to constrain well the radio SED of this particular source. We show the SED from 60 MHz all the way to 1.4 GHz, indicating measured spectral indices between all available frequencies in Figure \ref{fig:uss410_radiosed}. Clearly, USS410 remains an ultra-steep spectrum source from 150 to 1400 MHz and undergoing a dramatic turnover at frequencies between 60 and 150 MHz. 

\begin{figure*}
	\centering
	\includegraphics[scale=0.145]{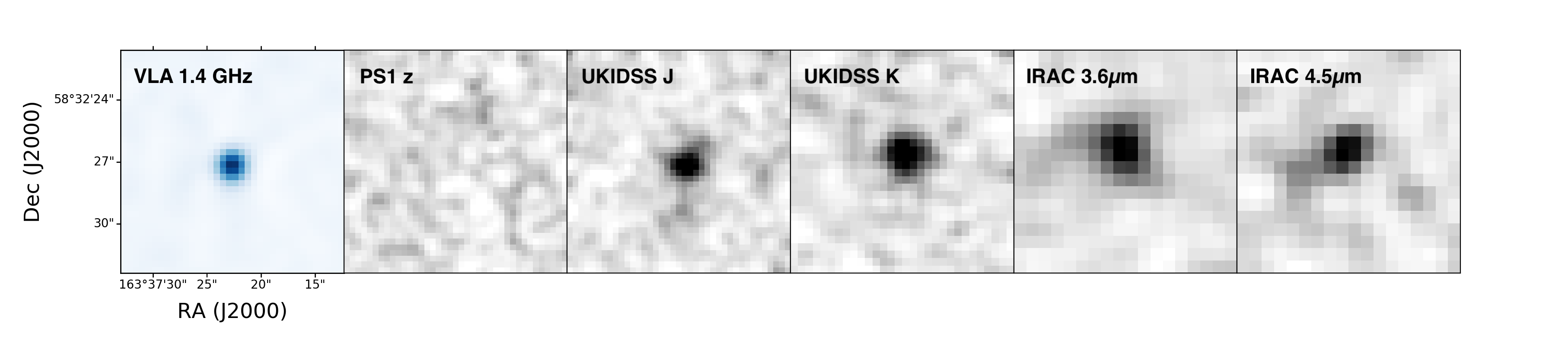}
	\caption{VLA 1.4 GHz images of the Lockman Hole source USS410, with PS1 $z$, UKIDSS $J$ and $K$ and IRAC 3.6 and 4.5 \SI{}{\micro\meter} bands. USS410 remains unresolved even with a beam size of 1.1$''$. There is no detection in PS1 $z$ band down to 22.3 AB. The best-fit photometric redshift is calculated to be $4.8 \pm 2.0$ (Figure \ref{fig:uss410_photoz}). Further, following the $K-z$ relation for radio galaxies \citep{wil03}, the $K$-band magnitude is consistent with a $z\ge4$ host galaxy}
	\label{fig:uss410_mw}
\end{figure*}

USS410 was detected in the UKIDSS Deep Extragalactic Survey (DXS) carried out using the UKIRT telescope \citep{dye06}, with faint detections in both $J$ and $K$ bands with magnitudes of 22.8 $\pm$ 0.1 and 22.0 $\pm$ 0.1 AB, respectively. There are also faint detections in the Spitzer IRAC 3.6 and 4.5 mircon bands from the SWIRE coverage in this field \citep{sur05}. The 3.6 micron flux density is $10.5 \pm 0.6$ $\mu$Jy and 4.5 micron flux density is $11.3 \pm 0.8$ $\mu$Jy. Fitting an SED to non-detections in the PS1 \textit{i,z,y} bands and detections in $J$, $K$, IRAC1 and IRAC2 bands, we derive a photometric redshift of $z_{\textrm{phot}} = 4.8 \pm 2.0$, although solutions at lower redshifts and at $z>5$ are not formally excluded (Figure \ref{fig:uss410_photoz}). Further, following the tight correlation that exists between the $K$ magnitude and redshift of high redshift radio galaxies (the $K-z$ relation; \citealt{lil84, jar01, wil03}), the redshift of this particular radio galaxy is consistent with $z>4$.

USS410 is unresolved in the VLA image, constraining the size to less than 8 kpc. Further, the turnover of the radio SED between 60 and 150 MHz makes it a possible megahertz-peaked spectrum (MPS) source at high-z. Follow-up spectroscopy is essential to reveal the nature of the host galaxy. The 1.4 GHz VLA image along with PS1 $z$ UKIDSS $J$ and $K$ and the two IRAC bands are shown in Figure \ref{fig:uss410_mw}.

\subsubsection{USS7 in ELAIS-N1 field}
\begin{figure}
	\centering
    \vspace{-10pt}
    \includegraphics[scale=0.45]{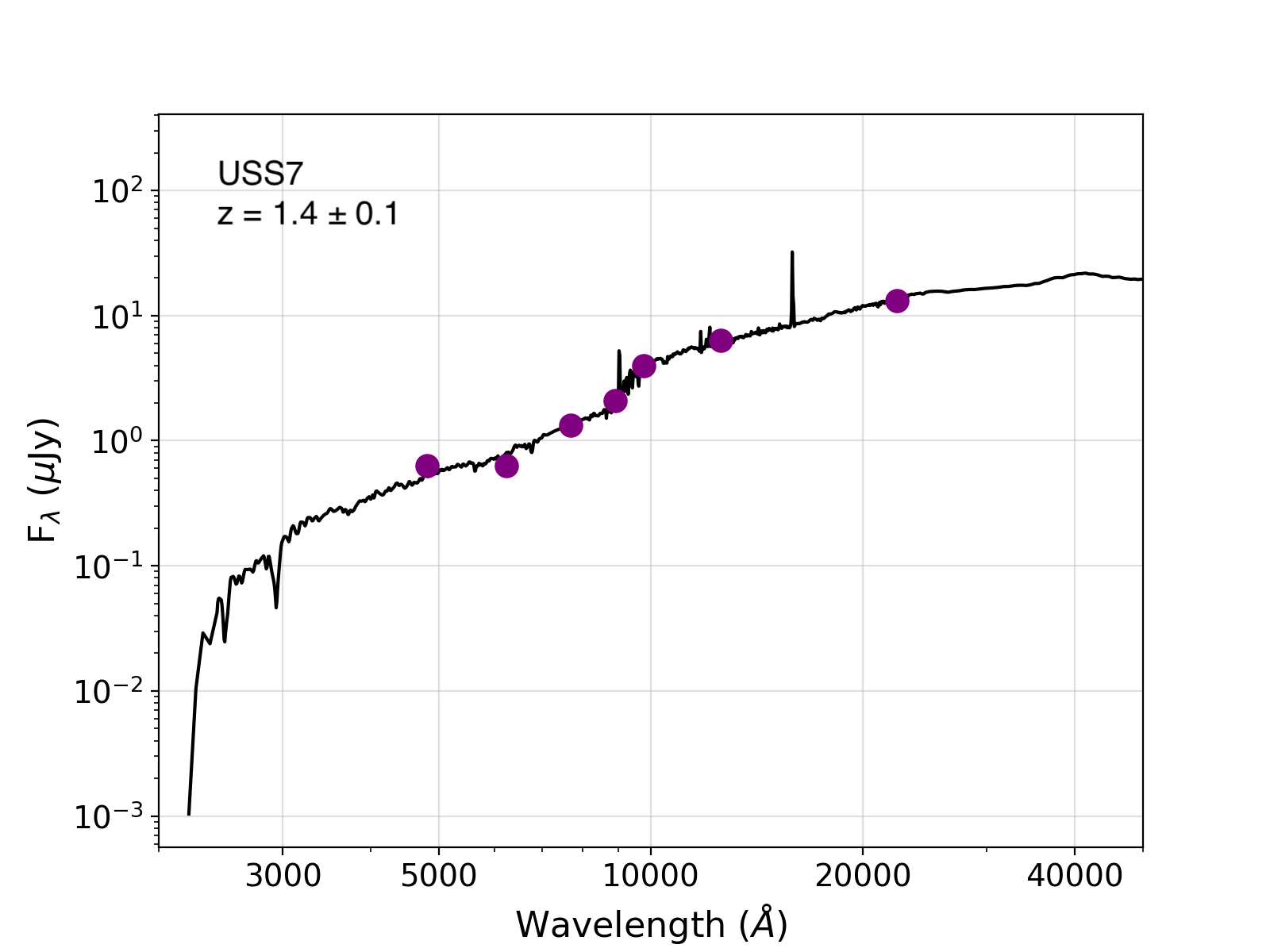}
    \caption{Best-fit SED (observed frame) with a photometric redshift of $z=1.4 \pm 0.1$ obtained for USS7 in the ELAIS-N1 field. SED templates by \citet{bro14} were used for the fitting. The photometry is obtained from HSC-SSP ($g, r, i, z, y$) and UKIDSS DXS ($J, K$). The error bars on the detections are smaller than the symbol size and therefore, not visible.}
    \label{fig:uss7_photoz}
\end{figure}
\begin{figure*}
	\centering
	\includegraphics[scale=0.145]{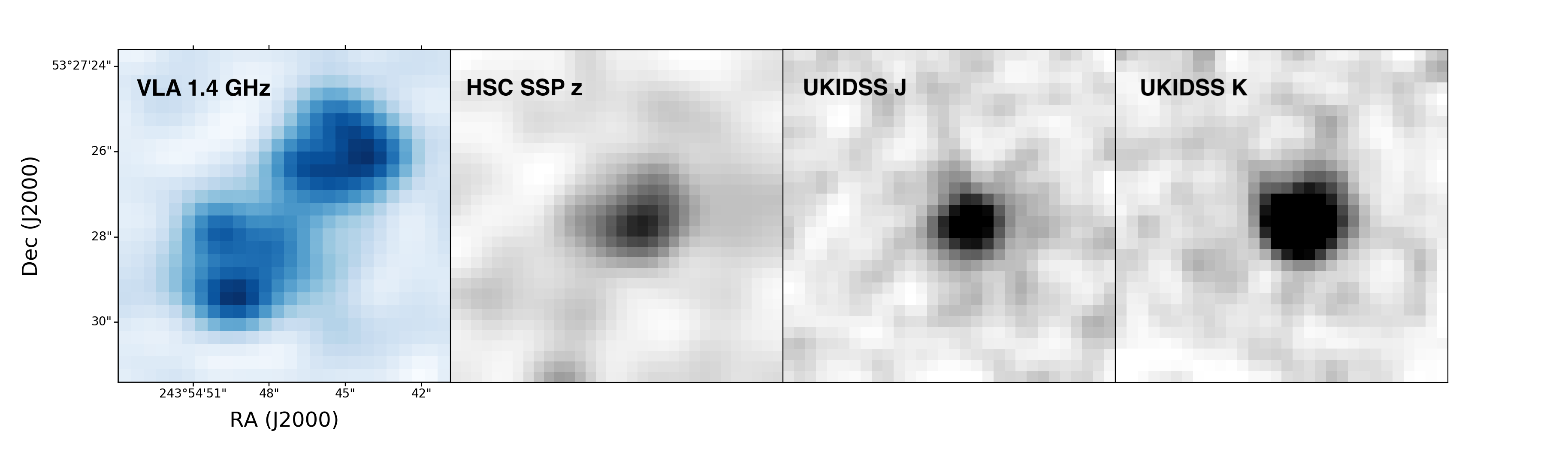}
	\caption{VLA 1.4 GHz image of the ELAIS-N1 source USS7, with HSC SSP $z$, and UKIDSS $J$ and $K$ bands. USS7 is resolved in the VLA map with a 2-component morphology. There is no detection in PS1 $z$ band but the source is detected in UKIDSS $J$ and $K$ bands. Further detections for this galaxy are found in the HSC-SSP (Table \ref{tab:USS7}), with a photometric redshift of 1.4 $\pm$ 0.1. The extent of the radio emission is around 6$''$.}
	\label{fig:uss7_mw}
\end{figure*}
\begin{figure*}
	\centering
	\includegraphics[scale=0.34]{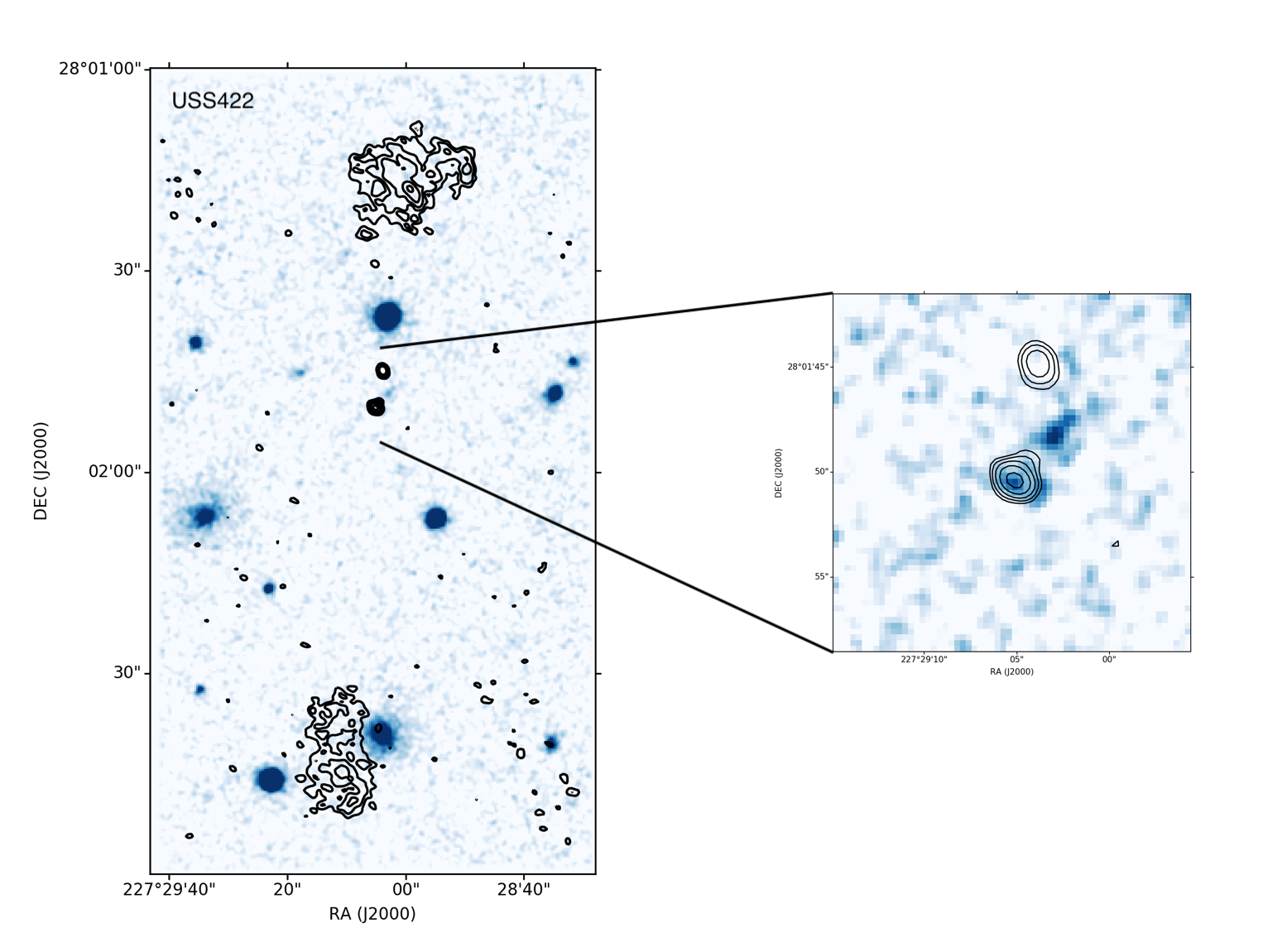}
	\caption{VLA 1.4 GHz contours of the giant radio galaxy (GRG) candidate USS422 overlaid on PanSTARRS1 $z$-band image. The southern lobe (A) has an ultra-steep spectrum when compared with FIRST ($\alpha^{\textrm{TGSS}}_{\textrm{FIRST}}=-1.62$), where the source is likely resolved out. The contours begin at \SI{60}{\micro\jansky}, which is roughly equivalent to $2\sigma$ and are a geometric progression of $\sqrt{2}$. The largest angular size of the radio galaxy is measured to be $1.7'$. There is no clear optical counterpart in the $z$-band image, however faint emission in the optical image is visible corresponding to the component `B' that could potentially be the host galaxy (inset). With a best-fit photometric redshift of $1.8\pm0.5$ (Figure \ref{fig:uss422_photoz}), the total extent of this radio galaxy is 875 kpc.}
	\label{fig:uss422}
\end{figure*}

Another source, USS7, was identified in the ELAIS-N1 field. With the recent data release of the Hyper Suprime-Cam Subaru Strategic Program (HSC SSP; \citealt{aih17}) this source was found to be detected in all of the HSC-SSP bands. Further, there were detections in the $J$ and $K$ bands in the UKIDSS DXS. Using the available photometric information, we derive a fairly robust photometric redshift of $1.4 \pm 0.1$, using SED templates from \citet{bro14}, which is shown in Figure \ref{fig:uss7_photoz}. The available photometry for this source along with USS410 is shown in Table \ref{tab:USS7}.

\begin{table}
	\centering
	\caption{Known optical and infrared photometry (in AB mags) for USS410 in the Lockman Hole field and USS7 in the ELIAS-N1 field (photometric redshift of $z=1.4 \pm 0.1$).}
	\begin{tabular}{l c c}
	\hline
	Target	&	Filter	&	Magnitude (AB)/Flux	density	\\
	\hline \hline
    USS410	&	$J$	&	22.8 $\pm$ 0.1 \\ 
    			&	$K$	& 	22.0 $\pm$ 0.1 \\
            		&	IRAC \SI{3.6}{\micro\meter} & 10.5 $\pm$ 0.6 \SI{}{\micro\jansky} \\
            		& 	IRAC \SI{4.5}{\micro\meter} & 11.3 $\pm$ 0.8 \SI{}{\micro\jansky} \\
	USS7 	& 	$g$	&	24.4 $\pm$ 0.1 \\
    			&	$r$	&	24.4 $\pm$ 0.2 \\
            		&	$i$	&	23.6 $\pm$ 0.1 \\
            		&	$z$	&	23.1 $\pm$ 0.2 \\
           		&	$y$	&	22.4 $\pm$ 0.1 \\
            		&   	$J$	&	21.9 $\pm$ 0.1 \\
            		&	$K$	&	21.1 $\pm$ 0.1 \\
	\hline
	\end{tabular}
	\label{tab:USS7}
\end{table}
 
The VLA image revealed a two-component radio morphology, with the near IR counterparts lying right between the two radio components, thereby increasing the likelihood of it being the true host of this radio galaxy. The angular separation between the two radio components is 3.7$''$. GMRT observations at 610 MHz with a resolution of $6.1''$ in this field have led to the detection of USS7 with a flux density of $18.6\pm0.1$ mJy (A. R. Taylor, private communication). This additional point on the radio SED reveals its spectral index to be consistently ultra-steep, with $\alpha^{150}_{610} = -1.34$ and $\alpha^{610}_{1400} = -1.30$. In contrast to what is observed for USS410, there is evidence of a slight steepening of the spectral index at lower radio frequencies. The K-corrected radio luminosities at 150, 610 and 1400 MHz are $\log L_{150}=27.8$, $\log L_{610} = 27.0$ and $\log L_{1400} = 26.5$ \SI{}{\watt\per\hertz}, respectively.

We show the radio image, along with HSC SSP $z$, and UKIDSS $J$ and $K$ images in Figure \ref{fig:uss7_mw}. The faint optical and infrared photometry combined with bright radio flux density at low frequencies and an ultra-steep spectral index suggest that USS7 could be part of an optically-faint, radio-loud, obscured AGN population. The optical faintness could also be attributed to the presence of a low accreting AGN.

\subsubsection{USS422 - candidate giant radio galaxy}
This particular source appears to be point-like and very faint in the FIRST image. However, in the lower resolution NVSS image it appears to be much brighter, indicating presence of extended emission. Due to the very steep spectral index determined using flux density from FIRST, we decided to observe this source with the VLA, even if it did not adhere to our HzRG sample selection criteria. 

The wide field VLA image revealed the presence of an additional northern lobe of what seems like a giant radio galaxy. Giant radio galaxies represent an extreme class of active galaxies with typical sizes in the range $\sim 0.7 - 5$ Mpc \citep[][and references therein]{dab17}. Figure \ref{fig:uss422} shows the VLA 1.4 GHz radio contours overlaid on the $z$-band image from PanSTARRS1. Blob `A' was initially selected in our sample as a USS source. We believe `B' to be emission from the core of this radio galaxy, as it aligns well with the two lobes. It is not entirely clear whether `C' is actually part of this system or is a background source. If it is indeed part of the system then it could be classified as a `knot' in the radio jet. `D' is the northern lobe which was not identified as an USS source. The two lobes `A' and `D' are separated by more than an arcminute and thus, make this galaxy a candidate giant radio galaxy (GRG). The total extent, measured by drawling a straight line from the edge of the northern lobe to the edge of the southern lobe passing through the core is 1.7$'$. 

\begin{figure}
	\centering
    \vspace{-10pt}
    \includegraphics[scale=0.45]{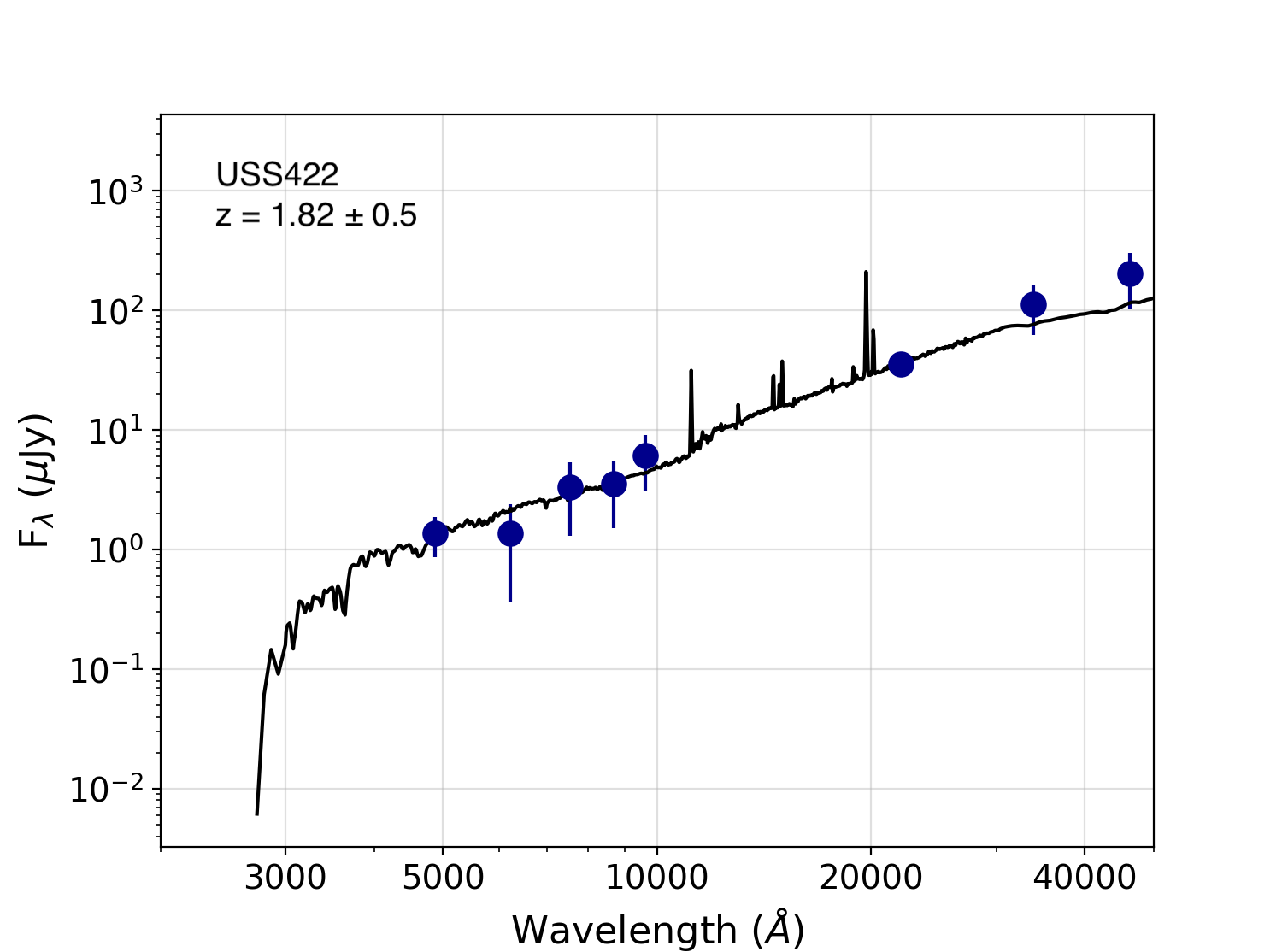}
    \caption{Best-fit SED for the candidate giant radio galaxy USS422 (observed frame) with a photometric redshift of $z=1.82 \pm 0.5$. The photometry used to fit the SED was obtained from PS1 $g, r, i, z, y$ bands, $K$-band from the Large Binocular Telescope and WISE $W1$ and $W2$ bands. Spectroscopic confirmation of the photo-z would make USS422 one of the highest-z giant radio galaxies currently known.}
    \label{fig:uss422_photoz}
\end{figure}

We performed aperture photometry ($2.5''$ diameter) in all PS1 bands at the position of component B. We also find counterparts in the \textit{WISE} \textit{W1} and \textit{W2} bands. For the \textit{WISE} images, we use cutouts from the unblurred coadds (unWISE; \citealt{lan14}), which are better suited for aperture photometry. The associated galaxy in these images, however, is blended with another source $\sim3''$ away. To get over the potentially overestimated flux density, we set the aperture size to be the PSF of the WISE images ($\sim6''$) and add a 50\% flux uncertainty. We further obtained a $K$-band image of the host galaxy using the Large Binocular Telescope (PI: Prandoni, Program ID 66; Saxena et al. in prep), and find it to have a magnitude of $K=20.02$ AB. We then use these flux densities to fit an SED using \textsc{eazy} and \citet{bro14} templates and derive a photometric redshift of $z=1.8 \pm 0.5$. The best-fit SED is shown in Figure \ref{fig:uss422_photoz}. 

If the photo-z estimate is correct, this would make USS422 one of the highest-$z$ giant radio galaxies currently known, with a total extent of $\sim875$ kpc. To achieve such a linear size, the galaxy must harbour an extremely massive black hole and/or have had an AGN accreting close to the Eddington limit for a long time, i.e. $>10^7$ yr \citep[see for example][]{sax17}. Follow-up spectroscopy is essential to truly confirm the redshift of the host galaxy.

\section{Discussion}
\label{sec:discussion}
Compact ultra-steep spectrum radio sources have been found to be excellent tracers for high-redshift radio galaxies. However, HzRGs are not the only sources that have an ultra-steep spectrum. Although our selection has been designed to maximise the chances of detecting high redshift radio galaxies, here we discuss some other classes of objects that could be present in our sample.

\subsection{Radio pulsars}
Radio emission generated in pulsar magnetospheres has been shown to have a spectral index ranging from $-0.7$ to $-3.3$ with a median spectral index of $-1.6$ \citep{sie73}. Radio pulsars are also extremely compact (angular size $<<1''$). As all known pulsars are galactic sources, the majority lies in or near the galactic plane \citep{man05}. Since in our analysis we exclude the galactic plane, there is a reduced chance of contamination by radio pulsars. \citet{fra16} searched for emission at 150 MHz from TGSS ADR towards all currently known and well-localised radio pulsars covered in the survey. We find that no radio source in our final sample lies within 10$''$ of the \citet{fra16} pulsar catalog. Additionally, pulse-averaged emission from radio pulsars is known to be highly variable in continuum surveys due to interstellar scattering, with changes in peak flux density varying by as much as $>50\%$ \citep{bro16}. Since we suppress the presence of highly variable sources in our sample by ensuring comparable flux densities in both FIRST and NVSS at 1.4 GHz, which were taken at different epochs, and the fact that we record consistent flux density measurements from the two VLA observations that were carried out 30 days apart for each target, we further minimise the presence of pulsars in our sample.

\subsection{Dust obscured radio galaxies}
A class of objects that could have strong radio luminosities but be extremely faint at optical and infrared wavelengths are highly obscured radio AGN. These could be similar to hot dust obscured galaxies (DOGs; \citealt{wu12}) that host a radio AGN. Such galaxies are hard to detect even at 3.6 and 4.5 microns (with WISE $W1$ and $W2$) and show significantly higher submillimeter ratios than normal galaxies, thereby suggesting the presence of very hot dust. Dedicated follow-up observations at submillimeter wavelengths are necessary to identify such objects in our sample.

\subsection{Diffuse sources: radio halos and fading radio galaxies}
Radio halos are diffuse radio sources that are generally found in the centres of clusters and are not identified with any particular host galaxy. They are known to have ultra-steep spectral indices, moderate radio luminosities and large projected sizes \citep{rot94}. The last property is taken care of in our sample as we ensure true compactness of a USS source and this should in principle exclude the presence of radio halos in our sample.  

The lobes of fading radio galaxies also appear to be diffuse, with their spectral index steepened due to extreme radiation and expansion losses. Due to their steep spectral indices, USS searches are often used to select fading radio sources \citep[see for example][]{par07}. We note that USS422 (Figure \ref{fig:uss422}) in our sample may be an example of a fading radio source discovered using USS selection. Follow-up observations at various wavelengths are essential to truly determine the nature of this galaxy.

To separate out radio galaxies at the highest redshifts from the plethora of ultra-steep spectrum radio sources that could be present in our sample, we again stress upon the importance of follow-up spectroscopy and deep imaging.

\section{Summary}
\label{sec:summary}

Here we summarise the methods, analysis and findings of this paper.

\begin{enumerate}
\item We have defined a sample of ultra-steep spectrum (USS) radio sources from the TGSS ADR1 at 150 MHz to search for high-redshift radio galaxies (HzRGs). We have used the TGSS ADR1 along with FIRST and NVSS at 1.4 GHz to select sources with spectral indices steeper than $-1.3$ ($\alpha^{150}_{1400}<-1.3$). We have employed strict size restrictions: deconvolved major axis $< 28''$ in TGSS ADR1 and $<  10''$ in FIRST to select compact sources and maximise chances of detecting radio sources at high-z. We have ensured that none of the USS radio sources have detections in available all-sky optical (SDSS, Pan-STARRS1) and infrared (ALLWISE) surveys to eliminate low-redshift interlopers. Next, we apply flux density limits that ensure that we probe a completely new parameter space by excluding sources that could have been detected in previous searches for USS sources. This leaves us with sources with flux densities $50 < S_{150} <200$ mJy. The final sample resulting from these selection criteria consists of 32 sources.

\item We have followed up 31 of these sources with the VLA in A-configuration at 1.4 GHz (L-band) and 14 sources at 370 MHz (P-band). From our L-band observations with an average beam size of $1.3''$, we note that 15 out of 32 sources have a single radio component. The angular sizes of a large majority of sources $<16''$, in line with predictions and expectations of radio sizes at high redshifts, increasing the likelihood of our sample probing HzRGs. P-band (370 MHz) imaging of 17 of these sources revealed ten of them to show signs of flattening of the radio SED at frequencies between 150 and 370 MHz, which is expected from compact HzRGs. 

\item We then analyse stacked photometry in optical and infrared bands of all sources in the sample. Although stacking analysis is most effective for sources that have similar properties, we do this anyway to gain insights into the median properties of our sample. We stack the optical and infrared (non-)detections from SDSS, Pan-STARRS1 and WISE centred on expected positions of the host galaxy and fit an SED to derive a median photometric redshift of 2.93 for the sample with $1\sigma$ limits ranging from $z=1.3-4.75$. Fits with $z>6$ are within $3\sigma$. 

\item Two sources in our sample lie in well-studied extragalactic fields that have deeper multiwavelength data available. One of these sources lies in the Lockman Hole field. With the availability of LOFAR LBA data at 60 MHz, we note the spectral turnover in the radio SED between 60 and 150 MHz, which has also been reported before in the literature. Additionally, the radio source has counterparts in $J$ and $K$ bands in the UKIDSS Deep Extragalactic Survey (DXS) and in Spitzer IRAC 3.6 and 4.5 micron bands from the SWIRE survey. Using these, we derive the best-fit photometric redshift to be $z=4.8 \pm 2.0$, although lower-z solutions are not formally excluded. Another source lies in the ELAIS-N1 field, with optical and infrared counterparts identified from the deep HSC-SSP optical data and $J$ and $K$ bands from the DXS. We derive the photo-$z$ for this source to be $z=1.4 \pm 0.1$.

\item In addition to the HzRG candidates, we observed a USS source that was not a candidate HzRG but showed signs of potential large scale diffuse emission. The wide-field VLA image revealed this source to be a candidate giant radio galaxy with a total extent of 1.7 arcmin. We derived the photo-$z$ of the associated host galaxy to be $z=1.8\pm0.5$. If the redshift is accurate, the total extent of this radio galaxy would be $\sim875$ kpc, making it one of the largest known giant radio galaxies at these redshifts.

\item Lastly, we discuss some other classes of objects that could be present in our sample. We note that chances of radio pulsars contaminating our sample are minimal, as our sample does not cover the sky area of the galactic plane and we also take measures to rule out variable sources. However, there is a possibility of the presence of dust obscured radio AGN at lower redshifts, fading radio galaxies and also bent, head-tail radio galaxies at lower redshifts to be present in our sample. Follow-up observations are essential to separate our the various classes of objects present in our sample.
\end{enumerate}

The current highest redshift radio galaxy is at $z=5.12$ with an ultra-steep spectral index of $\alpha = -1.6$ \citep{vbr99}. With new and improved radio surveys coming online, there is potential to push the highest known redshift for a radio galaxy to $z>6$. Radio galaxies at $z>6$ can be indispensable probes of the epoch of reionisation and studying even a single radio galaxy at $z>6$ could have tremendous implications on cosmology. We have attempted to extend the USS search technique at fainter flux densities to be potentially employed as a robust tool to search for HzRGs at the highest redshifts. We have now started an extensive follow-up campaign to spectroscopically determine the redshifts of our HzRG candidates and also study the underlying stellar populations in the host galaxies.

\section*{Acknowledgments}
We thank the referee for useful comments and suggestions that helped improve this paper. AS and HJR gratefully acknowledge support from the European Research Council under the European Unions Seventh Framework Programme (FP/2007-2013)/ERC Advanced Grant NEWCLUSTERS-321271. PNB is grateful for support from STFC via grant ST/M001229/1. This paper uses observational material taken with the Very Large Array, a National Radio Astronomy Observatory (NRAO) instrument. The NRAO is a facility of the National Science Foundation operated under cooperative agreement by Associated Universities, Inc. This paper uses data from the Giant Metrewave Radio Telescope (GMRT). We thank the staff of the GMRT that made these observations possible. GMRT is run by the National Centre for Radio Astrophysics of the Tata Institute of Fundamental Research.

This work has made extensive use of \textsc{IPython} \citep{per07}. This research made use of \textsc{Astropy}, a community-developed core \textsc{Python} package for Astronomy \citep{ast13}. This research made use of \textsc{APLpy}, an open-source plotting package for Python \citep{apl} and \textsc{matplotlib} \citep{plt} for generating figures and plots. The data analysis and visualisation were done using \textsc{topcat} \citep{top05}. This work would not have been possible without the countless hours put in by members of the open-source developing community all around the world. 



\renewcommand{\bibname}{References}

\bibliographystyle{mnras}
\bibliography{hzrg_paper}

\newcommand{\noop}[1]{}
\begin{thebibliography}{}
\makeatletter
\relax
\def\mn@urlcharsother{\let\do\@makeother \do\$\do\&\do\#\do\^\do\_\do\%\do\~}
\def\mn@doi{\begingroup\mn@urlcharsother \@ifnextchar [ {\mn@doi@}
  {\mn@doi@[]}}
\def\mn@doi@[#1]#2{\def\@tempa{#1}\ifx\@tempa\@empty \href
  {http://dx.doi.org/#2} {doi:#2}\else \href {http://dx.doi.org/#2} {#1}\fi
  \endgroup}
\def\mn@eprint#1#2{\mn@eprint@#1:#2::\@nil}
\def\mn@eprint@arXiv#1{\href {http://arxiv.org/abs/#1} {{\tt arXiv:#1}}}
\def\mn@eprint@dblp#1{\href {http://dblp.uni-trier.de/rec/bibtex/#1.xml}
  {dblp:#1}}
\def\mn@eprint@#1:#2:#3:#4\@nil{\def\@tempa {#1}\def\@tempb {#2}\def\@tempc
  {#3}\ifx \@tempc \@empty \let \@tempc \@tempb \let \@tempb \@tempa \fi \ifx
  \@tempb \@empty \def\@tempb {arXiv}\fi \@ifundefined
  {mn@eprint@\@tempb}{\@tempb:\@tempc}{\expandafter \expandafter \csname
  mn@eprint@\@tempb\endcsname \expandafter{\@tempc}}}

\bibitem[\protect\citeauthoryear{{Afonso} et~al.,}{{Afonso}
  et~al.}{2011}]{afo11}
{Afonso} J.,  et~al., 2011, \mn@doi [\apj] {10.1088/0004-637X/743/2/122}, \href
  {http://adsabs.harvard.edu/abs/2011ApJ...743..122A} {743, 122}

\bibitem[\protect\citeauthoryear{{Aihara} et~al.,}{{Aihara}
  et~al.}{2017}]{aih17}
{Aihara} H.,  et~al., 2017, preprint, \href
  {http://ads.nao.ac.jp/abs/2017arXiv170208449A} {} (\mn@eprint {arXiv}
  {1702.08449})

\bibitem[\protect\citeauthoryear{{Alam} et~al.,}{{Alam} et~al.}{2015}]{ala15}
{Alam} S.,  et~al., 2015, \mn@doi [\apjs] {10.1088/0067-0049/219/1/12}, \href
  {http://adsabs.harvard.edu/abs/2015ApJS..219...12A} {219, 12}

\bibitem[\protect\citeauthoryear{{Archibald}, {Dunlop}, {Hughes}, {Rawlings},
  {Eales}  \& {Ivison}}{{Archibald} et~al.}{2001}]{arc01}
{Archibald} E.~N.,  {Dunlop} J.~S.,  {Hughes} D.~H.,  {Rawlings} S.,  {Eales}
  S.~A.,   {Ivison} R.~J.,  2001, \mn@doi [\mnras]
  {10.1046/j.1365-8711.2001.04188.x}, \href
  {http://adsabs.harvard.edu/abs/2001MNRAS.323..417A} {323, 417}

\bibitem[\protect\citeauthoryear{{Astropy Collaboration} et~al.,}{{Astropy
  Collaboration} et~al.}{2013}]{ast13}
{Astropy Collaboration} et~al., 2013, \mn@doi [\aap]
  {10.1051/0004-6361/201322068}, \href
  {http://adsabs.harvard.edu/abs/2013A%26A...558A..33A} {558, A33}

\bibitem[\protect\citeauthoryear{Becker, White  \& Helfand}{Becker
  et~al.}{1995}]{Becker1995}
Becker R.~H.,  White R.~L.,   Helfand D.~J.,  1995, \mn@doi [\apj]
  {10.1086/176166}, 450, 559

\bibitem[\protect\citeauthoryear{{Best}, {Longair}  \& {R{\"o}ttgering}}{{Best}
  et~al.}{1998a}]{bes98}
{Best} P.~N.,  {Longair} M.~S.,   {R{\"o}ttgering} H.~J.~A.,  1998a, \mn@doi
  [\mnras] {10.1046/j.1365-8711.1998.01245.x}, \href
  {http://adsabs.harvard.edu/abs/1998MNRAS.295..549B} {295, 549}

\bibitem[\protect\citeauthoryear{{Best} et~al.,}{{Best} et~al.}{1998b}]{bes98b}
{Best} P.~N.,  et~al., 1998b, \mn@doi [\mnras]
  {10.1046/j.1365-8711.1998.02155.x}, \href
  {http://adsabs.harvard.edu/abs/1998MNRAS.301L..15B} {301, L15}

\bibitem[\protect\citeauthoryear{{Bilicki}, {Jarrett}, {Peacock}, {Cluver}  \&
  {Steward}}{{Bilicki} et~al.}{2014}]{bil14}
{Bilicki} M.,  {Jarrett} T.~H.,  {Peacock} J.~A.,  {Cluver} M.~E.,   {Steward}
  L.,  2014, \mn@doi [\apjs] {10.1088/0067-0049/210/1/9}, \href
  {http://adsabs.harvard.edu/abs/2014ApJS..210....9B} {210, 9}

\bibitem[\protect\citeauthoryear{{Blumenthal} \& {Miley}}{{Blumenthal} \&
  {Miley}}{1979}]{blu79}
{Blumenthal} G.,  {Miley} G.,  1979, \aap, \href
  {http://adsabs.harvard.edu/abs/1979A%26A....80...13B} {80, 13}

\bibitem[\protect\citeauthoryear{{Blundell} \& {Rawlings}}{{Blundell} \&
  {Rawlings}}{1999}]{blu99b}
{Blundell} K.~M.,  {Rawlings} S.,  1999, \mn@doi [\nat] {10.1038/20612}, \href
  {http://adsabs.harvard.edu/abs/1999Natur.399..330B} {399, 330}

\bibitem[\protect\citeauthoryear{{Blundell}, {Rawlings}  \&
  {Willott}}{{Blundell} et~al.}{1999}]{blu99}
{Blundell} K.~M.,  {Rawlings} S.,   {Willott} C.~J.,  1999, \mn@doi [\aj]
  {10.1086/300721}, \href {http://adsabs.harvard.edu/abs/1999AJ....117..677B}
  {117, 677}

\bibitem[\protect\citeauthoryear{{Brammer}, {van Dokkum}  \& {Coppi}}{{Brammer}
  et~al.}{2008}]{bra08}
{Brammer} G.~B.,  {van Dokkum} P.~G.,   {Coppi} P.,  2008, \mn@doi [\apj]
  {10.1086/591786}, \href {http://adsabs.harvard.edu/abs/2008ApJ...686.1503B}
  {686, 1503}

\bibitem[\protect\citeauthoryear{{Brook}, {Karastergiou}, {Johnston}, {Kerr},
  {Shannon}  \& {Roberts}}{{Brook} et~al.}{2016}]{bro16}
{Brook} P.~R.,  {Karastergiou} A.,  {Johnston} S.,  {Kerr} M.,  {Shannon}
  R.~M.,   {Roberts} S.~J.,  2016, \mn@doi [\mnras] {10.1093/mnras/stv2715},
  \href {http://adsabs.harvard.edu/abs/2016MNRAS.456.1374B} {456, 1374}

\bibitem[\protect\citeauthoryear{{Brown} et~al.,}{{Brown} et~al.}{2014}]{bro14}
{Brown} M.~J.~I.,  et~al., 2014, \mn@doi [\apjs] {10.1088/0067-0049/212/2/18},
  \href {http://adsabs.harvard.edu/abs/2014ApJS..212...18B} {212, 18}

\bibitem[\protect\citeauthoryear{{Callingham} et~al.,}{{Callingham}
  et~al.}{2015}]{cal15}
{Callingham} J.~R.,  et~al., 2015, \mn@doi [\apj]
  {10.1088/0004-637X/809/2/168}, \href
  {http://adsabs.harvard.edu/abs/2015ApJ...809..168C} {809, 168}

\bibitem[\protect\citeauthoryear{{Callingham} et~al.,}{{Callingham}
  et~al.}{2017}]{cal17}
{Callingham} J.~R.,  et~al., 2017, \mn@doi [\apj]
  {10.3847/1538-4357/836/2/174}, \href
  {http://adsabs.harvard.edu/abs/2017ApJ...836..174C} {836, 174}

\bibitem[\protect\citeauthoryear{{Carilli}, {Harris}, {Pentericci},
  {R{\"o}ttgering}, {Miley}, {Kurk}  \& {van Breugel}}{{Carilli}
  et~al.}{2002a}]{car02b}
{Carilli} C.~L.,  {Harris} D.~E.,  {Pentericci} L.,  {R{\"o}ttgering} H.~J.~A.,
   {Miley} G.~K.,  {Kurk} J.~D.,   {van Breugel} W.,  2002a, \mn@doi [\apj]
  {10.1086/338669}, \href {http://adsabs.harvard.edu/abs/2002ApJ...567..781C}
  {567, 781}

\bibitem[\protect\citeauthoryear{{Carilli}, {Gnedin}  \& {Owen}}{{Carilli}
  et~al.}{2002b}]{car02}
{Carilli} C.~L.,  {Gnedin} N.~Y.,   {Owen} F.,  2002b, \mn@doi [\apj]
  {10.1086/342179}, \href {http://adsabs.harvard.edu/abs/2002ApJ...577...22C}
  {577, 22}

\bibitem[\protect\citeauthoryear{{Chambers}, {Miley}, {van Breugel}  \&
  {Huang}}{{Chambers} et~al.}{1996}]{cha96}
{Chambers} K.~C.,  {Miley} G.~K.,  {van Breugel} W.~J.~M.,   {Huang} J.-S.,
  1996, \mn@doi [\apjs] {10.1086/192337}, \href
  {http://adsabs.harvard.edu/abs/1996ApJS..106..215C} {106, 215}

\bibitem[\protect\citeauthoryear{{Chambers} et~al.,}{{Chambers}
  et~al.}{2016a}]{ps1}
{Chambers} K.~C.,  et~al., 2016a, preprint, \href
  {http://adsabs.harvard.edu/abs/2016arXiv161205560C} {} (\mn@eprint {arXiv}
  {1612.05560})

\bibitem[\protect\citeauthoryear{{Chambers} et~al.,}{{Chambers}
  et~al.}{2016b}]{cha16}
{Chambers} K.~C.,  et~al., 2016b, preprint, \href
  {http://adsabs.harvard.edu/abs/2016arXiv161205560C} {} (\mn@eprint {arXiv}
  {1612.05560})

\bibitem[\protect\citeauthoryear{{Ciardi} et~al.,}{{Ciardi}
  et~al.}{2015}]{cia15}
{Ciardi} B.,  et~al., 2015, \mn@doi [\mnras] {10.1093/mnras/stv1640}, \href
  {http://adsabs.harvard.edu/abs/2015MNRAS.453..101C} {453, 101}

\bibitem[\protect\citeauthoryear{{Condon}, {Cotton}, {Greisen}, {Yin},
  {Perley}, {Taylor}  \& {Broderick}}{{Condon} et~al.}{1998}]{con98}
{Condon} J.~J.,  {Cotton} W.~D.,  {Greisen} E.~W.,  {Yin} Q.~F.,  {Perley}
  R.~A.,  {Taylor} G.~B.,   {Broderick} J.~J.,  1998, \mn@doi [\aj]
  {10.1086/300337}, \href {http://adsabs.harvard.edu/abs/1998AJ....115.1693C}
  {115, 1693}

\bibitem[\protect\citeauthoryear{{Dabhade}, {Gaikwad}, {Bagchi},
  {Pandey-Pommier}, {Sankhyayan}  \& {Raychaudhury}}{{Dabhade}
  et~al.}{2017}]{dab17}
{Dabhade} P.,  {Gaikwad} M.,  {Bagchi} J.,  {Pandey-Pommier} M.,  {Sankhyayan}
  S.,   {Raychaudhury} S.,  2017, \mn@doi [\mnras] {10.1093/mnras/stx860},
  \href {http://adsabs.harvard.edu/abs/2017MNRAS.469.2886D} {469, 2886}

\bibitem[\protect\citeauthoryear{{Daly} \& {Guerra}}{{Daly} \&
  {Guerra}}{2002}]{dal02}
{Daly} R.~A.,  {Guerra} E.~J.,  2002, \mn@doi [\aj] {10.1086/342741}, \href
  {http://adsabs.harvard.edu/abs/2002AJ....124.1831D} {124, 1831}

\bibitem[\protect\citeauthoryear{{De Breuck}, {van Breugel}, {R{\"o}ttgering}
  \& {Miley}}{{De Breuck} et~al.}{2000}]{deb00}
{De Breuck} C.,  {van Breugel} W.,  {R{\"o}ttgering} H.~J.~A.,   {Miley} G.,
  2000, \mn@doi [\aaps] {10.1051/aas:2000181}, \href
  {http://adsabs.harvard.edu/abs/2000A%26AS..143..303D} {143, 303}

\bibitem[\protect\citeauthoryear{{De Breuck} et~al.,}{{De Breuck}
  et~al.}{2010}]{deb10}
{De Breuck} C.,  et~al., 2010, \mn@doi [\apj] {10.1088/0004-637X/725/1/36},
  \href {http://adsabs.harvard.edu/abs/2010ApJ...725...36D} {725, 36}

\bibitem[\protect\citeauthoryear{{Dewdney}, {Hall}, {Schilizzi}  \&
  {Lazio}}{{Dewdney} et~al.}{2009}]{ska}
{Dewdney} P.~E.,  {Hall} P.~J.,  {Schilizzi} R.~T.,   {Lazio} T.~J.~L.~W.,
  2009, \mn@doi [IEEE Proceedings] {10.1109/JPROC.2009.2021005}, \href
  {http://adsabs.harvard.edu/abs/2009IEEEP..97.1482D} {97, 1482}

\bibitem[\protect\citeauthoryear{{Dye} et~al.,}{{Dye} et~al.}{2006}]{dye06}
{Dye} S.,  et~al., 2006, \mn@doi [\mnras] {10.1111/j.1365-2966.2006.10928.x},
  \href {http://adsabs.harvard.edu/abs/2006MNRAS.372.1227D} {372, 1227}

\bibitem[\protect\citeauthoryear{{Ewall-Wice}, {Dillon}, {Mesinger}  \&
  {Hewitt}}{{Ewall-Wice} et~al.}{2014}]{ewa14}
{Ewall-Wice} A.,  {Dillon} J.~S.,  {Mesinger} A.,   {Hewitt} J.,  2014, \mn@doi
  [\mnras] {10.1093/mnras/stu666}, \href
  {http://adsabs.harvard.edu/abs/2014MNRAS.441.2476E} {441, 2476}

\bibitem[\protect\citeauthoryear{{Fanti}, {Fanti}, {Dallacasa}, {Schilizzi},
  {Spencer}  \& {Stanghellini}}{{Fanti} et~al.}{1995}]{fan95}
{Fanti} C.,  {Fanti} R.,  {Dallacasa} D.,  {Schilizzi} R.~T.,  {Spencer} R.~E.,
    {Stanghellini} C.,  1995, \aap, \href
  {http://adsabs.harvard.edu/abs/1995A%26A...302..317F} {302, 317}

\bibitem[\protect\citeauthoryear{{Frail}, {Jagannathan}, {Mooley}  \&
  {Intema}}{{Frail} et~al.}{2016}]{fra16}
{Frail} D.~A.,  {Jagannathan} P.,  {Mooley} K.~P.,   {Intema} H.~T.,  2016,
  \mn@doi [\apj] {10.3847/0004-637X/829/2/119}, \href
  {http://adsabs.harvard.edu/abs/2016ApJ...829..119F} {829, 119}

\bibitem[\protect\citeauthoryear{{Furlanetto} \& {Loeb}}{{Furlanetto} \&
  {Loeb}}{2002}]{fur02}
{Furlanetto} S.~R.,  {Loeb} A.,  2002, \mn@doi [\apj] {10.1086/342757}, \href
  {http://adsabs.harvard.edu/abs/2002ApJ...579....1F} {579, 1}

\bibitem[\protect\citeauthoryear{{Garn} \& {Alexander}}{{Garn} \&
  {Alexander}}{2008}]{gar08}
{Garn} T.,  {Alexander} P.,  2008, \mn@doi [\mnras]
  {10.1111/j.1365-2966.2008.13980.x}, \href
  {http://adsabs.harvard.edu/abs/2008MNRAS.391.1000G} {391, 1000}

\bibitem[\protect\citeauthoryear{{Garn}, {Green}, {Riley}  \&
  {Alexander}}{{Garn} et~al.}{2010}]{gar10}
{Garn} T.~S.,  {Green} D.~A.,  {Riley} J.~M.,   {Alexander} P.,  2010, Bulletin
  of the Astronomical Society of India, \href
  {http://adsabs.harvard.edu/abs/2010BASI...38..103G} {38, 103}

\bibitem[\protect\citeauthoryear{{Hatch} et~al.,}{{Hatch} et~al.}{2011}]{hat11}
{Hatch} N.~A.,  et~al., 2011, \mn@doi [\mnras]
  {10.1111/j.1365-2966.2010.17538.x}, \href
  {http://adsabs.harvard.edu/abs/2011MNRAS.410.1537H} {410, 1537}

\bibitem[\protect\citeauthoryear{Hunter}{Hunter}{2007}]{plt}
Hunter J.~D.,  2007, \mn@doi [Computing In Science \& Engineering]
  {10.1109/MCSE.2007.55}, 9, 90

\bibitem[\protect\citeauthoryear{{Hurley-Walker}}{{Hurley-Walker}}{2017}]{hur17}
{Hurley-Walker} N.,  2017, preprint, \href
  {http://adsabs.harvard.edu/abs/2017arXiv170306635H} {} (\mn@eprint {arXiv}
  {1703.06635})

\bibitem[\protect\citeauthoryear{{Intema}, {Jagannathan}, {Mooley}  \&
  {Frail}}{{Intema} et~al.}{2017}]{int17}
{Intema} H.~T.,  {Jagannathan} P.,  {Mooley} K.~P.,   {Frail} D.~A.,  2017,
  \mn@doi [\aap] {10.1051/0004-6361/201628536}, \href
  {http://adsabs.harvard.edu/abs/2017A%26A...598A..78I} {598, A78}

\bibitem[\protect\citeauthoryear{{Jarvis}, {Rawlings}, {Eales}, {Blundell},
  {Bunker}, {Croft}, {McLure}  \& {Willott}}{{Jarvis} et~al.}{2001}]{jar01}
{Jarvis} M.~J.,  {Rawlings} S.,  {Eales} S.,  {Blundell} K.~M.,  {Bunker}
  A.~J.,  {Croft} S.,  {McLure} R.~J.,   {Willott} C.~J.,  2001, \mn@doi
  [\mnras] {10.1111/j.1365-2966.2001.04730.x}, \href
  {http://adsabs.harvard.edu/abs/2001MNRAS.326.1585J} {326, 1585}

\bibitem[\protect\citeauthoryear{{Jarvis}, {Teimourian}, {Simpson}, {Smith},
  {Rawlings}  \& {Bonfield}}{{Jarvis} et~al.}{2009}]{jar09}
{Jarvis} M.~J.,  {Teimourian} H.,  {Simpson} C.,  {Smith} D.~J.~B.,  {Rawlings}
  S.,   {Bonfield} D.,  2009, \mn@doi [\mnras]
  {10.1111/j.1745-3933.2009.00715.x}, \href
  {http://adsabs.harvard.edu/abs/2009MNRAS.398L..83J} {398, L83}

\bibitem[\protect\citeauthoryear{{Ker}, {Best}, {Rigby}, {R{\"o}ttgering}  \&
  {Gendre}}{{Ker} et~al.}{2012}]{ker12}
{Ker} L.~M.,  {Best} P.~N.,  {Rigby} E.~E.,  {R{\"o}ttgering} H.~J.~A.,
  {Gendre} M.~A.,  2012, \mn@doi [\mnras] {10.1111/j.1365-2966.2011.20235.x},
  \href {http://adsabs.harvard.edu/abs/2012MNRAS.420.2644K} {420, 2644}

\bibitem[\protect\citeauthoryear{{Lang}}{{Lang}}{2014}]{lan14}
{Lang} D.,  2014, \mn@doi [\aj] {10.1088/0004-6256/147/5/108}, \href
  {http://adsabs.harvard.edu/abs/2014AJ....147..108L} {147, 108}

\bibitem[\protect\citeauthoryear{{Lilly} \& {Longair}}{{Lilly} \&
  {Longair}}{1984}]{lil84}
{Lilly} S.~J.,  {Longair} M.~S.,  1984, \mn@doi [\mnras]
  {10.1093/mnras/211.4.833}, \href
  {http://adsabs.harvard.edu/abs/1984MNRAS.211..833L} {211, 833}

\bibitem[\protect\citeauthoryear{{Mack} \& {Wyithe}}{{Mack} \&
  {Wyithe}}{2012}]{mac12}
{Mack} K.~J.,  {Wyithe} J.~S.~B.,  2012, \mn@doi [\mnras]
  {10.1111/j.1365-2966.2012.21561.x}, \href
  {http://adsabs.harvard.edu/abs/2012MNRAS.425.2988M} {425, 2988}

\bibitem[\protect\citeauthoryear{{Mahony} et~al.,}{{Mahony}
  et~al.}{2016}]{mah16}
{Mahony} E.~K.,  et~al., 2016, \mn@doi [\mnras] {10.1093/mnras/stw2225}, \href
  {http://adsabs.harvard.edu/abs/2016MNRAS.463.2997M} {463, 2997}

\bibitem[\protect\citeauthoryear{{Manchester}, {Hobbs}, {Teoh}  \&
  {Hobbs}}{{Manchester} et~al.}{2005}]{man05}
{Manchester} R.~N.,  {Hobbs} G.~B.,  {Teoh} A.,   {Hobbs} M.,  2005, \mn@doi
  [\aj] {10.1086/428488}, \href
  {http://adsabs.harvard.edu/abs/2005AJ....129.1993M} {129, 1993}

\bibitem[\protect\citeauthoryear{{McLure}, {Willott}, {Jarvis}, {Rawlings},
  {Hill}, {Mitchell}, {Dunlop}  \& {Wold}}{{McLure} et~al.}{2004}]{mcl04}
{McLure} R.~J.,  {Willott} C.~J.,  {Jarvis} M.~J.,  {Rawlings} S.,  {Hill}
  G.~J.,  {Mitchell} E.,  {Dunlop} J.~S.,   {Wold} M.,  2004, \mn@doi [\mnras]
  {10.1111/j.1365-2966.2004.07793.x}, \href
  {http://adsabs.harvard.edu/abs/2004MNRAS.351..347M} {351, 347}

\bibitem[\protect\citeauthoryear{{Middelberg}, {Norris}, {Hales}, {Seymour},
  {Johnston-Hollitt}, {Huynh}, {Lenc}  \& {Mao}}{{Middelberg}
  et~al.}{2011}]{mid11}
{Middelberg} E.,  {Norris} R.~P.,  {Hales} C.~A.,  {Seymour} N.,
  {Johnston-Hollitt} M.,  {Huynh} M.~T.,  {Lenc} E.,   {Mao} M.~Y.,  2011,
  \mn@doi [\aap] {10.1051/0004-6361/201014926}, \href
  {http://adsabs.harvard.edu/abs/2011A%26A...526A...8M} {526, A8}

\bibitem[\protect\citeauthoryear{{Miley}}{{Miley}}{1968}]{mil68}
{Miley} G.~K.,  1968, \mn@doi [\nat] {10.1038/218933a0}, \href
  {http://adsabs.harvard.edu/abs/1968Natur.218..933M} {218, 933}

\bibitem[\protect\citeauthoryear{{Miley} \& {De Breuck}}{{Miley} \& {De
  Breuck}}{2008}]{mil08}
{Miley} G.,  {De Breuck} C.,  2008, \mn@doi [\aapr]
  {10.1007/s00159-007-0008-z}, \href
  {http://adsabs.harvard.edu/abs/2008A%26ARv..15...67M} {15, 67}

\bibitem[\protect\citeauthoryear{{Miley} et~al.,}{{Miley} et~al.}{2004}]{mil04}
{Miley} G.~K.,  et~al., 2004, \mn@doi [\nat] {10.1038/nature02125}, \href
  {http://adsabs.harvard.edu/abs/2004Natur.427...47M} {427, 47}

\bibitem[\protect\citeauthoryear{{Miley} et~al.,}{{Miley} et~al.}{2006}]{mil06}
{Miley} G.~K.,  et~al., 2006, \mn@doi [\apjl] {10.1086/508534}, \href
  {http://adsabs.harvard.edu/abs/2006ApJ...650L..29M} {650, L29}

\bibitem[\protect\citeauthoryear{{Minkowski}}{{Minkowski}}{1960}]{min60}
{Minkowski} R.,  1960, \mn@doi [\apj] {10.1086/146994}, \href
  {http://adsabs.harvard.edu/abs/1960ApJ...132..908M} {132, 908}

\bibitem[\protect\citeauthoryear{{Morabito} et~al.,}{{Morabito}
  et~al.}{2017}]{mor17}
{Morabito} L.~K.,  et~al., 2017, submitted

\bibitem[\protect\citeauthoryear{{Neeser}, {Eales}, {Law-Green}, {Leahy}  \&
  {Rawlings}}{{Neeser} et~al.}{1995}]{nee95}
{Neeser} M.~J.,  {Eales} S.~A.,  {Law-Green} J.~D.,  {Leahy} J.~P.,
  {Rawlings} S.,  1995, \mn@doi [\apj] {10.1086/176201}, \href
  {http://adsabs.harvard.edu/abs/1995ApJ...451...76N} {451, 76}

\bibitem[\protect\citeauthoryear{{Nilsson}, {Valtonen}, {Kotilainen}  \&
  {Jaakkola}}{{Nilsson} et~al.}{1993}]{nil93}
{Nilsson} K.,  {Valtonen} M.~J.,  {Kotilainen} J.,   {Jaakkola} T.,  1993,
  \mn@doi [\apj] {10.1086/173016}, \href
  {http://adsabs.harvard.edu/abs/1993ApJ...413..453N} {413, 453}

\bibitem[\protect\citeauthoryear{{Orsi}, {Fanidakis}, {Lacey}  \&
  {Baugh}}{{Orsi} et~al.}{2016}]{ors16}
{Orsi} {\'A}.~A.,  {Fanidakis} N.,  {Lacey} C.~G.,   {Baugh} C.~M.,  2016,
  \mn@doi [\mnras] {10.1093/mnras/stv2919}, \href
  {http://adsabs.harvard.edu/abs/2016MNRAS.456.3827O} {456, 3827}

\bibitem[\protect\citeauthoryear{{Parma}, {Murgia}, {de Ruiter}, {Fanti},
  {Mack}  \& {Govoni}}{{Parma} et~al.}{2007}]{par07}
{Parma} P.,  {Murgia} M.,  {de Ruiter} H.~R.,  {Fanti} R.,  {Mack} K.-H.,
  {Govoni} F.,  2007, \mn@doi [\aap] {10.1051/0004-6361:20077592}, \href
  {http://adsabs.harvard.edu/abs/2007A%26A...470..875P} {470, 875}

\bibitem[\protect\citeauthoryear{{Pentericci} et~al.,}{{Pentericci}
  et~al.}{2000}]{pen00}
{Pentericci} L.,  et~al., 2000, \aap, \href
  {http://adsabs.harvard.edu/abs/2000A%26A...361L..25P} {361, L25}

\bibitem[\protect\citeauthoryear{P\'erez \& Granger}{P\'erez \&
  Granger}{2007}]{per07}
P\'erez F.,  Granger B.~E.,  2007, \mn@doi [Computing in Science and
  Engineering] {10.1109/MCSE.2007.53}, 9, 21

\bibitem[\protect\citeauthoryear{{Planck Collaboration} et~al.,}{{Planck
  Collaboration} et~al.}{2014}]{pla14}
{Planck Collaboration} et~al., 2014, \mn@doi [\aap]
  {10.1051/0004-6361/201321591}, \href
  {http://adsabs.harvard.edu/abs/2014A%26A...571A..16P} {571, A16}

\bibitem[\protect\citeauthoryear{{Rengelink}, {Tang}, {de Bruyn}, {Miley},
  {Bremer}, {R{\"o}ttgering}  \& {Bremer}}{{Rengelink} et~al.}{1997}]{ren97}
{Rengelink} R.~B.,  {Tang} Y.,  {de Bruyn} A.~G.,  {Miley} G.~K.,  {Bremer}
  M.~N.,  {R{\"o}ttgering} H.~J.~A.,   {Bremer} M.~A.~R.,  1997, \mn@doi
  [\aaps] {10.1051/aas:1997358}, \href
  {http://adsabs.harvard.edu/abs/1997A%26AS..124..259R} {124}

\bibitem[\protect\citeauthoryear{{Reuland}, {R{\"o}ttgering}, {van Breugel}  \&
  {De Breuck}}{{Reuland} et~al.}{2004}]{reu04}
{Reuland} M.,  {R{\"o}ttgering} H.,  {van Breugel} W.,   {De Breuck} C.,  2004,
  \mn@doi [\mnras] {10.1111/j.1365-2966.2004.08063.x}, \href
  {http://adsabs.harvard.edu/abs/2004MNRAS.353..377R} {353, 377}

\bibitem[\protect\citeauthoryear{{Robitaille} \& {Bressert}}{{Robitaille} \&
  {Bressert}}{2012}]{apl}
{Robitaille} T.,  {Bressert} E.,  2012, {APLpy: Astronomical Plotting Library
  in Python}, Astrophysics Source Code Library (\mn@eprint {ascl} {1208.017})

\bibitem[\protect\citeauthoryear{{Rocca-Volmerange}, {Le Borgne}, {De Breuck},
  {Fioc}  \& {Moy}}{{Rocca-Volmerange} et~al.}{2004}]{roc04}
{Rocca-Volmerange} B.,  {Le Borgne} D.,  {De Breuck} C.,  {Fioc} M.,   {Moy}
  E.,  2004, \mn@doi [\aap] {10.1051/0004-6361:20031717}, \href
  {http://adsabs.harvard.edu/abs/2004A%26A...415..931R} {415, 931}

\bibitem[\protect\citeauthoryear{{R{\"o}ttgering}, {Lacy}, {Miley}, {Chambers}
  \& {Saunders}}{{R{\"o}ttgering} et~al.}{1994}]{rot94}
{R{\"o}ttgering} H.~J.~A.,  {Lacy} M.,  {Miley} G.~K.,  {Chambers} K.~C.,
  {Saunders} R.,  1994, \aaps, \href
  {http://adsabs.harvard.edu/abs/1994A%26AS..108...79R} {108}

\bibitem[\protect\citeauthoryear{{R{\"o}ttgering}, {Daddi}, {Overzier}  \&
  {Wilman}}{{R{\"o}ttgering} et~al.}{2003}]{rot03}
{R{\"o}ttgering} H.,  {Daddi} E.,  {Overzier} R.,   {Wilman} R.,  2003, \mn@doi
  [\nar] {10.1016/S1387-6473(03)00129-5}, \href
  {http://adsabs.harvard.edu/abs/2003NewAR..47..309R} {47, 309}

\bibitem[\protect\citeauthoryear{{Saxena}, {R{\"o}ttgering}  \&
  {Rigby}}{{Saxena} et~al.}{2017}]{sax17}
{Saxena} A.,  {R{\"o}ttgering} H.~J.~A.,   {Rigby} E.~E.,  2017, \mn@doi
  [\mnras] {10.1093/mnras/stx1150}, \href
  {http://adsabs.harvard.edu/abs/2017MNRAS.469.4083S} {469, 4083}

\bibitem[\protect\citeauthoryear{{Seymour} et~al.,}{{Seymour}
  et~al.}{2007}]{sey07}
{Seymour} N.,  et~al., 2007, \mn@doi [\apjs] {10.1086/517887}, \href
  {http://adsabs.harvard.edu/abs/2007ApJS..171..353S} {171, 353}

\bibitem[\protect\citeauthoryear{{Seymour} et~al.,}{{Seymour}
  et~al.}{2008}]{sey08}
{Seymour} N.,  et~al., 2008, \mn@doi [\apjl] {10.1086/590081}, \href
  {http://adsabs.harvard.edu/abs/2008ApJ...681L...1S} {681, L1}

\bibitem[\protect\citeauthoryear{{Sheldon}, {Cunha}, {Mandelbaum}, {Brinkmann}
  \& {Weaver}}{{Sheldon} et~al.}{2012}]{she12}
{Sheldon} E.~S.,  {Cunha} C.~E.,  {Mandelbaum} R.,  {Brinkmann} J.,   {Weaver}
  B.~A.,  2012, \mn@doi [\apjs] {10.1088/0067-0049/201/2/32}, \href
  {http://adsabs.harvard.edu/abs/2012ApJS..201...32S} {201, 32}

\bibitem[\protect\citeauthoryear{{Sieber}}{{Sieber}}{1973}]{sie73}
{Sieber} W.,  1973, \aap, \href
  {http://adsabs.harvard.edu/abs/1973A%26A....28..237S} {28, 237}

\bibitem[\protect\citeauthoryear{{Smith} \& {Spinrad}}{{Smith} \&
  {Spinrad}}{1980}]{spi80}
{Smith} H.~E.,  {Spinrad} H.,  1980, \mn@doi [\apj] {10.1086/157758}, \href
  {http://adsabs.harvard.edu/abs/1980ApJ...236..419S} {236, 419}

\bibitem[\protect\citeauthoryear{{Spinrad}, {Westphal}, {Kristian}  \&
  {Sandage}}{{Spinrad} et~al.}{1977}]{spi77}
{Spinrad} H.,  {Westphal} J.,  {Kristian} J.,   {Sandage} A.,  1977, \mn@doi
  [\apjl] {10.1086/182517}, \href
  {http://adsabs.harvard.edu/abs/1977ApJ...216L..87S} {216, L87}

\bibitem[\protect\citeauthoryear{{Surace}, {Shupe}, {Fang}, {Evans}, {Alexov},
  {Frayer}, {Lonsdale}  \& {SWIRE Team}}{{Surace} et~al.}{2005}]{sur05}
{Surace} J.~A.,  {Shupe} D.~L.,  {Fang} F.,  {Evans} T.,  {Alexov} A.,
  {Frayer} D.,  {Lonsdale} C.~J.,   {SWIRE Team} 2005, in American Astronomical
  Society Meeting Abstracts. p.~1246

\bibitem[\protect\citeauthoryear{{Taylor}}{{Taylor}}{2005}]{top05}
{Taylor} M.~B.,  2005, in {Shopbell} P.,  {Britton} M.,   {Ebert} R.,  eds,
  Astronomical Society of the Pacific Conference Series Vol. 347, Astronomical
  Data Analysis Software and Systems XIV. p.~29

\bibitem[\protect\citeauthoryear{{Tielens}, {Miley}  \& {Willis}}{{Tielens}
  et~al.}{1979}]{tie79}
{Tielens} A.~G.~G.~M.,  {Miley} G.~K.,   {Willis} A.~G.,  1979, \aaps, \href
  {http://adsabs.harvard.edu/abs/1979A%26AS...35..153T} {35, 153}

\bibitem[\protect\citeauthoryear{{Tingay} et~al.,}{{Tingay}
  et~al.}{2013}]{tin13}
{Tingay} S.~J.,  et~al., 2013, \mn@doi [\pasa] {10.1017/pasa.2012.007}, \href
  {http://adsabs.harvard.edu/abs/2013PASA...30....7T} {30, e007}

\bibitem[\protect\citeauthoryear{{Venemans} et~al.,}{{Venemans}
  et~al.}{2002}]{ven02}
{Venemans} B.~P.,  et~al., 2002, \mn@doi [\apjl] {10.1086/340563}, \href
  {http://adsabs.harvard.edu/abs/2002ApJ...569L..11V} {569, L11}

\bibitem[\protect\citeauthoryear{{Villar-Mart{\'{\i}}n}, {Humphrey}, {De
  Breuck}, {Fosbury}, {Binette}  \& {Vernet}}{{Villar-Mart{\'{\i}}n}
  et~al.}{2007}]{vil07}
{Villar-Mart{\'{\i}}n} M.,  {Humphrey} A.,  {De Breuck} C.,  {Fosbury} R.,
  {Binette} L.,   {Vernet} J.,  2007, \mn@doi [\mnras]
  {10.1111/j.1365-2966.2006.11371.x}, \href
  {http://adsabs.harvard.edu/abs/2007MNRAS.375.1299V} {375, 1299}

\bibitem[\protect\citeauthoryear{White, Becker, Helfand  \& Gregg}{White
  et~al.}{1997}]{White1997}
White R.~L.,  Becker R.~H.,  Helfand D.~J.,   Gregg M.~D.,  1997, \mn@doi
  [\apj] {10.1086/303564}, 475, 479

\bibitem[\protect\citeauthoryear{{Willott}, {Rawlings}, {Jarvis}  \&
  {Blundell}}{{Willott} et~al.}{2003}]{wil03}
{Willott} C.~J.,  {Rawlings} S.,  {Jarvis} M.~J.,   {Blundell} K.~M.,  2003,
  \mn@doi [\mnras] {10.1046/j.1365-8711.2003.06172.x}, \href
  {http://adsabs.harvard.edu/abs/2003MNRAS.339..173W} {339, 173}

\bibitem[\protect\citeauthoryear{{Wright} et~al.,}{{Wright}
  et~al.}{2010}]{wri10}
{Wright} E.~L.,  et~al., 2010, \mn@doi [\aj] {10.1088/0004-6256/140/6/1868},
  \href {http://adsabs.harvard.edu/abs/2010AJ....140.1868W} {140, 1868}

\bibitem[\protect\citeauthoryear{{Wu} et~al.,}{{Wu} et~al.}{2012}]{wu12}
{Wu} J.,  et~al., 2012, \mn@doi [\apj] {10.1088/0004-637X/756/1/96}, \href
  {http://adsabs.harvard.edu/abs/2012ApJ...756...96W} {756, 96}

\bibitem[\protect\citeauthoryear{{Xu}, {Chen}, {Fan}, {Trac}  \& {Cen}}{{Xu}
  et~al.}{2009}]{xu09}
{Xu} Y.,  {Chen} X.,  {Fan} Z.,  {Trac} H.,   {Cen} R.,  2009, \mn@doi [\apj]
  {10.1088/0004-637X/704/2/1396}, \href
  {http://adsabs.harvard.edu/abs/2009ApJ...704.1396X} {704, 1396}

\bibitem[\protect\citeauthoryear{{de Gasperin}, {Intema}  \& {Frail}}{{de
  Gasperin} et~al.}{2017}]{deg17}
{de Gasperin} F.,  {Intema} H.~T.,   {Frail} D.~A.,  2017, preprint, \href
  {http://adsabs.harvard.edu/abs/2017arXiv171111367D} {} (\mn@eprint {arXiv}
  {1711.11367})

\bibitem[\protect\citeauthoryear{{van Breugel}, {De Breuck}, {Stanford},
  {Stern}, {R{\"o}ttgering}  \& {Miley}}{{van Breugel} et~al.}{1999}]{vbr99}
{van Breugel} W.,  {De Breuck} C.,  {Stanford} S.~A.,  {Stern} D.,
  {R{\"o}ttgering} H.,   {Miley} G.,  1999, \mn@doi [\apjl] {10.1086/312080},
  \href {http://adsabs.harvard.edu/abs/1999ApJ...518L..61V} {518, L61}

\bibitem[\protect\citeauthoryear{{van Haarlem} et~al.,}{{van Haarlem}
  et~al.}{2013}]{lofar}
{van Haarlem} M.~P.,  et~al., 2013, \mn@doi [\aap]
  {10.1051/0004-6361/201220873}, \href
  {http://adsabs.harvard.edu/abs/2013A%26A...556A...2V} {556, A2}

\makeatother
\end{thebibliography}




\appendix
\section{Source catalog}
Here we report the measured flux densities and angular sizes of all sources imaged with the VLA, with an average beam size of $1.3''$.

\begin{landscape}
\begin{table}
\centering
\footnotesize
\caption{Flux densities and angular sizes from VLA L-band observations of 32 HzRG candidates with a detection threshold of at least $5\sigma$. The sources are listed in order of Right Ascension. All source names in the column Catalogue ID have `TGSSADR' as prefix. The column LAS stands for largest angular size. The source USS255 was not observed with the VLA, but we include it here nonetheless to present out complete sample. We also present the source ID from VLA FIRST and the reported flux density in the FIRST source catalogue. The average error in the reported flux density from FIRST is roughly $5-10$\%.}
\begin{tabular}{c c c c c c c c H c H c c c}
\hline
\\
  \multicolumn{1}{c}{Catalogue ID} &
  \multicolumn{1}{c}{Name} &
  \multicolumn{1}{c}{Component} &
  \multicolumn{1}{c}{RA (J2000)} &
  \multicolumn{1}{c}{DEC (J2000)} &
  \multicolumn{1}{c}{S$_{\textrm{1.4}}$ (mJy)} &
  \multicolumn{1}{c}{S$_{\textrm{1.4}}^{\textrm{tot}}$ (mJy)} &
  \multicolumn{1}{c}{S$_{\textrm{150}}$ (mJy)} &
  \multicolumn{1}{H}{S$_{\textrm{F}}$ (mJy)} &
  \multicolumn{1}{c}{$\alpha_{\textrm{1.4}}^{\textrm{150}}$} &
  \multicolumn{1}{H}{S$_{\textrm{N}}$ (mJy)} &
  \multicolumn{1}{c}{LAS ($''$)} &
  \multicolumn{1}{c}{FIRST ID} &
  \multicolumn{1}{c}{S$_{\textrm{FIRST}}^{\textrm{tot}}$ (mJy)} \\
\\
\hline \hline
\\
J004828.6+115143 & USS255 & & 00:48:28.6 & +11:51:43.9 & 3.8 $\pm$ 0.2 & $-$ &  84 $\pm$ 16 & 3.8 $\pm$ 0.2 & $-1.4 $ $\pm$ 0.1 & 5.0 $\pm$ 0.5 & $-$ & J004828.4+115144 & 3.7 \\

J070319.0+515157 & USS206 & A & 07:03:18.5 & +51:52:01.2 & 4.6 $\pm$ 0.09 & 7.35 $\pm$ 0.1 & 102 $\pm$ 20 & 2.6 $\pm$ 0.1 & $-1.6 $ $\pm$ 0.1 & 7.1 $\pm$ 0.5 & 8.46 & J070319.0+515155 & 2.6 \\
& & B & 07:03:19.0 & +51:51:54.0 & 2.75 $\pm$ 0.1 & &  &  &  &  & & &  \\

J071622.5+370338 & USS182 & A  & 07:16:22.6 & +37:03:36.0 & 5.02 $\pm$ 0.07 & 8.54 $\pm$ 0.1 & 146 $\pm$ 30 & 6.4 $\pm$ 0.3 & $-1.4 $ $\pm$ 0.1 & 6.5 $\pm$ 0.5 & 3.17 & J071622.5+370337 & 6.4 \\
& & B & 07:16:22.3 & +37:03:39.6 & 3.52 $\pm$ 0.05 & & &  &  &  & & & \\

J071737.9+592248 & USS31 & A  & 07:17:37.9 & +59:22:44.4 & 3.91 $\pm$ 0.03 & 4.84 $\pm$ 0.03 & 149 $\pm$ 30 & 4.4 $\pm$ 0.1 & $-1.7$ $\pm$ 0.1 & 4.4 $\pm$ 0.4 & 7.55 & J071738.2+592245 & 1.4 \\
& & B & 07:17:37.2 & +59:22:51.6 & 0.93 $\pm$ 0.03 & & & & & & & & \\

J073036.8+481103 & USS188 & A & 07:30:36.7 & +48:11:06.0 & 5.67 $\pm$ 0.09 & 9.2 $\pm$ 0.1 & 166 $\pm$ 33 & 8.71 $\pm$ 0.4 & $-1.3$ $\pm$ 0.1 & 7.3 $\pm$ 0.5 & 4.20 & J073036.6+481104 & 8.7 \\
& & B & 07:30:36.5 & +48:11:02.4 & 3.53 $\pm$ 0.02 & &  &  &  &  & & & \\

J075751.8+611620 & USS309 & A & 07:57:51.6 & +61:16:19.2 & 4.61 $\pm$ 0.12 & 8.4 $\pm$ 0.1 & 83 $\pm$ 17 & 4.7 $\pm$ 0.1 & $-1.3 $ $\pm$ 0.1 & 7.5 $\pm$ 0.5 & 6.10 & J075751.7+611624 & 2.8 \\
& & B & 07:57:51.6 & +61:16:26.4 & 3.79 $\pm$ 0.07 & & & & & & & & \\

J080849.9+382603 & USS27 & & 08:08:49.9 & +38:26:02.4 & 7.63 $\pm$ 0.06 & & 111 $\pm$ 22 & 6.1 $\pm$ 0.3 & $-1.3 $ $\pm$ 0.1 & 6.0 $\pm$ 0.4 & 2.63 & J080849.9+382604 & 6.1 \\
 
J081103.2+511443 & USS36 & & 08:11:03.1 & +51:14:42.0 & 6.01 $\pm$ 0.1 & &  123 $\pm$ 25 & 5.4 $\pm$ 0.3 & $-1.4 $ $\pm$ 0.1 & 4.6 $\pm$ 0.4 & 1.33 & J081103.2+511442 & 5.4\\

J085746.8+175756 & USS273 & & 08:57:46.8 & +17:57:57.6 & 7.37 $\pm$ 0.05 & &  154 $\pm$ 31 & 7.8 $\pm$ 0.4 & $-1.3 $ $\pm$ 0.1 & 6.5 $\pm$ 0.5 & 1.37 & J085746.8+175757 & 7.8 \\

J095635.3+270823 & USS72 & & 09:56:35.3 & +27:08:24.0 & 3.85 $\pm$ 0.05 & &  52 $\pm$ 10 & 2.7  $\pm$ 0.1 & $-1.4 $ $\pm$ 0.1 & 3.1 $\pm$ 0.4 & 1.29 & J095635.4+270822 & 3.2 \\
  
J095946.1+520906 & USS483 & A & 09:59:46.1 & +52:09:07.2 & 2.11 $\pm$ 0.06 & 3.3 $\pm$ 0.1 & 89 $\pm$ 18 & 3.1  $\pm$ 0.2 & $-1.5 $ $\pm$ 0.1 & 3.6 $\pm$ 0.4 & 2.77 & J095946.1+520907 & 3.1 \\
& & B & 09:59:45.8 & +52:09:03.6 & 1.19 $\pm$ 0.06 & &  &  &  &  & & & \\

J101813.4+111746 & USS46 & & 10:18:13.4 & +11:17:45.6 & 4.91 $\pm$ 0.1 & & 87 $\pm$ 17 & 4.1 $\pm$ 0.2 & $-1.4 $ $\pm$ 0.1 & 5.3 $\pm$ 0.4 & 1.62 & J101813.4+111745 & 4.2 \\

J105429.5+583226 & USS410 & & 10:54:29.5 & +58:32:27.6 & 3.75 $\pm$ 0.05 & & 96 $\pm$ 19 & 2.8  $\pm$ 0.1 & $-1.6 $ $\pm$ 0.1 & 3.8 $\pm$ 0.5 & 1.29 & J105429.5+583226 & 3.1 \\

J110638.7+545018 & USS18 & & 11:06:38.6 & +54:50:20.4 & 5.72 $\pm$ 0.1 & & 131 $\pm$ 26 & 5.4 $\pm$ 0.3 & $-1.4 $ $\pm$ 0.1 & 5.6 $\pm$ 0.4 & 1.73 & J110638.8+545020 & 5.4 \\

J111054.9+610200 & USS172 & A & 11:10:55.4 & +61:02:02.4 & 3.77 $\pm$ 0.06 & 7.1 $\pm$ 0.1 & 124 $\pm$ 25 & 5.1  $\pm$ 0.2 & $-1.4 $ $\pm$ 0.1 & 6.6 $\pm$ 0.5 & 9.67 & J111055.4+610202 & 3.2 \\
& & B & 11:10:54.2 & +61:01:58.8 & 3.31 $\pm$ 0.04 & & & & & & & & \\

J113517.5+465906 & USS159 & & 11:35:17.5 & +46:59:06.0 & 8.01 $\pm$ 0.05 & & 149 $\pm$ 30 & 7.0 $\pm$ 0.3 & $-1.4 $ $\pm$ 0.1 & 6.3 $\pm$ 0.5 & 1.22 & J113517.6+465905 & 6.9 \\

J114006.3+464451 & USS253 & A & 11:40:06.2 & +46:44:49.2 & 6.36 $\pm$ 0.1 & 8.13 $\pm$ 0.1 & 156 $\pm$ 31 & 5.9 $\pm$ 0.3 & $-1.5 $ $\pm$ 0.1 & 6.9 $\pm$ 0.5 & 10.0 & J114006.2+464449 & 5.9 \\
& & B & 11:40:06.7 & +46:44:52.8 & 0.87 $\pm$ 0.07 & & & & & & & & \\
& & C & 11:40:06.7 & +46:44:56.4 & 0.90 $\pm$ 0.07 & & & & & & & & \\
 \\
\hline\end{tabular}
\label{tab:vlaflux}
\end{table}
\end{landscape}

\begin{landscape}
\begin{table}
\centering
\ContinuedFloat
\captionsetup{list=off,format=cont}
\caption{Continued.}
\begin{tabular}{c c c c c c c c H c H c c c}
\hline
\\
  \multicolumn{1}{c}{Catalogue ID} &
  \multicolumn{1}{c}{Name} &
  \multicolumn{1}{c}{Component} &
  \multicolumn{1}{c}{RA (J2000)} &
  \multicolumn{1}{c}{DEC (J2000)} &
  \multicolumn{1}{c}{S$_{\textrm{1.4}}$ (mJy)} &
  \multicolumn{1}{c}{S$_{\textrm{1.4}}^{\textrm{tot}}$ (mJy)} &
  \multicolumn{1}{c}{S$_{\textrm{150}}$ (mJy)} &
  \multicolumn{1}{H}{S$_{\textrm{F}}$ (mJy)} &
  \multicolumn{1}{c}{$\alpha_{\textrm{1.4}}^{\textrm{150}}$} &
  \multicolumn{1}{H}{S$_{\textrm{N}}$ (mJy)} &
  \multicolumn{1}{c}{LAS ($''$)} &
  \multicolumn{1}{c}{FIRST ID} &
  \multicolumn{1}{c}{S$_{\textrm{FIRST}}^{\textrm{tot}}$ (mJy)} \\
\\
\hline \hline
\\
J122043.6+611906 & USS312 & & 12:20:43.9 & +61:19:04.8 & 2.52 $\pm$ 0.05 & & 79 $\pm$ 16 & 3.4  $\pm$ 0.2 & $-1.4 $ $\pm$ 0.1 & 4.6 $\pm$ 0.4 & 1.26 & J122043.9+611907 & 3.4 \\

J124158.9+204009 & USS439 & A & 12:41:59.0 & +20:40:08.4 & 6.82 $\pm$ 0.06 & 7.95 $\pm$ 0.1 & 136 $\pm$ 27 & 7.3 $\pm$ 0.4 & $-1.3 $ $\pm$ 0.1 & 9.2 $\pm$ 0.5 & 7.76 & J124158.9+204009 & 7.3 \\
& & B & 12:41:58.8 & +20:40:08.4 & 0.30 $\pm$ 0.04 & & & & & & & & \\
& & C & 12:41:58.6 & +20:40:12.0 & 0.32 $\pm$ 0.04 & & & & & & & & \\
& & D & 12:41:58.6 & +20:40:12.0 & 0.51 $\pm$ 0.04 & & & & & & & & \\

J131206.0+452257 & USS268 & & 13:12:06.0 & +45:22:55.2 & 4.62 $\pm$ 0.05 & & 92 $\pm$ 18 & 3.8 $\pm$ 0.2 & $-1.4 $ $\pm$ 0.1 & 4.9 $\pm$ 0.5 & 2.02 & J131206.0+452257 & 3.9 \\

J132801.4+202858 & USS67 & A & 13:28:01.2 & +20:28:51.6 & 2.88 $\pm$ 0.05 & 5.1 $\pm$ 0.1 & 59 $\pm$ 12 & 3.5 $\pm$ 0.1 & $-1.3 $ $\pm$ 0.1 & 5.0 $\pm$ 0.4 & 9.40 & J132801.6+202901 & 1.9 \\
& & B & 13:28:01.7 & +20:28:58.8 & 2.20 $\pm$ 0.05 & & & & & & & & \\
  
J133911.8+203348 & USS32 & & 13:39:11.8 & +20:33:50.4 & 3.99 $\pm$ 0.07 & & 76 $\pm$ 15 & 3.9  $\pm$ 0.2 & $-1.4 $ $\pm$ 0.1 & 4.2 $\pm$ 0.5 & 2.09 & J133911.7+203349 & 3.9 \\ 

J135045.7+221716 & USS415 & A & 13:50:45.4 & +22:17:16.8 & 3.86 $\pm$ 0.1 & 6.62 $\pm$ 0.1 & 188 $\pm$ 38 & 3.8 $\pm$ 0.1 & $-1.7 $ $\pm$ 0.1 & 6.7 $\pm$ 0.5 & 8.12 & J135045.8+221713 & 2.1 \\ 
& & B & 13:50:45.8 & +22:17:13.2 & 2.76 $\pm$ 0.05 & & & & & & & & \\

J145154.4+400656 & USS320 & A & 14:51:54.5 & +40:06:57.6 & 2.15 $\pm$ 0.03 & 3.03 $\pm$ 0.1 & 86 $\pm$ 17 & 2.7  $\pm$ 0.1 & $-1.7 $ $\pm$ 0.1 & 3.2 $\pm$ 0.5 & 2.70 & J145154.5+400656 & 2.7  \\
& & B & 14:51:54.2 & +40:06:54.0 & 0.88 $\pm$ 0.03 & & & & & & & & \\

J153049.9+104933 & USS202 & & 15:30:49.9 & +10:49:30.0 & 7.50 $\pm$ 0.07 & & 170 $\pm$ 34 & 7.3 $\pm$ 0.4 & $-1.4 $ $\pm$ 0.1 & 7.4 $\pm$ 0.5 & 1.69 & J153049.8+104931 & 7.3 \\ 

J153813.3+134153 & USS43 & & 15:38:13.4 & +13:41:52.8 & 12.5 $\pm$ 0.25 & & 200 $\pm$ 40 & 9.6 $\pm$ 0.5 & $-1.4 $ $\pm$ 0.1 & 7.9 $\pm$ 0.5 & 1.47 & J153813.4+134153 & 9.6 \\

J160819.1+123951 & USS51 & A & 16:08:19.2 & +12:39:50.4 & 6.52 $\pm$ 0.04 & 7.88 $\pm$ 0.1 & 59 $\pm$ 12 & 3.2  $\pm$ 0.2 & $-1.3 $ $\pm$ 0.1 & 4.4 $\pm$ 0.4 & 5.69 & J160819.1+123950 & 3.2 \\
& & B & 16:08:19.2 & +12:39:54.0 & 1.36 $\pm$ 0.02 & & & & & & & & \\

J161026.5+103924 & USS450 & A & 16:10:26.6 & +10:39:21.6 & 6.55 $\pm$ 0.07 & 9.65 $\pm$ 0.2 & 131 $\pm$ 26 & 7.2 $\pm$ 0.4 & $-1.4 $ $\pm$ 0.1 & 9.7 $\pm$ 0.5 & 15.78 & J161026.1+103937 & 3.0 \\
& & B & 16:10:26.2 & +10:39:36.0 & 3.10 $\pm$ 0.05 & & & & & & & & \\ 

J161538.9+532726 & USS7 & A  & 16:15:38.9 & +53:27:25.2 & 2.80 $\pm$ 0.04 & 6.33 $\pm$ 0.1 & 122 $\pm$ 24 & 4.3 $\pm$ 0.2 & $-1.5 $ $\pm$ 0.1 & 4.1 $\pm$ 0.4 & 3.72 & J161539.1+532727 & 4.3 \\
& & B & 16:15:39.1 & +53:27:28.8 & 3.53 $\pm$ 0.03 & & & & & & & & \\ 

J161907.9+631250 & USS404 & & 16:19:08.2 & +63:12:46.8 & 9.09 $\pm$ 0.09 & & 153 $\pm$ 31 & 7.7 $\pm$ 0.4 & $-1.4 $ $\pm$ 0.1 & 8.5 $\pm$ 0.5 & 2.12 & J161908.3+631248 & 7.7 \\

J162521.8+254647 & USS337 & & 16:25:21.8 & +25:46:48.0 & 11.14 $\pm$ 0.33 & & 171 $\pm$ 34 & 8.8 $\pm$ 0.4 & $-1.3 $ $\pm$ 0.1 & 11.7 $\pm$ 0.9 & 1.66 & J162521.8+254648 & 8.8 \\  

J163643.1+195830 & USS98 & A & 16:35:43.2 & +19:58:30.0 & 1.47 $\pm$ 0.20 & 3.8 $\pm$ 0.3 & 178 $\pm$ 36 & 5.6  $\pm$ 0.1 & $-1.5 $ $\pm$ 0.1 & 7.4 $\pm$ 1.1 & 7.24 & J163543.2+195833 & 2.0 \\ 
& & B & 16:35:43.0 & +19:58:22.8 & 2.33 $\pm$ 0.22 & & & & & & & & \\ 
 \\
 \hline\end{tabular}
 \end{table}

 \end{landscape}

\section{VLA L-band images and contour maps}
Here we show VLA L-band (1.4 GHz) A-configuration images in greyscale, with contours overlaid for all 31 HzRG candidates that were observed. The average beam size is $\sim1.3''$ and the noise in the images varies from $50$ to \SI{70}{\micro\jansky}. The contours begin at \SI{250}{\micro\jansky}, which translates to roughly $3.5-5\sigma$. The images are presented in order of source Right Ascension.

\begin{figure*}
\centering
\includegraphics[width=.35\textwidth]{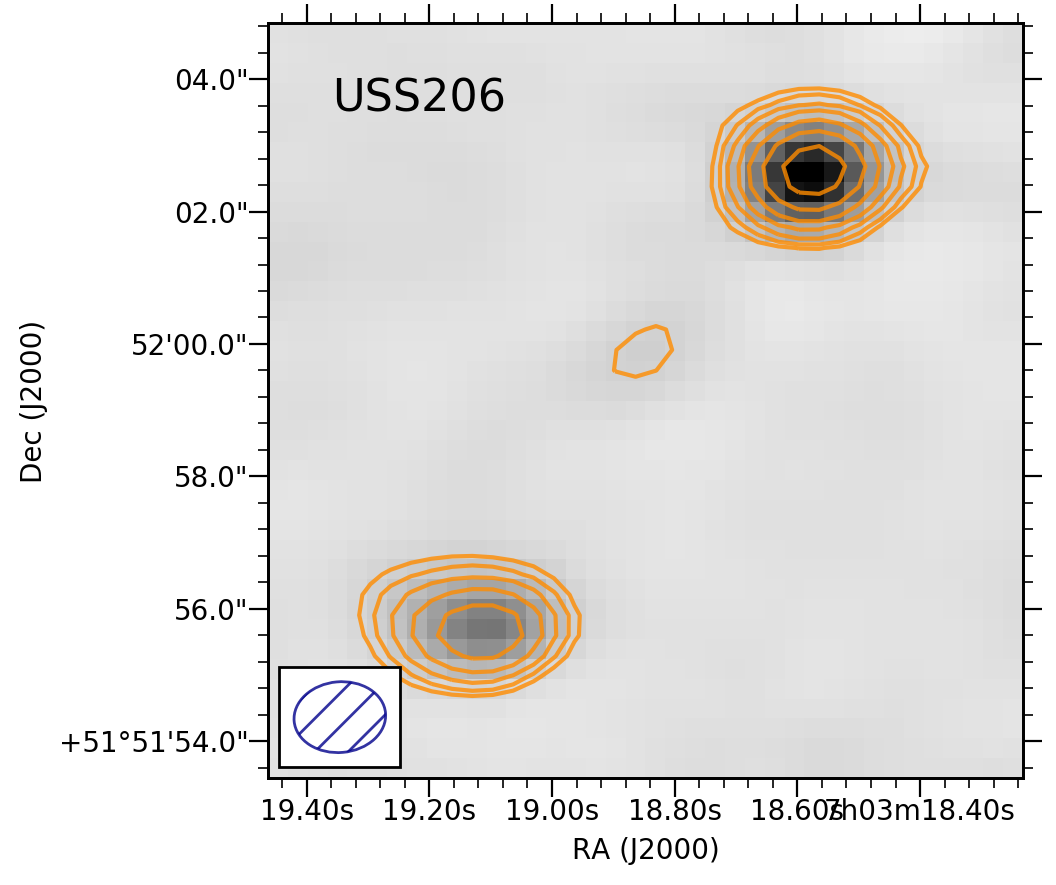}
\includegraphics[width=.36\textwidth]{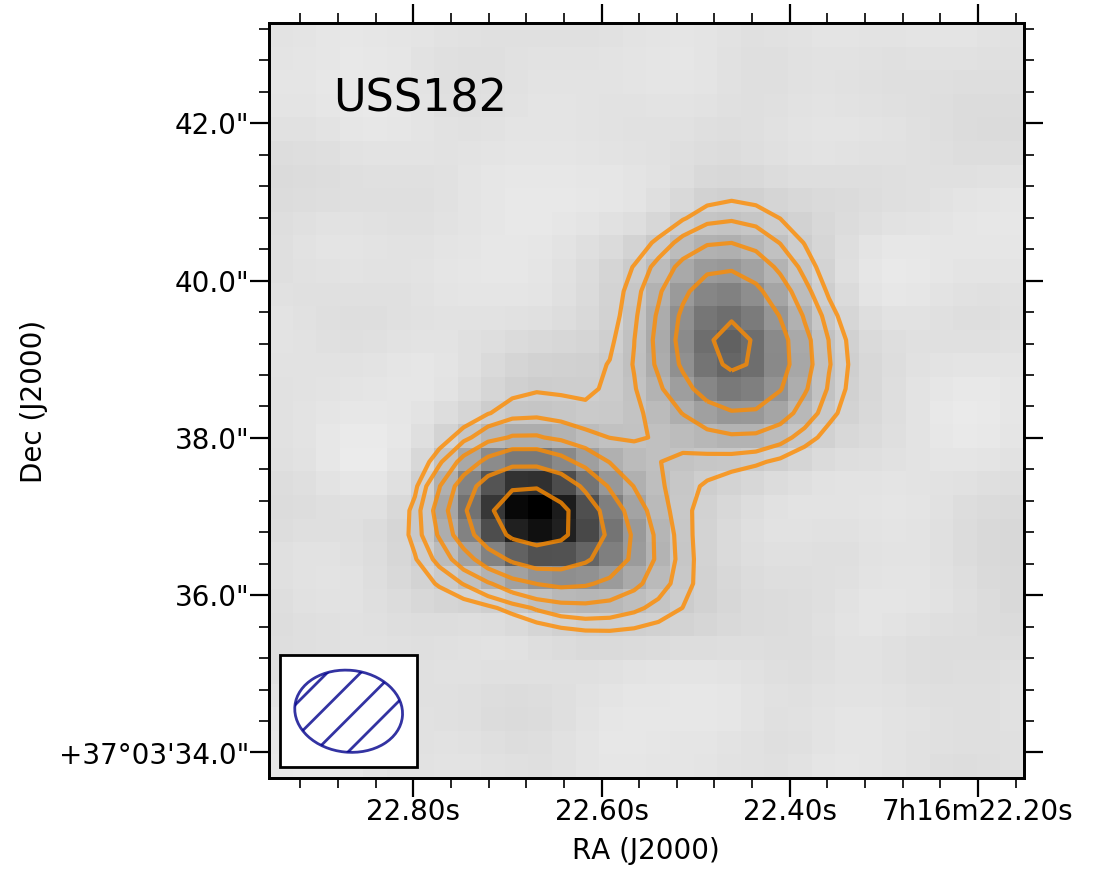}

\includegraphics[width=.355\textwidth]{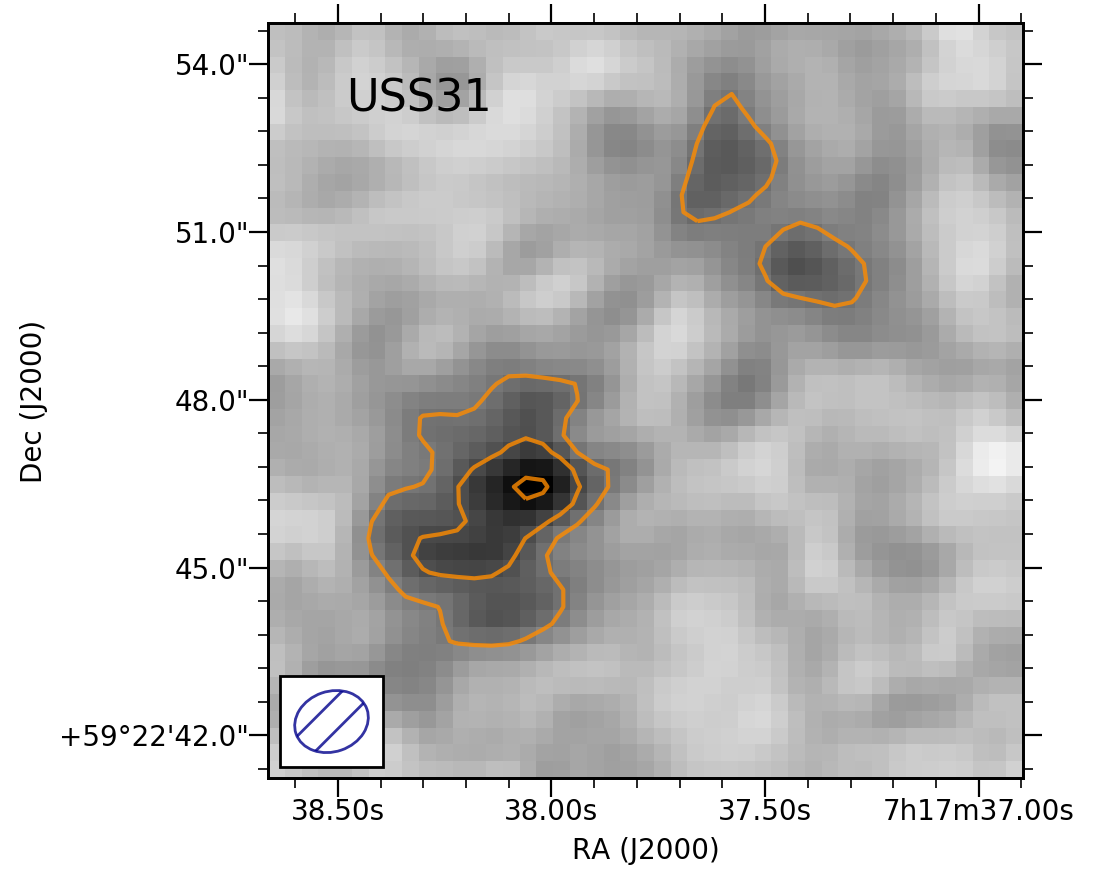}
\includegraphics[width=.35\textwidth]{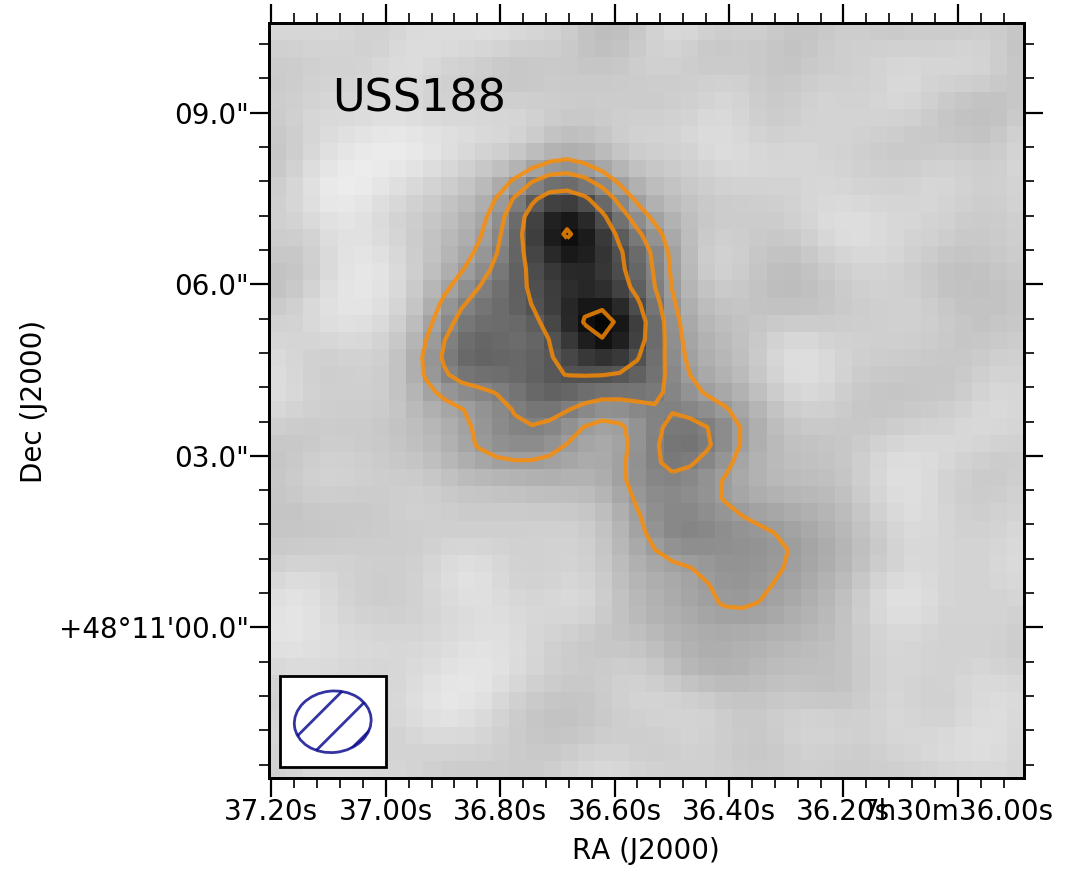}

\includegraphics[width=.4\textwidth]{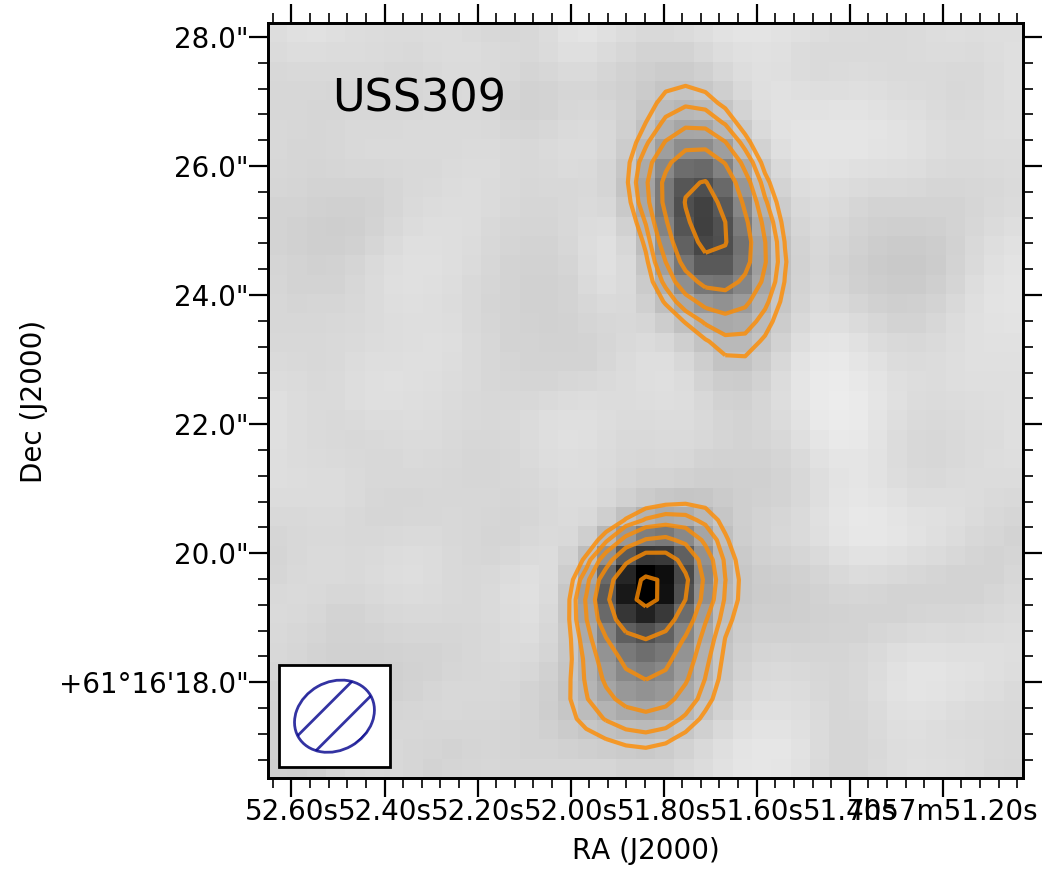}

\includegraphics[width=.32\textwidth]{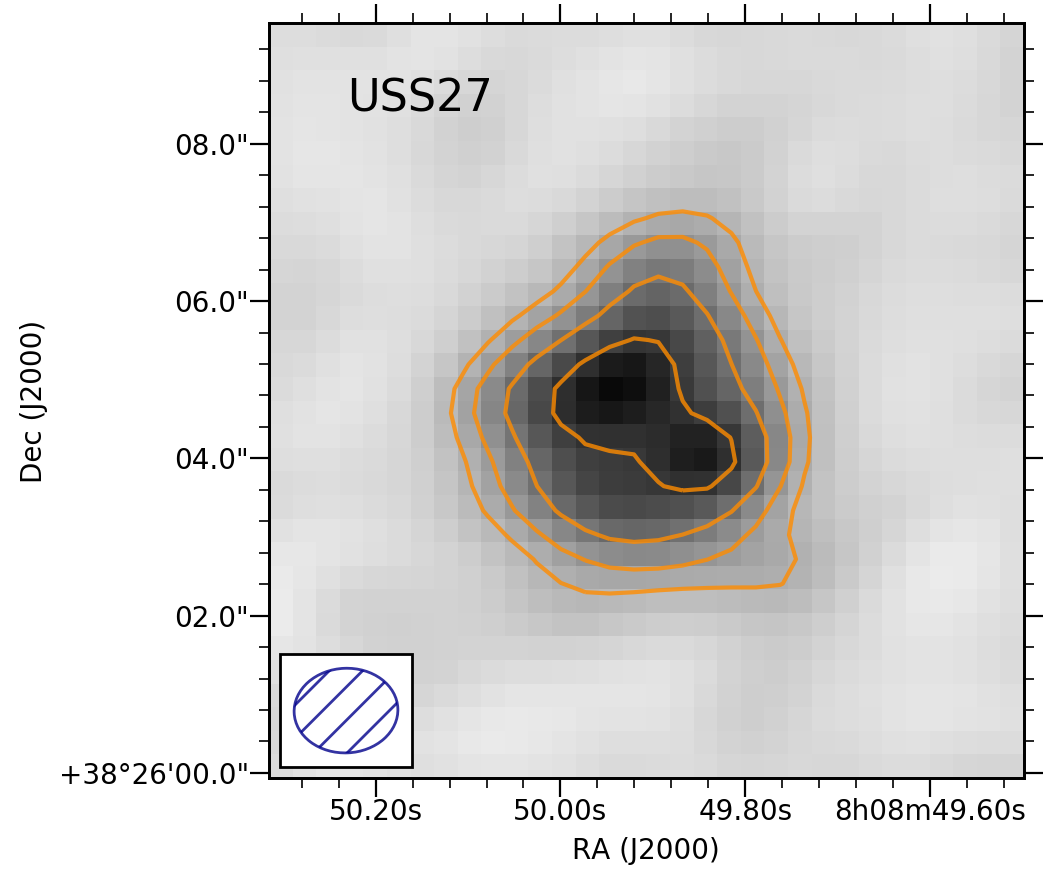}
\includegraphics[width=.32\textwidth]{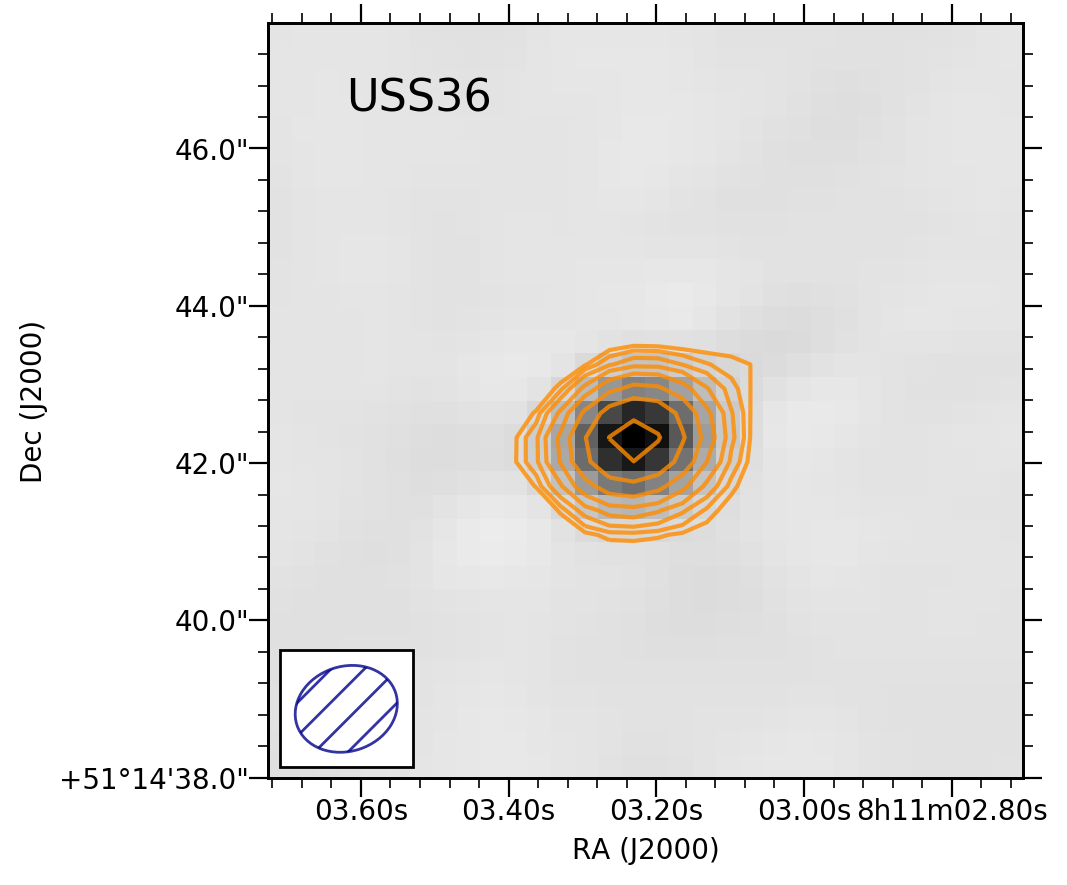}
\includegraphics[width=.32\textwidth]{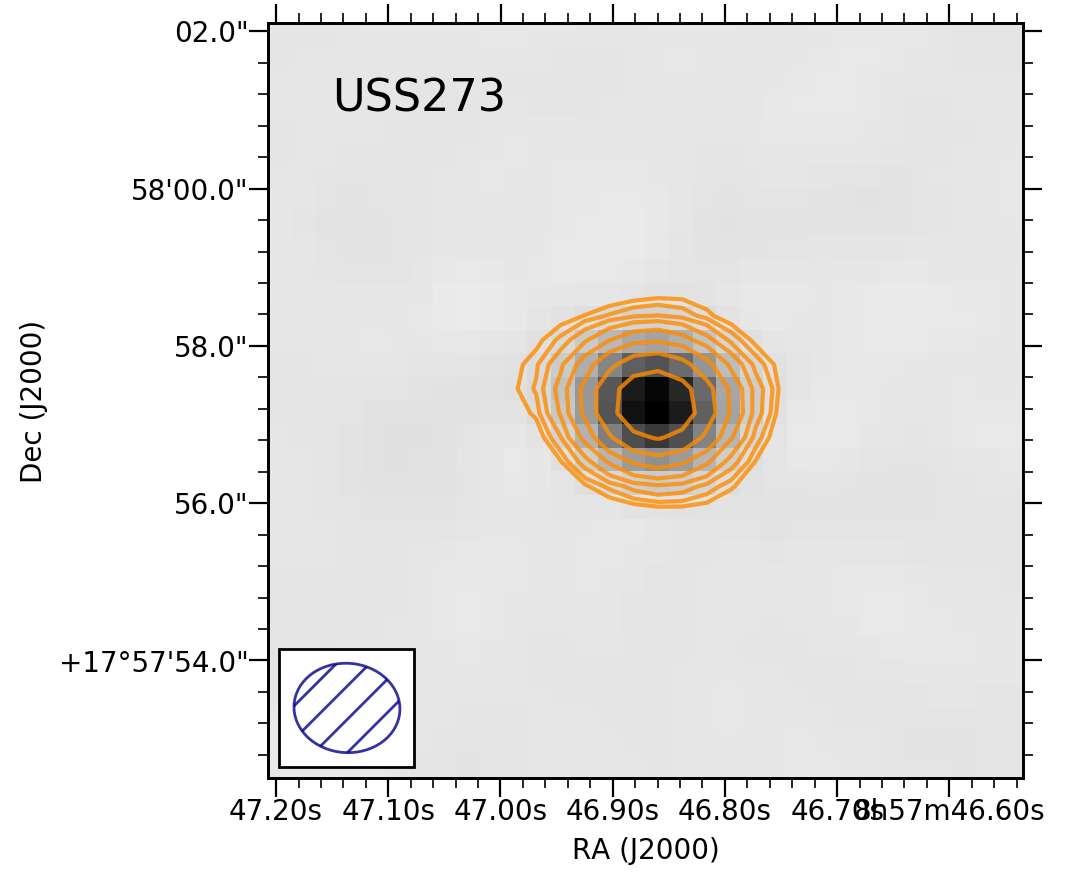}
\caption{VLA L-band images and contours for 31 sources observed. The contours shown begin at 0.25 mJy, which is on average $3.5-5\sigma$ and are a geometric progression of $\sqrt{2}$, such that the flux density increases by a factor of 2 for every 2 contours. The beam used to image each source is shown in the bottom-left.}
\label{fig:figure3}
\end{figure*}

\begin{figure*}
\ContinuedFloat
\captionsetup{list=off,format=cont}
\centering
\includegraphics[width=.4\textwidth]{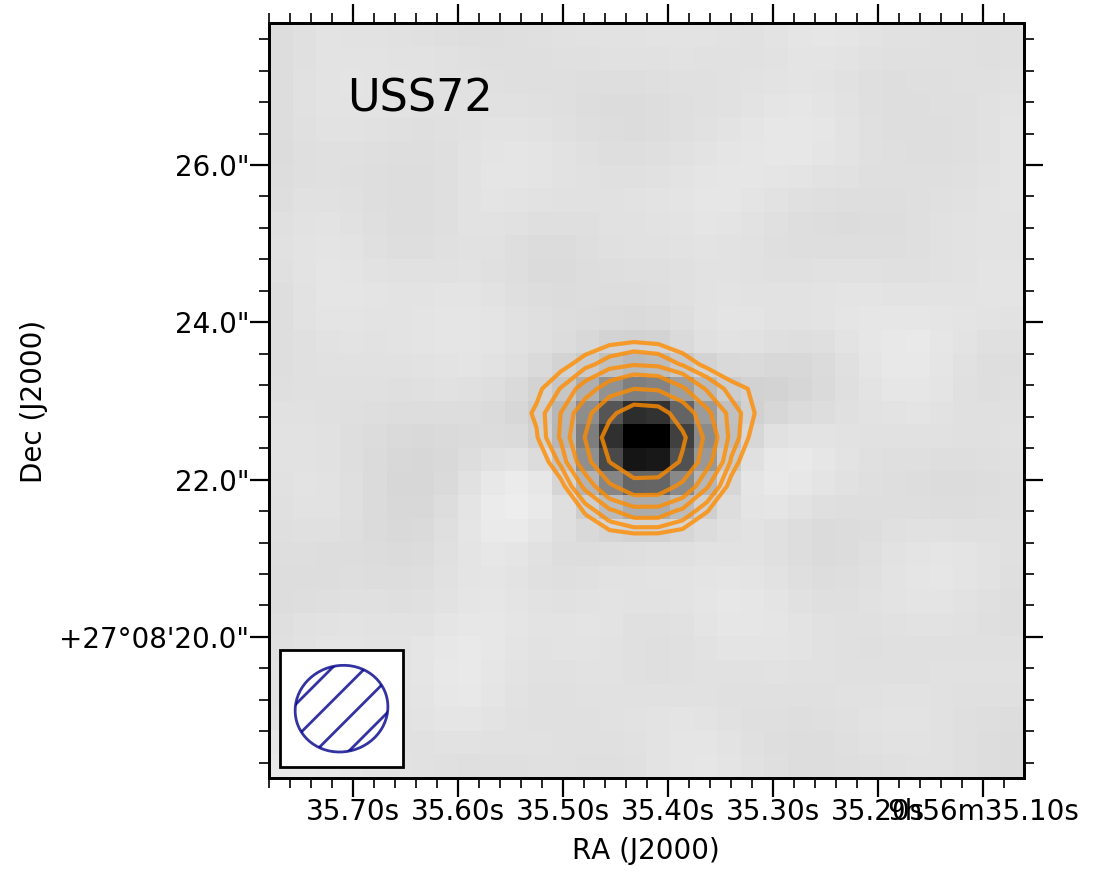}
\includegraphics[width=.41\textwidth]{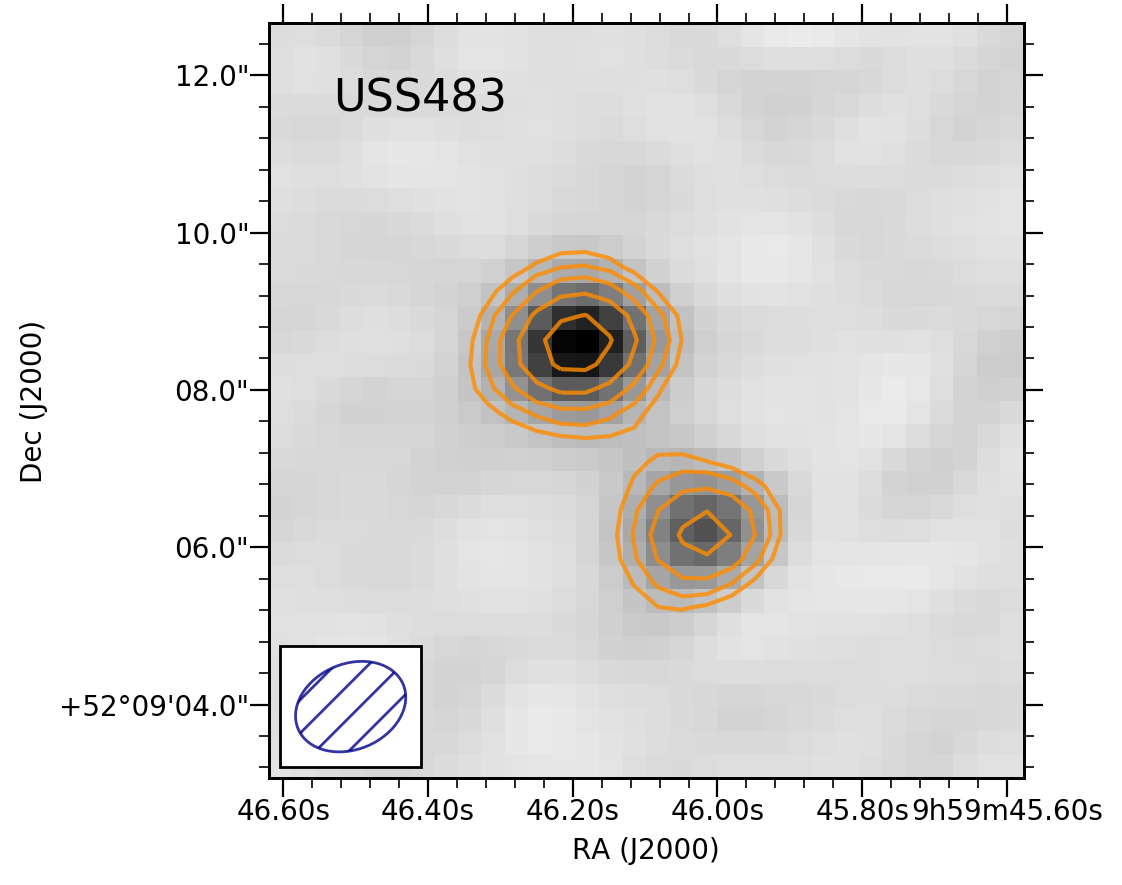}

\includegraphics[width=.33\textwidth]{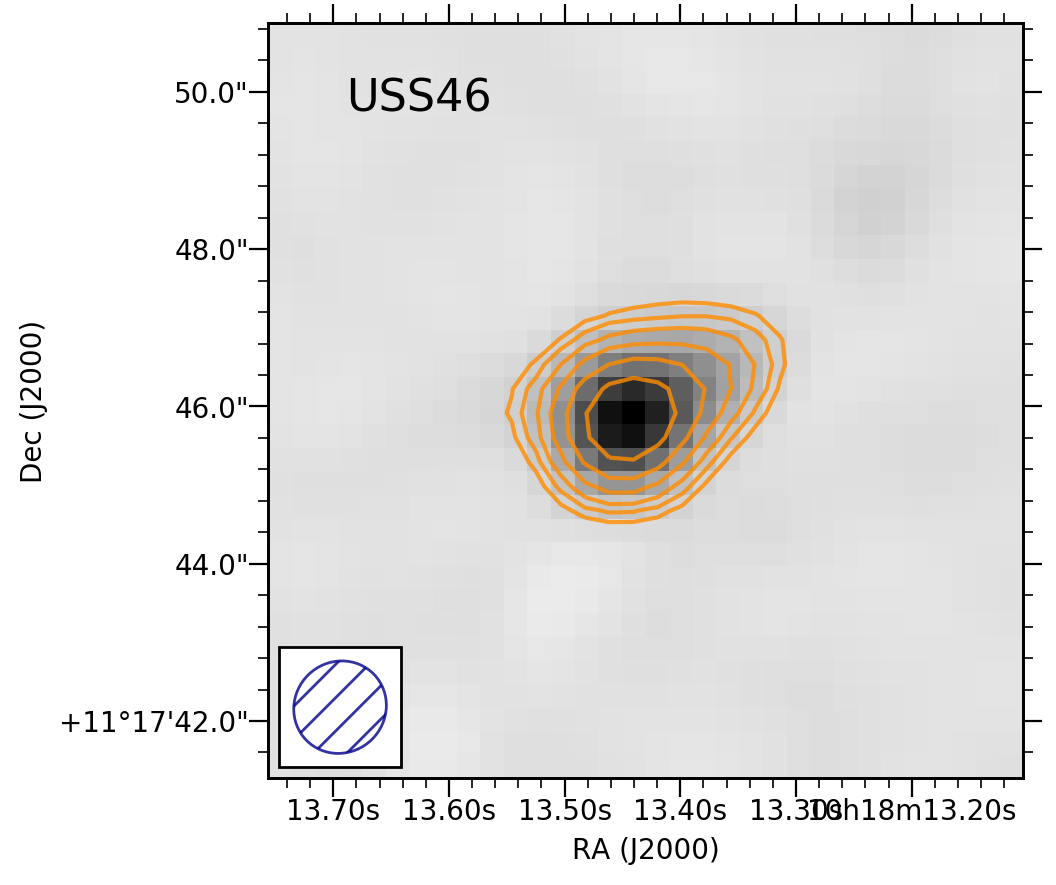}
\includegraphics[width=.33\textwidth]{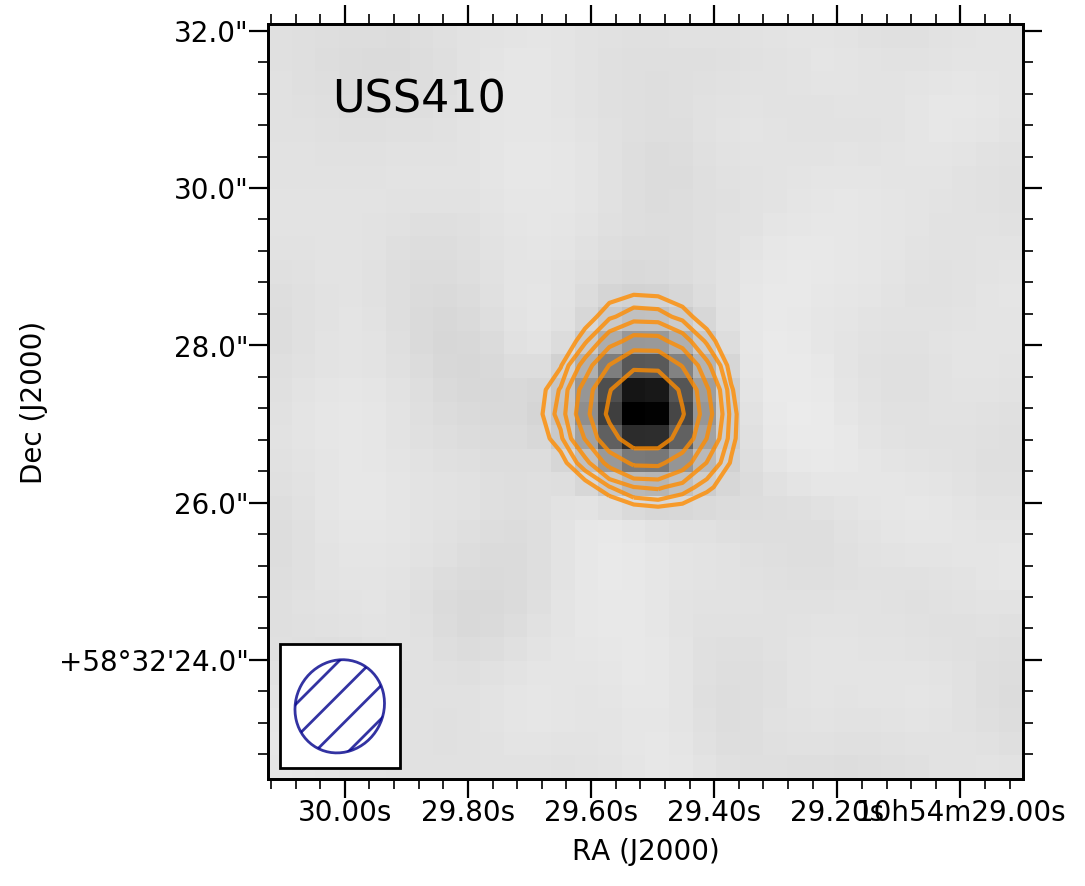}
\includegraphics[width=.328\textwidth]{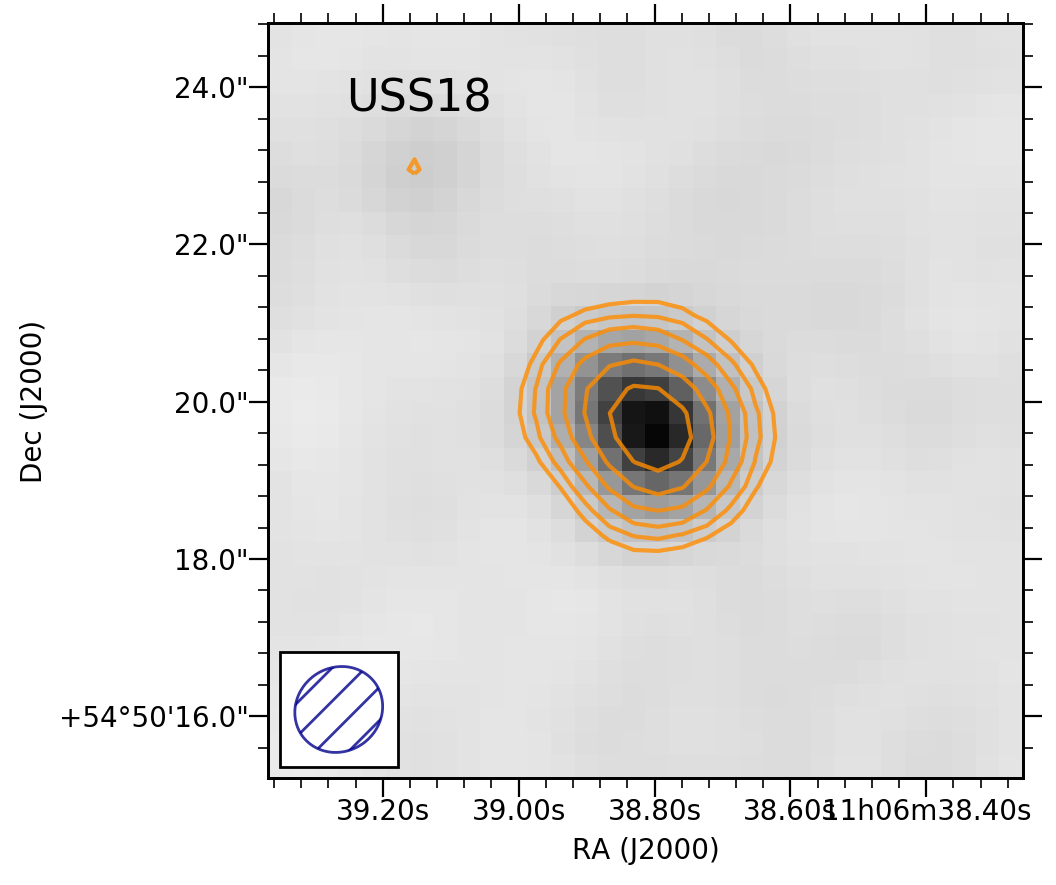}

\includegraphics[width=.4\textwidth]{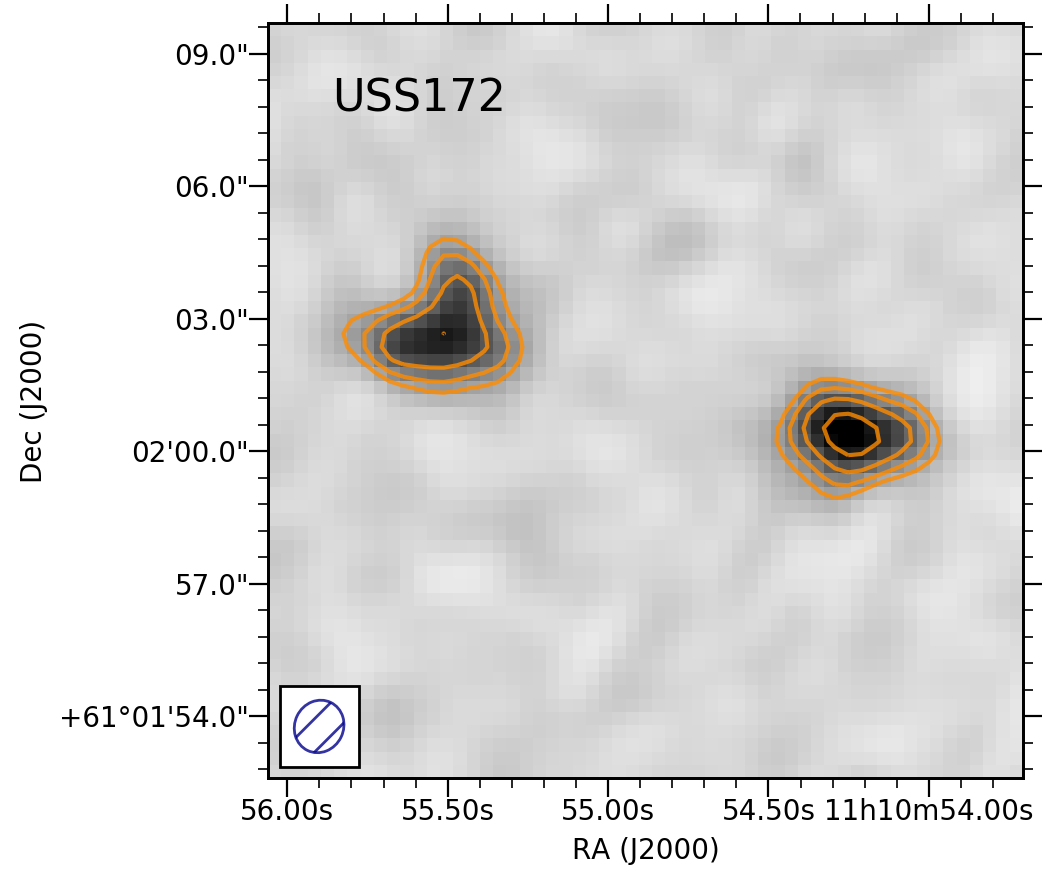}

\includegraphics[width=.3\textwidth]{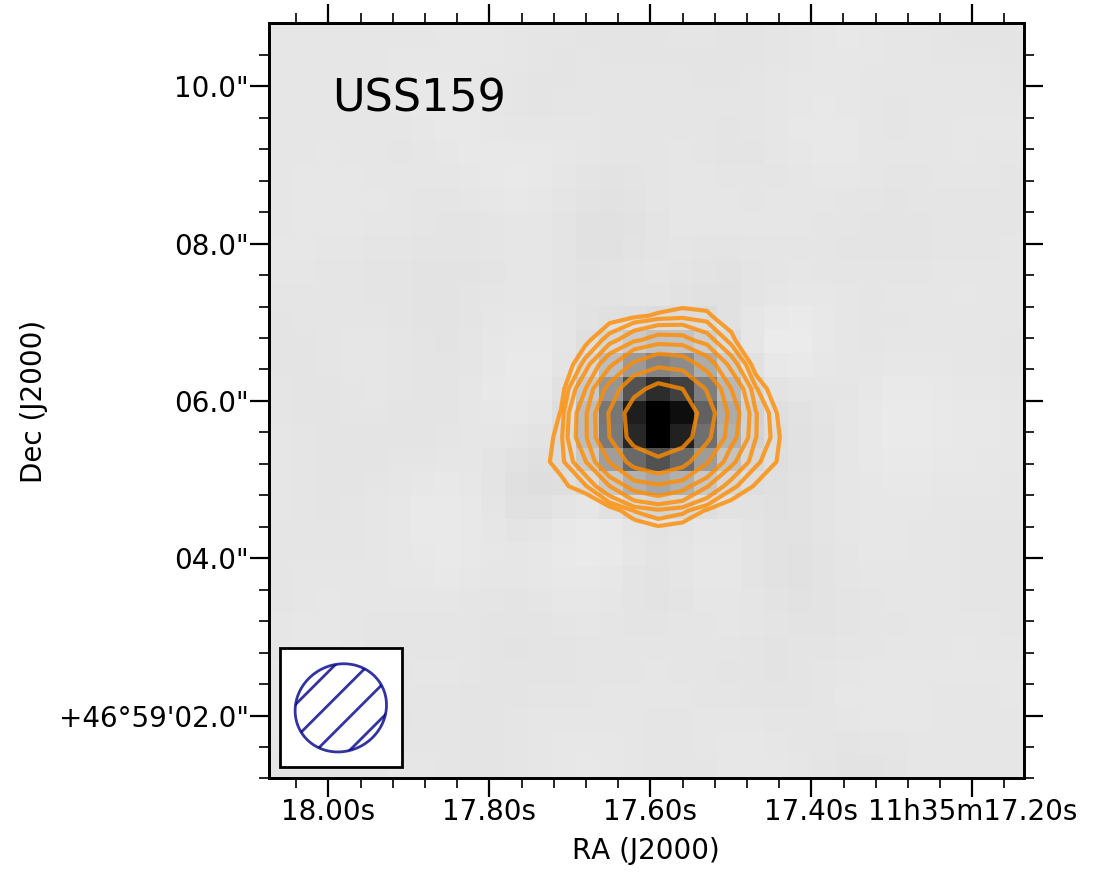}
\includegraphics[width=.35\textwidth]{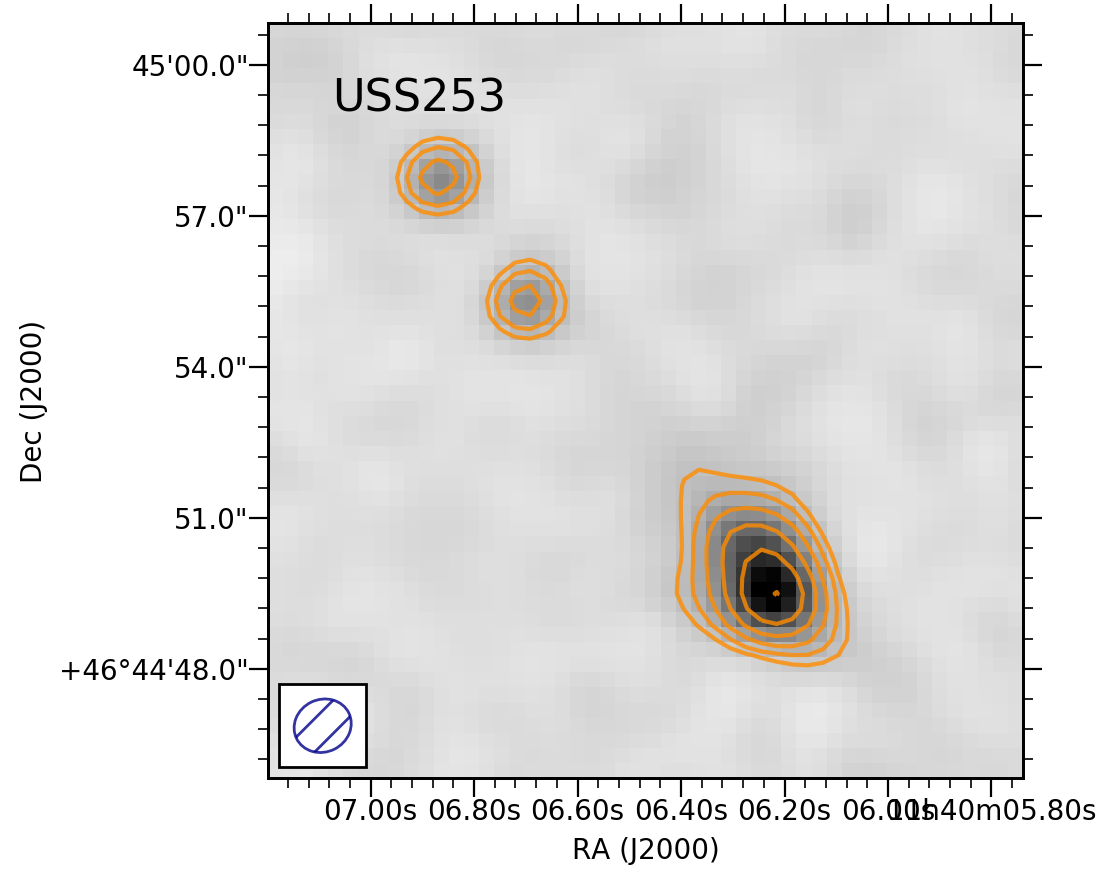}
\includegraphics[width=.3\textwidth]{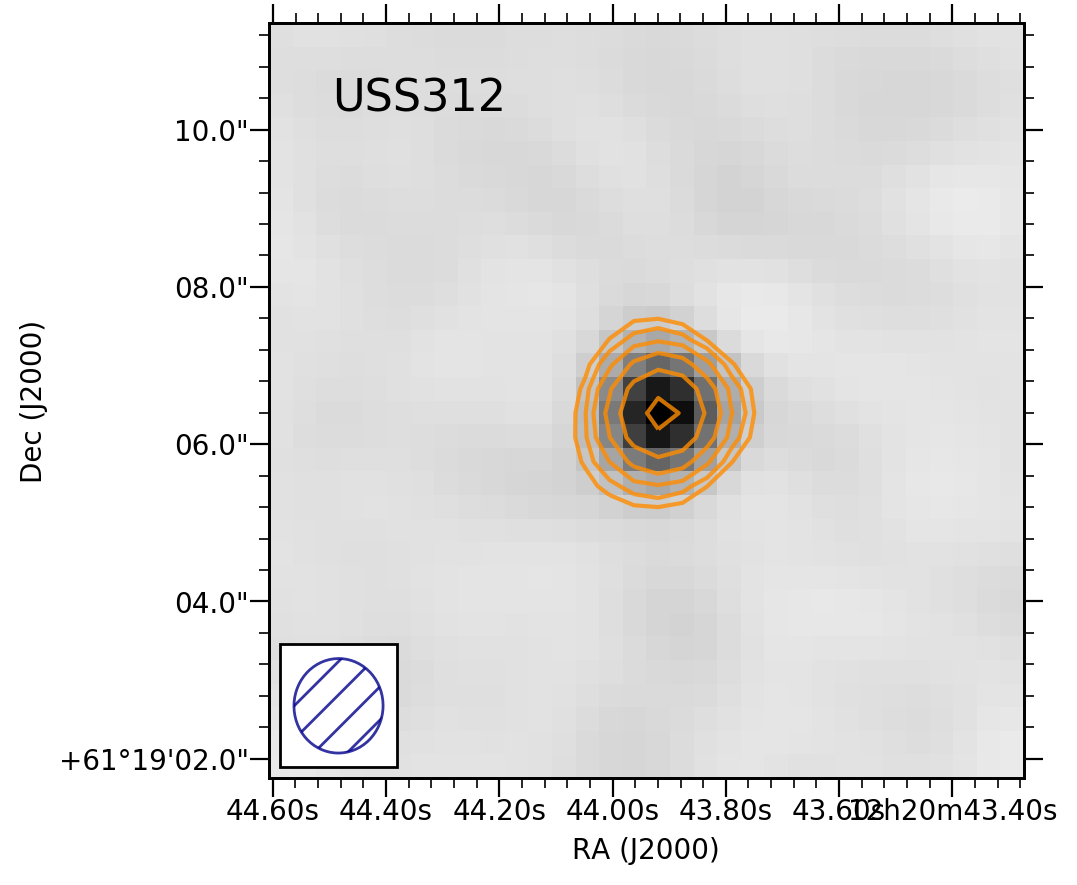}
\caption{Continued.}
\end{figure*}

\begin{figure*}
\ContinuedFloat
\captionsetup{list=off,format=cont}
\centering
\includegraphics[width=.4\textwidth]{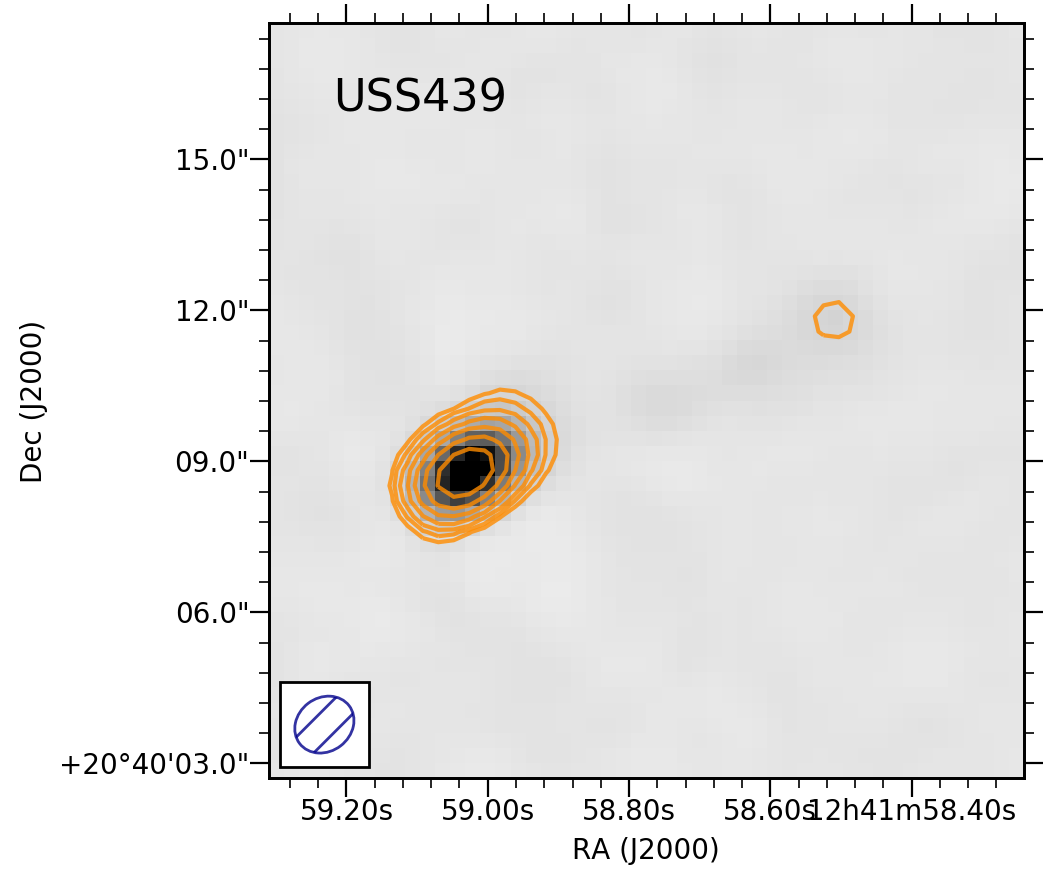}

\includegraphics[width=.34\textwidth]{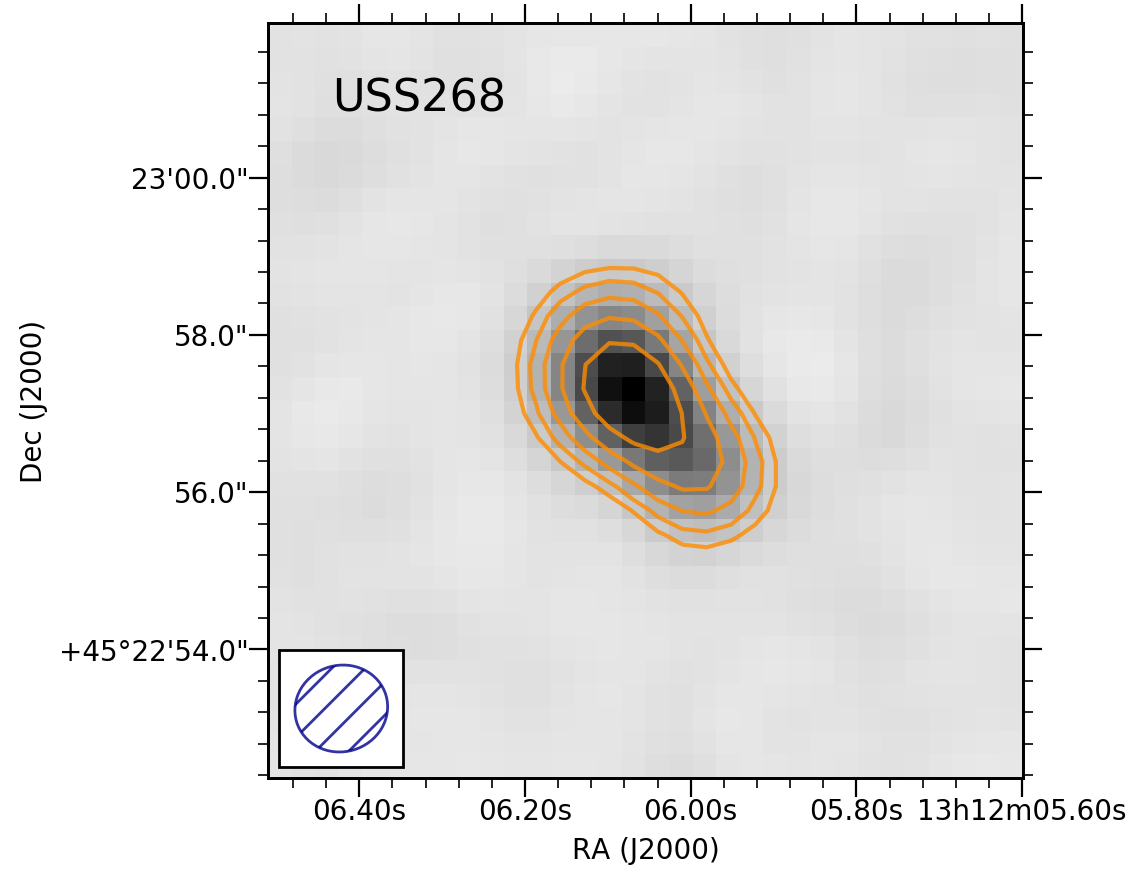}
\includegraphics[width=.322\textwidth]{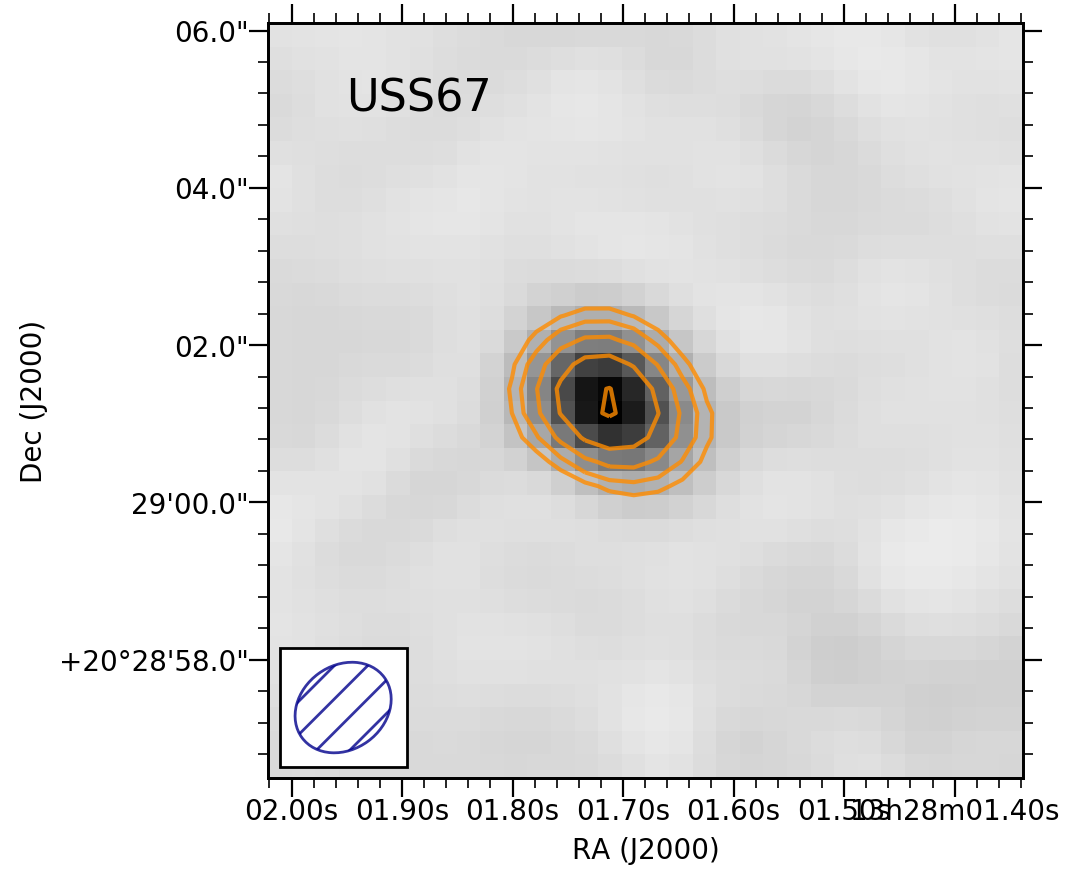}
\includegraphics[width=.325\textwidth]{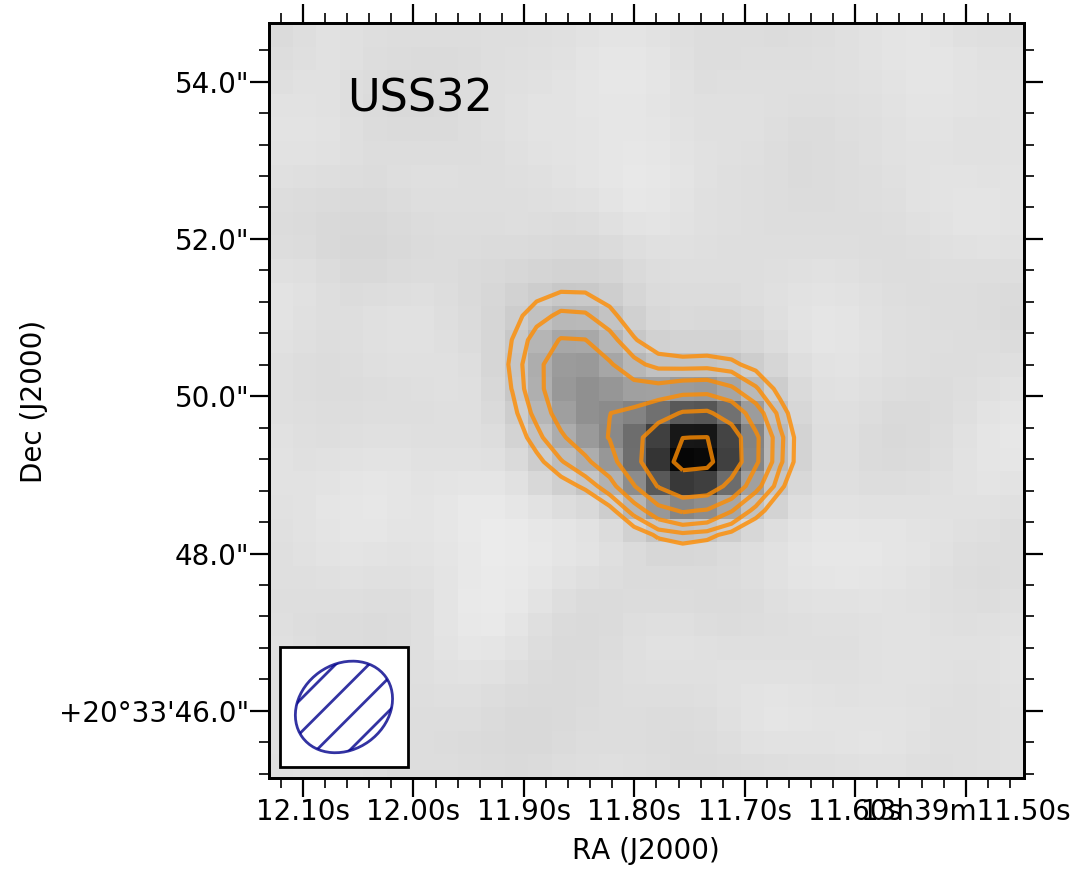}

\includegraphics[width=.4\textwidth]{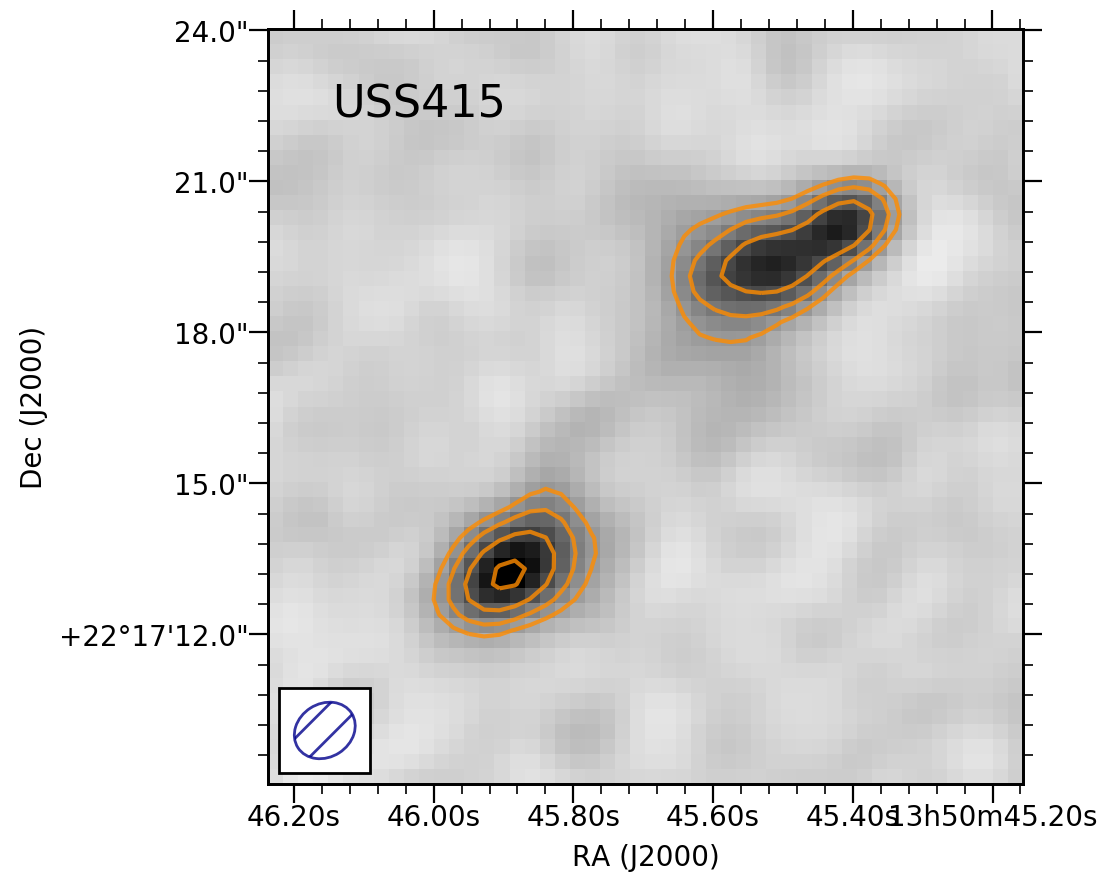}
\includegraphics[width=.4\textwidth]{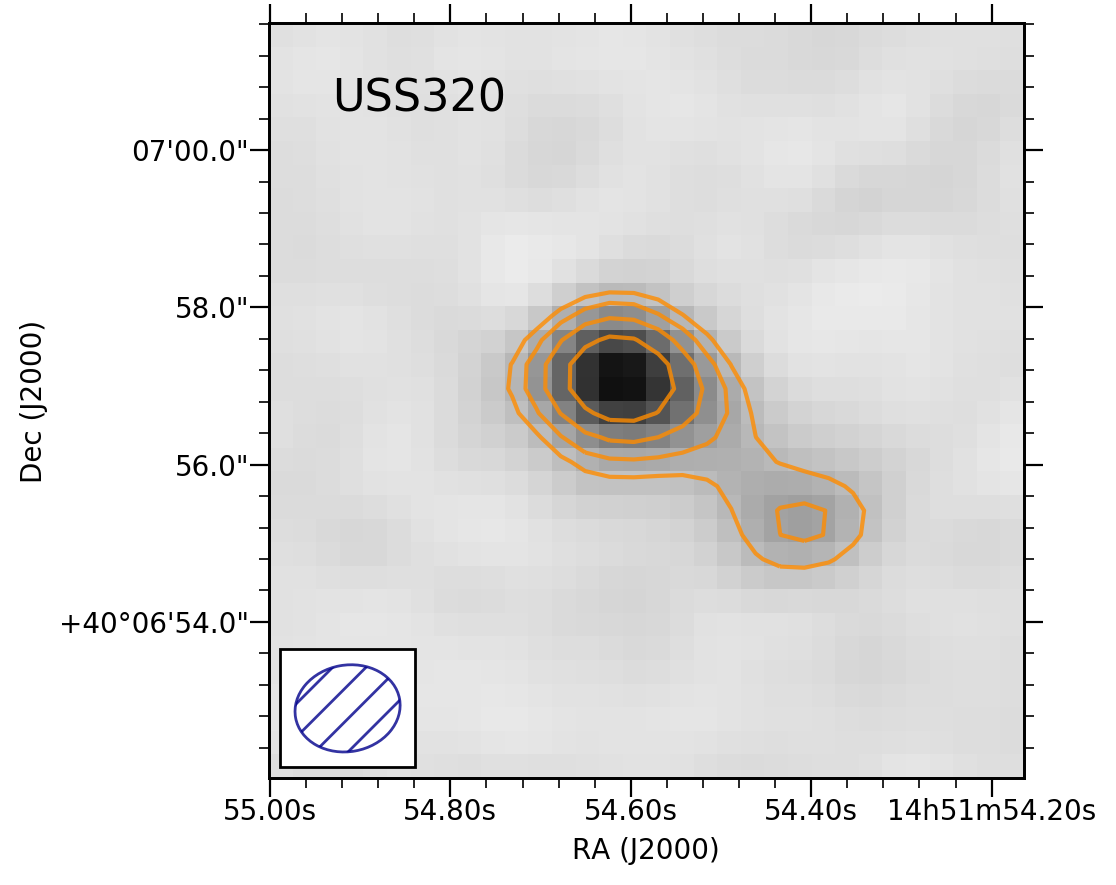}

\includegraphics[width=.32\textwidth]{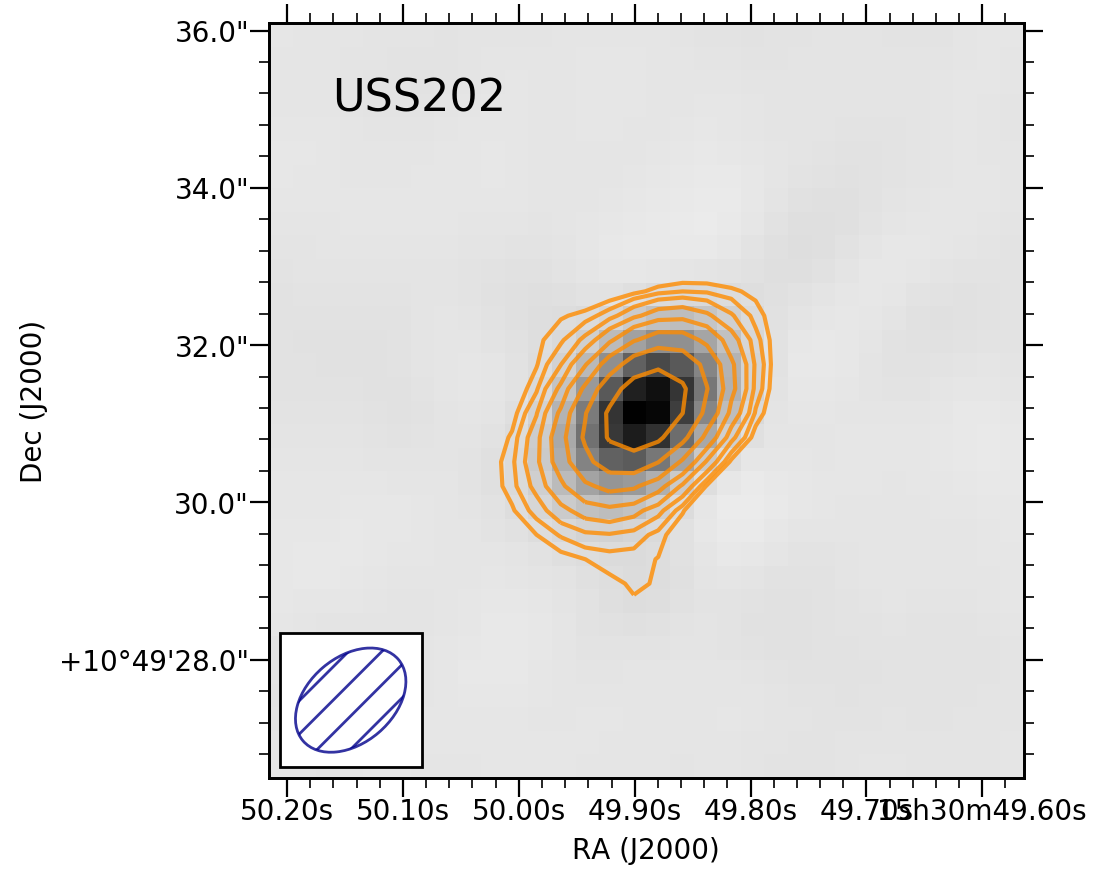}
\includegraphics[width=.32\textwidth]{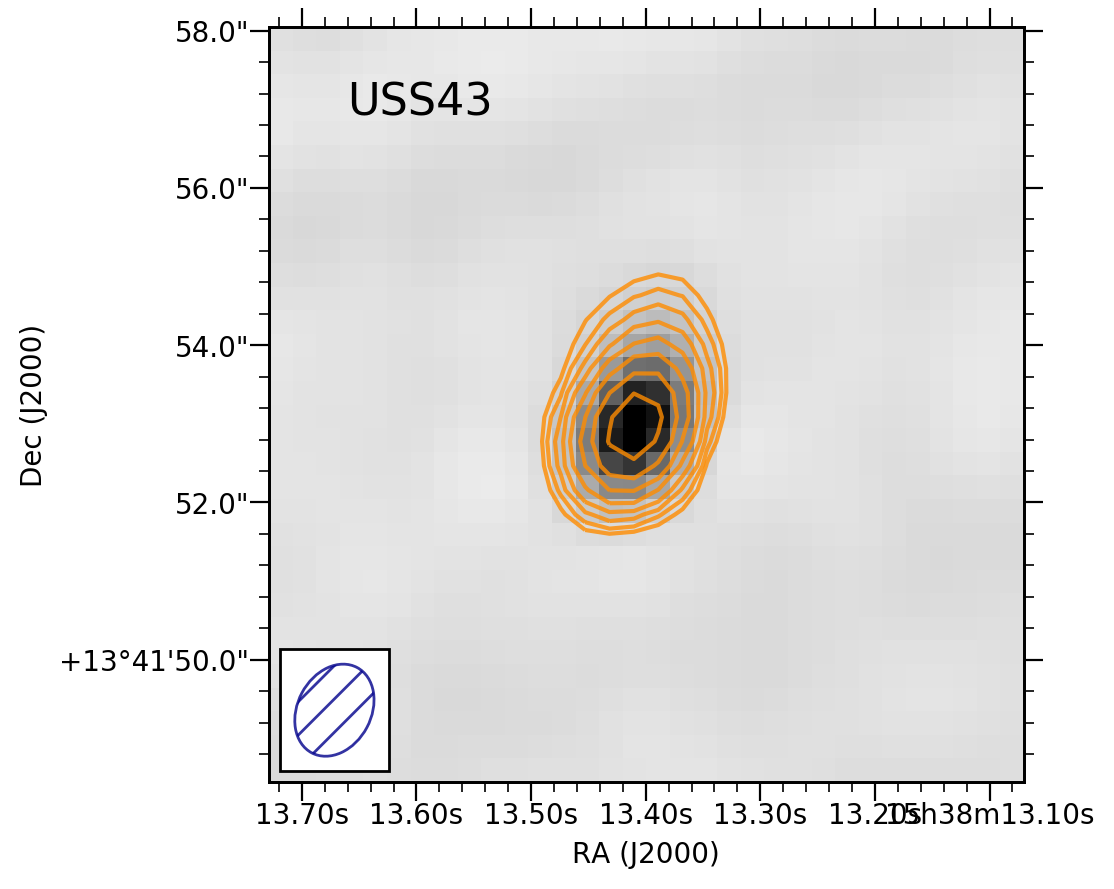}
\includegraphics[width=.32\textwidth]{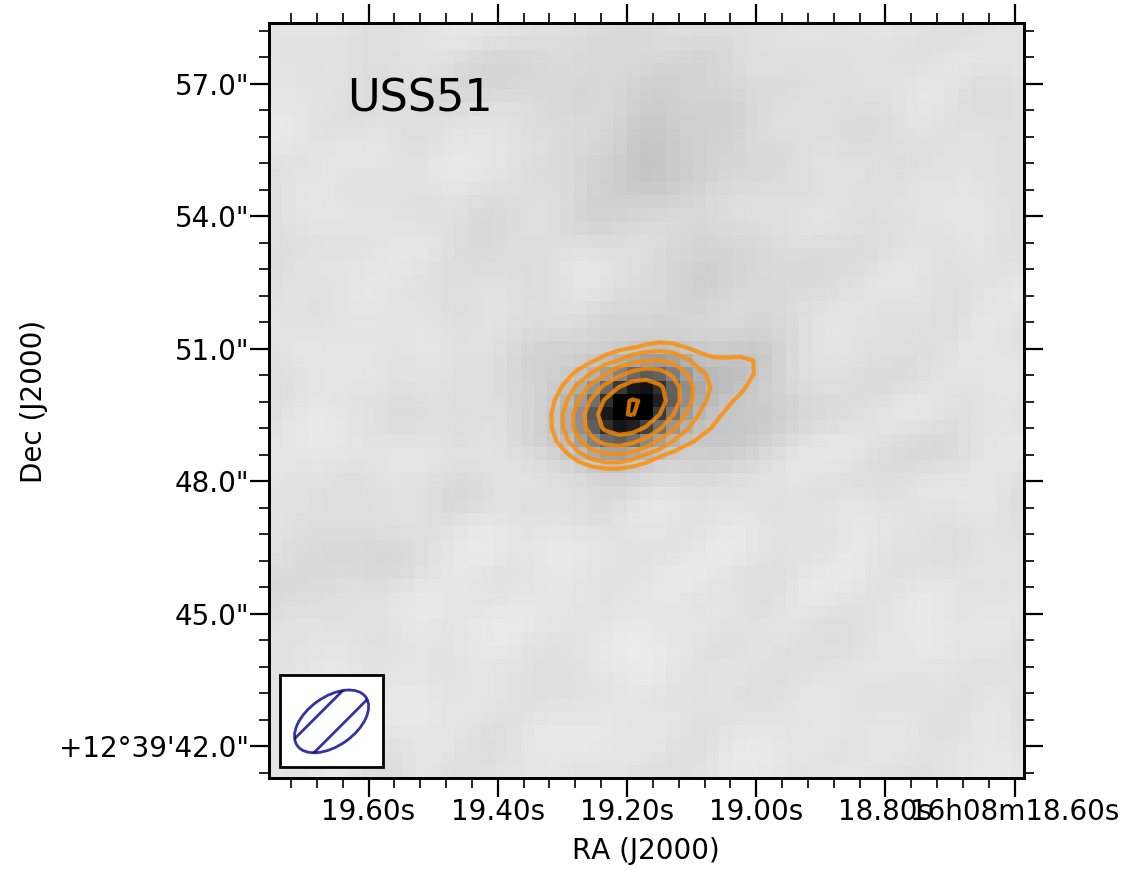}
\caption{Continued.}
\end{figure*}

\begin{figure*}
\ContinuedFloat
\captionsetup{list=off,format=cont}
\centering
\includegraphics[width=.41\textwidth]{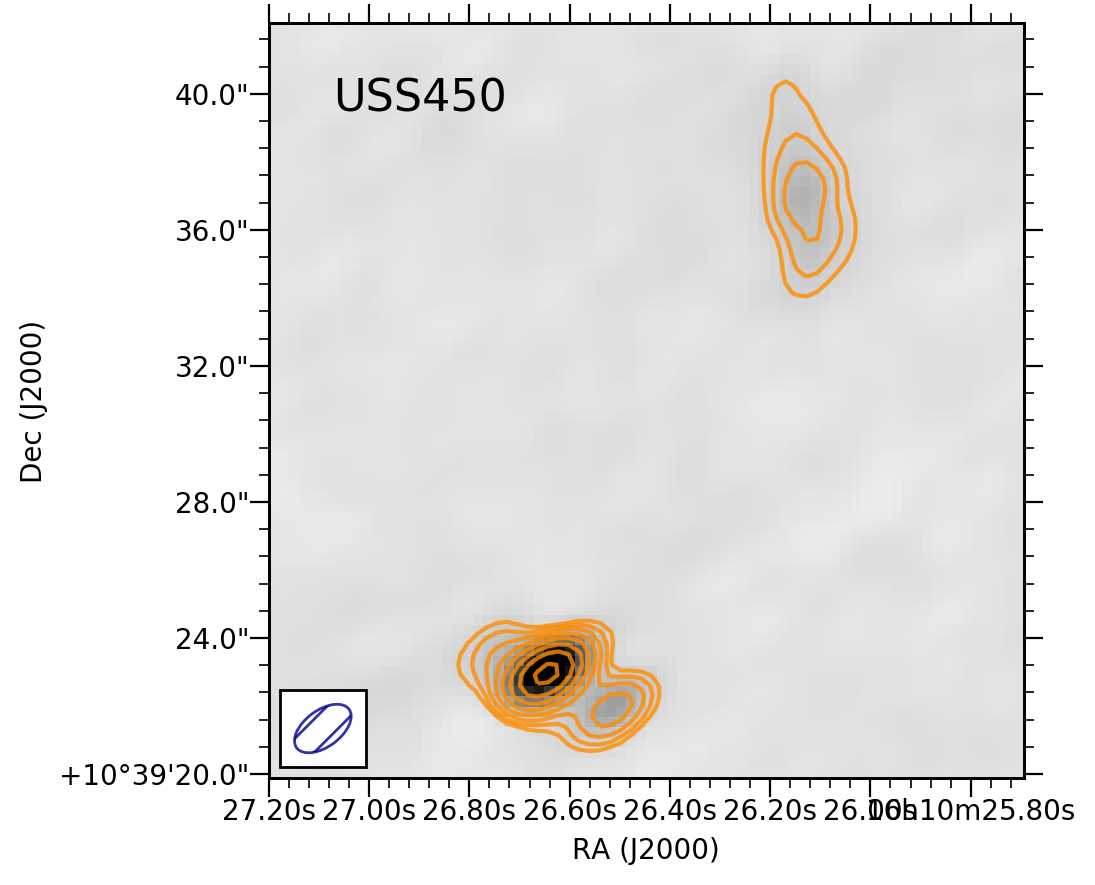}
\includegraphics[width=.4\textwidth]{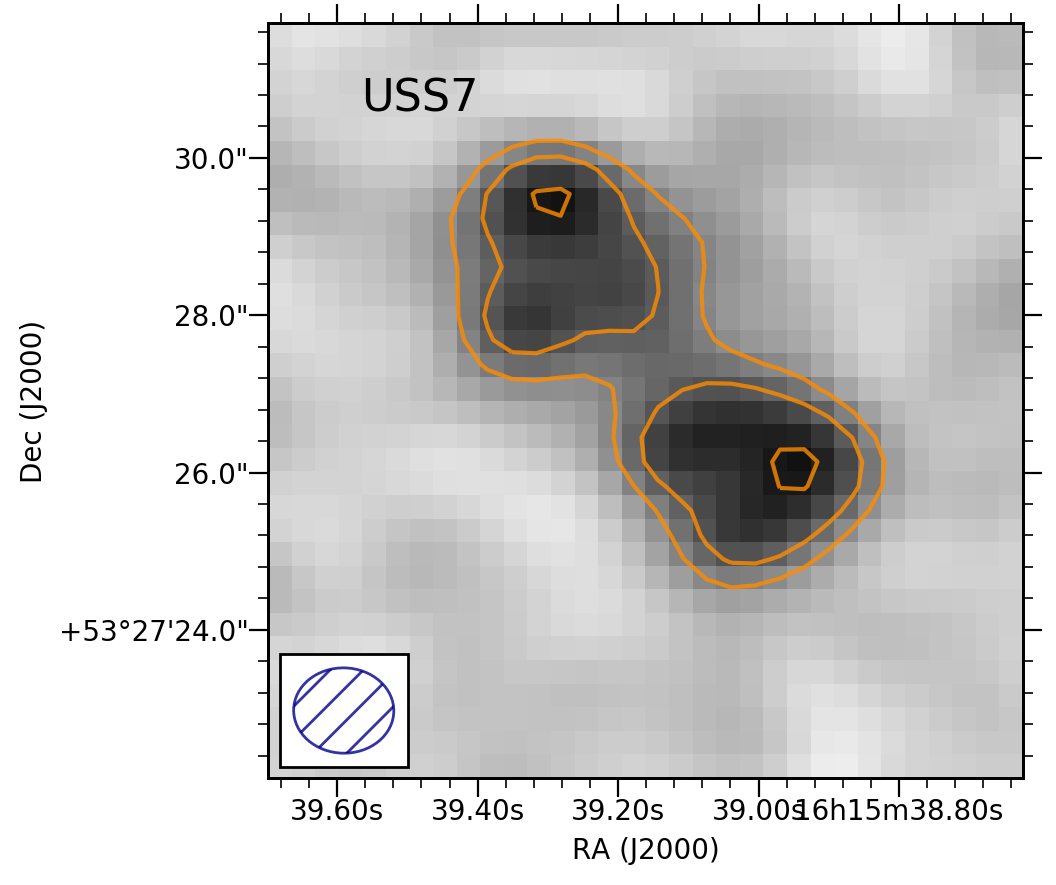}

\includegraphics[width=.34\textwidth]{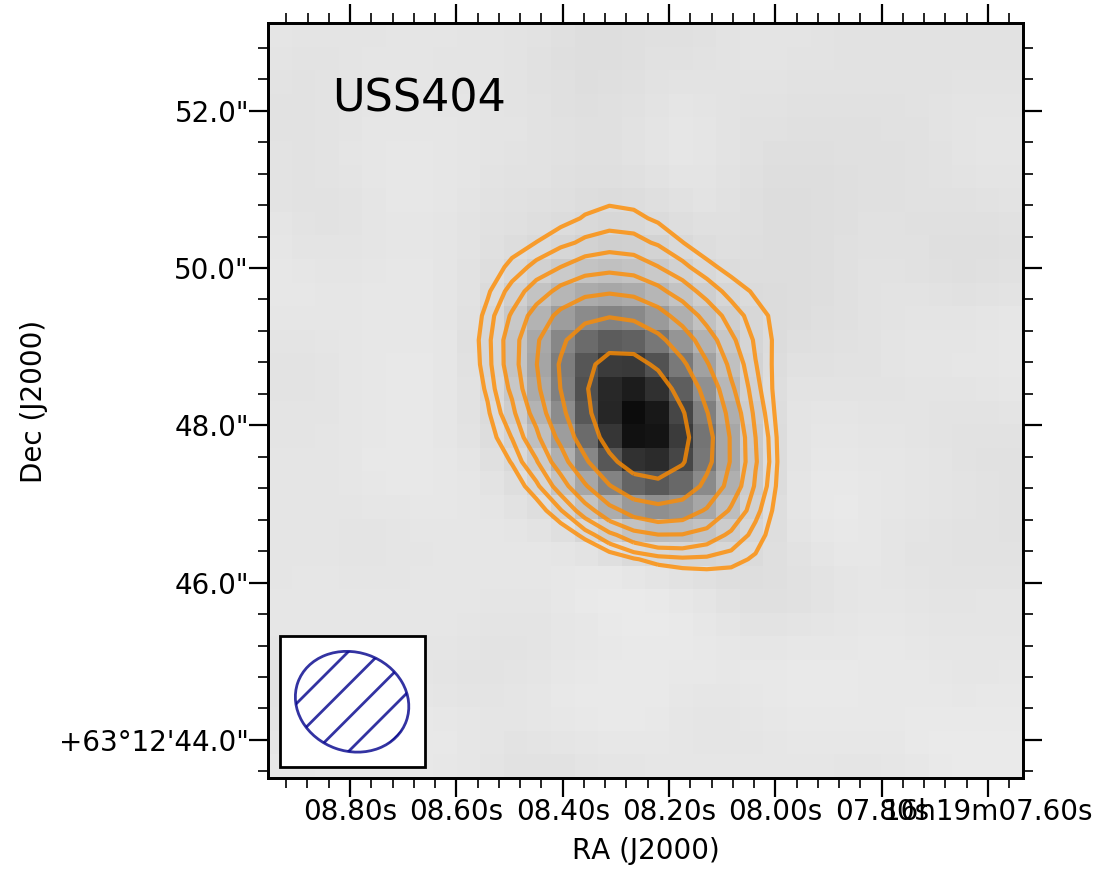}
\includegraphics[width=.33\textwidth]{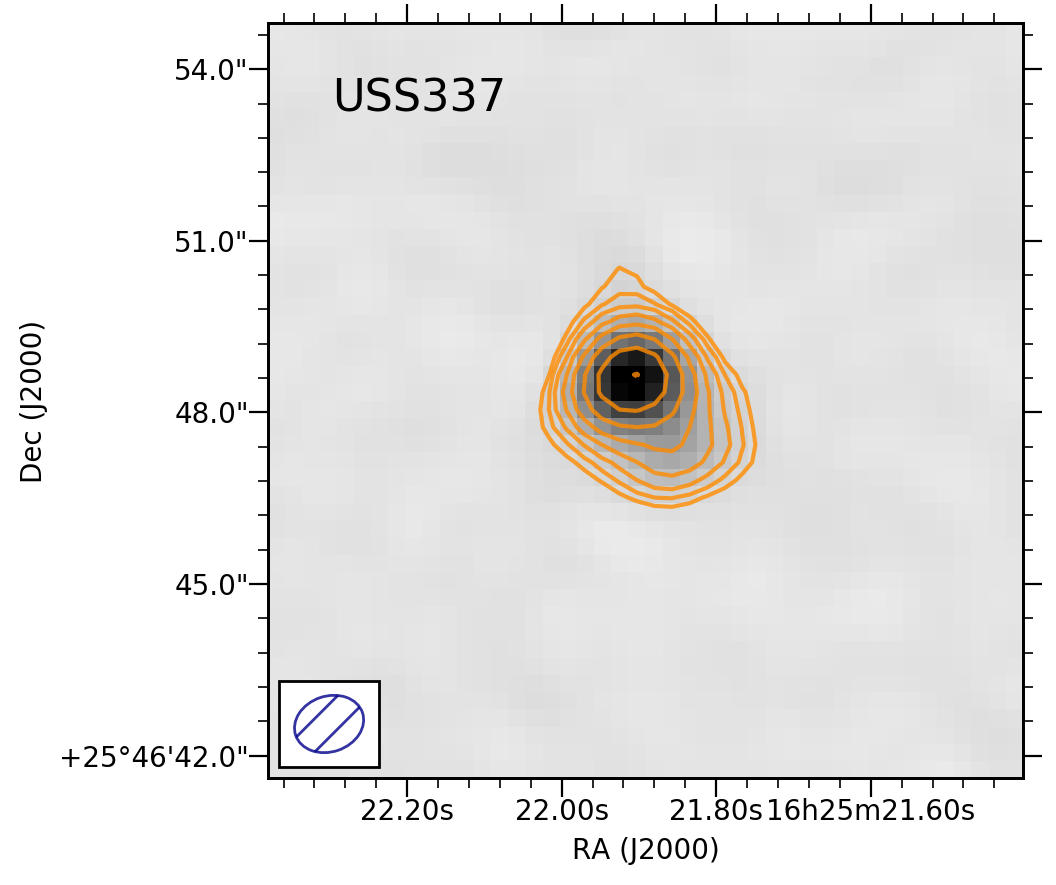}

\includegraphics[width=.33\textwidth]{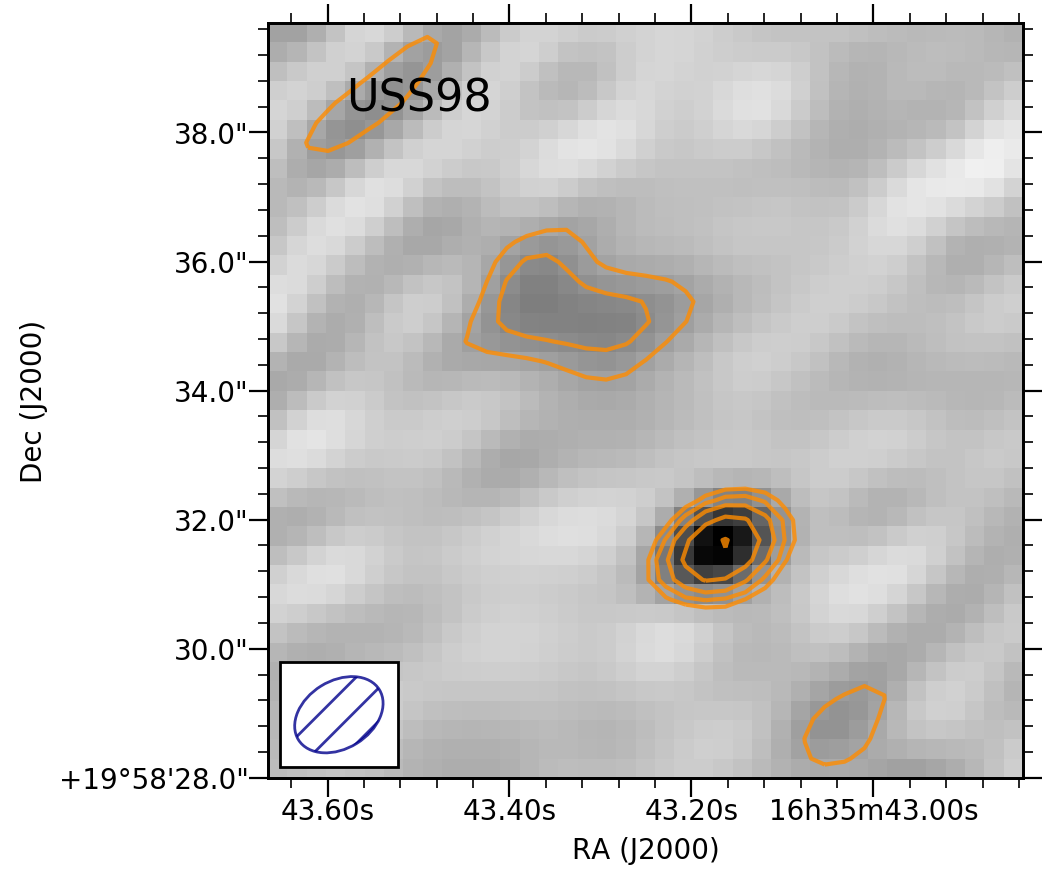}
\caption{Continued.}
\end{figure*}



\bsp	
\label{lastpage}
\end{document}